\documentclass{article}
\usepackage{amsmath}
\usepackage{amssymb}
\usepackage{amsthm}
\usepackage{cite}
\usepackage[arrow,curve,matrix]{xy}
\begin{document}
\def\e#1\e{\begin{equation}#1\end{equation}}
\def\ea#1\ea{\begin{align}#1\end{align}}
\def\eq#1{{\rm(\ref{#1})}}
\theoremstyle{plain}
\newtheorem{thm}{Theorem}[section]
\newtheorem{lem}[thm]{Lemma}
\newtheorem{prop}[thm]{Proposition}
\newtheorem{cor}[thm]{Corollary}
\theoremstyle{definition}
\newtheorem{dfn}[thm]{Definition}
\newtheorem{ex}[thm]{Example}
\newtheorem{ass}[thm]{Assumption}
\newtheorem{rem}[thm]{Remark}
\newtheorem{cond}[thm]{Condition}
\def\dim{\mathop{\rm dim}\nolimits}
\def\Ker{\mathop{\rm Ker}}
\def\Re{\mathop{\rm Re}}
\def\Im{\mathop{\rm Im}}
\def\Spec{\mathop{\rm Spec}\nolimits}
\def\Sch{\mathop{\rm Sch}\nolimits}
\def\Stab{\mathop{\rm Stab}\nolimits}
\def\coh{\mathop{\rm coh}}
\def\ad{\mathop{\rm ad}\nolimits}
\def\Hom{\mathop{\rm Hom}\nolimits}
\def\Aut{\mathop{\rm Aut}}
\def\End{\mathop{\rm End}}
\def\CF{\mathop{\rm CF}\nolimits}
\def\CFi{\mathop{\rm CF}\nolimits^{\rm ind}}
\def\SF{\mathop{\rm SF}\nolimits}
\def\SFa{\mathop{\rm SF}\nolimits_{\rm al}}
\def\SFai{\mathop{\rm SF}\nolimits_{\rm al}^{\rm ind}}
\def\oSFa{\mathop{\bar{\rm SF}}\nolimits_{\rm al}}
\def\Ht{{\mathcal H}{}^{\rm to}}
\def\bHt{\bar{\mathcal H}{}^{\rm to}}
\def\Lt{{\mathcal L}{}^{\rm to}}
\def\bLt{\bar{\mathcal L}{}^{\rm to}}
\def\Oss{\mathop{\rm Obj\kern .1em}\nolimits_{\rm ss}}
\def\Ext{\mathop{\rm Ext}\nolimits}
\def\id{\mathop{\rm id}\nolimits}
\def\fObj{\mathop{\mathfrak{Obj}\kern .05em}\nolimits}
\def\bdim{{\mathbin{\bf dim}\kern .1em}}
\def\modA{\text{\rm mod-$A$}}
\def\modKQ{\text{\rm mod-$\K Q$}}
\def\modKQI{\text{\rm mod-$\K Q/I$}}
\def\nilKQ{\text{\rm nil-$\K Q$}}
\def\nilKQI{\text{\rm nil-$\K Q/I$}}
\def\bs{\boldsymbol}
\def\ge{\geqslant}
\def\le{\leqslant\nobreak}
\def\pr{{\mathop{\preceq}\nolimits}}
\def\N{{\mathbin{\mathbb N}}}
\def\R{{\mathbin{\mathbb R}}}
\def\Z{{\mathbin{\mathbb Z}}}
\def\Q{{\mathbin{\mathbb Q}}}
\def\C{{\mathbin{\mathbb C}}}
\def\CP{{\mathbin{\mathbb{CP}}}}
\def\K{{\mathbin{\mathbb K\kern .05em}}}
\def\A{{\mathbin{\mathcal A}}}
\def\H{{\mathbin{\mathcal H}}}
\def\L{{\mathbin{\mathcal L}}}
\def\M{{\mathcal M}}
\def\P{{\mathbin{\mathcal P}}}
\def\T{{\mathbin{\mathcal T}}}
\def\fF{{\mathbin{\mathfrak F}}}
\def\sIp{{\smash{\sst(I,\pr)}}}
\def\dss{\de_{\rm ss}}
\def\bdss{\bar\de_{\rm ss}}
\def\Iss{I_{\rm ss}}
\def\Mss{{\mathcal M}_{\rm ss}}
\def\al{\alpha}
\def\be{\beta}
\def\ga{\gamma}
\def\de{\delta}
\def\bde{\bar\delta}
\def\io{\iota}
\def\ep{\epsilon}
\def\bep{\bar\epsilon}
\def\la{\lambda}
\def\ka{\kappa}
\def\th{\theta}
\def\ze{\zeta}
\def\up{\upsilon}
\def\vp{\varphi}
\def\si{\sigma}
\def\om{\omega}
\def\De{\Delta}
\def\La{\Lambda}
\def\Om{\Omega}
\def\Ga{\Gamma}
\def\Si{\Sigma}
\def\Th{\Theta}
\def\Up{\Upsilon}
\def\pd{\partial}
\def\ts{\textstyle}
\def\sst{\scriptscriptstyle}
\def\w{\wedge}
\def\sm{\setminus}
\def\bu{\bullet}
\def\op{\oplus}
\def\ot{\otimes}
\def\bigop{\bigoplus}
\def\bigot{\bigotimes}
\def\iy{\infty}
\def\ra{\rightarrow}
\def\ab{\allowbreak}
\def\longra{\longrightarrow}
\def\t{\times}
\def\ci{\circ}
\def\el{{\mathbin{\ell\kern .08em}}}
\def\ti{\tilde}
\def\d{{\rm d}}
\def\ha{{\ts\frac{1}{2}}}
\def\md#1{\vert #1 \vert}
\def\bmd#1{\big\vert #1 \big\vert}
\title{Holomorphic generating functions for invariants counting
coherent sheaves on Calabi--Yau 3-folds}
\author{Dominic Joyce}
\date{}
\maketitle

\begin{abstract} Let $X$ be a Calabi--Yau 3-fold, $\T=D^b(\coh(X))$
the derived category of coherent sheaves on $X$, and $\Stab(\T)$ the
complex manifold of Bridgeland stability conditions on $\T$. It is
conjectured that one can define invariants $J^\al(Z,\P)\in\Q$ for
$(Z,\P)\in\Stab(\T)$ and $\al\in K(\T)$ generalizing
Donaldson--Thomas invariants, which `count' $(Z,\P)$-semistable
(complexes of) coherent sheaves on $X$, and whose transformation law
under change of $(Z,\P)$ is known.

This paper explains how to combine such invariants $J^\al(Z,\P)$, if
they exist, into a family of holomorphic generating functions
$F^\al:\Stab(\T)\ra\C$ for $\al\in K(\T)$. Surprisingly, requiring
the $F^\al$ to be continuous and holomorphic determines them
essentially uniquely, and implies they satisfy a p.d.e., which can
be interpreted as the flatness of a connection over $\Stab(\T)$ with
values in an infinite-dimensional Lie algebra~$\L$.

The author believes that underlying this mathematics there should be
some new physics, in String Theory and Mirror Symmetry. String
Theorists are invited to work out and explain this new physics.
\end{abstract}

\baselineskip 11.8pt plus .15pt

\section{Introduction}
\label{gf1}

To set the scene we start with an analogy, which is explained by
McDuff and Salamon \cite{McSa}. If $(M,\om)$ is a compact symplectic
manifold, one can define the {\it Gromov--Witten invariants\/}
$\Phi_A(\al,\be,\ga)$ of $M$. It is natural to encode these in a
{\it holomorphic generating function} ${\cal S}:H^{\rm
ev}(M,\C)\ra\C$ called the {\it Gromov--Witten potential}, given by
a (formal) power series with coefficients the $\Phi_A(\al,\be,\ga)$.
Identities on the $\Phi_A(\al,\be,\ga)$ imply that ${\cal S}$
satisfies a p.d.e., the {\it WDVV equation}. This p.d.e.\ can be
interpreted as the flatness of a 1-parameter family of connections
defined using ${\cal S}$, which make $H^{\rm ev}(M,\C)$ into a {\it
Frobenius manifold}.

The goal of this paper (which we do not achieve) is to tell a story
with many similar features. Let $X$ be a Calabi--Yau 3-fold,
$\coh(X)$ the abelian category of coherent sheaves on $X$, and
$\T=D^b(\coh(X))$ its bounded derived category. Let $K(\T)$ be the
image of the Chern character map $K_0(\T)\ra H^{\rm even}(X,\Q)$, a
lattice of finite rank. Define the {\it Euler form\/}
$\bar\chi:K(\T)\t K(\T)\ra\Z$ by
\e
\ts\sum_{k\in\Z}(-1)^k\dim \Hom_\T(U,V[k])=\bar\chi([U],[V])
\quad\text{for all $U,V\in\T$.}
\label{gf1eq1}
\e
Then $\bar\chi$ is biadditive, antisymmetric, and nondegenerate.

Following Bridgeland \cite{Brid1} one can define {\it stability
conditions} $(Z,\P)$ on the triangulated category $\T$, consisting
of a group homomorphism $Z:K(\T)\ra\C$ called the {\it central
charge}, and extra data $\P$ encoding the $(Z,\P)$-{\it semistable
objects\/} in $\T$. The family of stability conditions $\Stab(\T)$
is a finite-dimensional complex manifold, with the map
$(Z,\P)\mapsto Z$ a local biholomorphism
$\Stab(\T)\ra\Hom(K(\T),\C)$. In String Theory terms, the `stringy
K\"ahler moduli space' of $X$ should be thought of as a complex
Lagrangian submanifold of $\Stab(\T)$, the subset of stability
conditions represented by Super Conformal Field Theories.

We would like to define invariants $J^\al(Z,\P)\in\Q$ `counting'
$(Z,\P)$-semistable objects in each class $\al\in K(\T)\sm\{0\}$, so
roughly counting semistable sheaves. In the final version of the
theory these should be extensions of {\it Donaldson--Thomas
invariants} \cite{DoTh,Thom} and invariant under deformations of
$X$, but for the present we may make do with the author's `motivic'
invariants defined in \cite{Joyc6} for the abelian category case,
which are not invariant under deformations of~$X$.

The important thing about the invariants $J^\al(Z,\P)$ is that their
transformation laws under change of stability condition are known
completely, and described in the abelian case in \cite{Joyc6}.
Basically $J^\al(Z,\P)$ is a locally constant function of $(Z,\P)$,
except that when $(Z,\P)$ crosses a locus in $\Stab(\T)$ where
$\al=\al_1+\cdots+\al_n$ for $\al_k\in K(\T)$ and
$Z(\al_1),\ldots,Z(\al_n)$ all have the same phase ${\rm e}^{i\phi}$
in $\C\sm\{0\}$, then $J^\al(Z,\P)$ jumps by a multiple
of~$J^{\al_1}(Z,\P)\cdots J^{\al_n}(Z,\P)$.

This paper studies the problem of how best to combine such
invariants $J^\al(Z,\P)$ into generating functions which should be
continuous, holomorphic functions of $(Z,\P)$ on $\Stab(\T)$, a bit
like the Gromov--Witten potential. In fact we shall define a
function $F^\al:\Stab(\T)\ra\C$ for each $\al\in K(\T)\sm\{0\}$,
given by
\e
\begin{split}
F^\al(Z,\P)=\!\!\!\!\!\!\sum_{\substack{n\ge
1,\;\al_1,\ldots,\al_n\in K(\T)\sm\{0\}:\\
\text{$\al_1+\cdots+\al_n=\al$, $Z(\al_k)\ne 0$ all
$k$}}}\!\!\!\!\!\!\!\!\!\!\!\!\!\!\! F_n\bigl(Z(\al_1),
\ldots,Z(\al_n)\bigr) \prod_{i=1}^nJ^{\al_i}(Z,\P)\cdot
\\
\raisebox{-6pt}{\begin{Large}$\displaystyle\biggl[$\end{Large}}
\frac{1}{2^{n-1}}\!\!\!\!\!
\sum_{\substack{\text{connected, simply-connected digraphs
$\Ga$:}\\
\text{vertices $\{1,\ldots,n\}$, edge $\mathop{\bu} \limits^{\sst
i}\ra\mathop{\bu}\limits^{\sst j}$ implies $i<j$}}} \,\,\,
\prod_{\substack{\text{edges}\\
\text{$\mathop{\bu}\limits^{\sst i}\ra\mathop{\bu}\limits^{\sst
j}$}\\ \text{in $\Ga$}}}\bar\chi(\al_i,\al_j)
\raisebox{-6pt}{\begin{Large}$\displaystyle\biggr]$\end{Large}},
\end{split}
\label{gf1eq2}
\e
where $F_n:(\C^\t)^n\ra\C$ are some functions to be determined, and
$\C^\t=\C\sm\{0\}$. Here the sum over graphs comes from the
transformation laws \eq{gf2eq21} below for the $J^\al(Z,\P)$,
determined in the abelian case in~\cite[\S 6.5]{Joyc6}.

Let us admit at once that there are two very major issues about
\eq{gf1eq2} that this paper does not even attempt to solve, which is
why the goals of the paper are not achieved. The first is that we do
not define the invariants $J^\al(Z,\P)$. In the abelian category
case $\A=\coh(X)$, for Gieseker type stability conditions
$(\tau,T,\le)$, we do define and study such invariants $J^\al(\tau)$
in \cite{Joyc6}. But the extension to Bridgeland stability
conditions on $D^b(\coh(X))$ still requires a lot of work.

The second issue is the {\it convergence\/} of the infinite sum
\eq{gf1eq2}, and of other infinite sums below. I am not at all
confident about this: it may be that \eq{gf1eq2} does not converge
at all, or does so only in special limiting corners of $\Stab(\T)$,
and I am not going to conjecture that \eq{gf1eq2} or other sums
converge. Instead, we shall simply treat our sums as convergent.
This means that the results of this paper are rigorous and the sums
known to converge only in rather restricted situations: working with
abelian categories $\A$ rather than triangulated categories $\T$,
and imposing finiteness conditions on $\A$ that do not hold for
coherent sheaves $\A=\coh(X)$, but do work for categories of quiver
representations~$\A=\modKQ$.

The question we do actually answer in this paper is the following.
Suppose for the moment that \eq{gf1eq2} converges in as strong a
sense as necessary. What are the conditions on the functions
$F_n(z_1,\ldots,z_n)$ for $F^\al$ to be both continuous and
holomorphic? Since the $J^\al(Z,\P)$ are not continuous in $(Z,\P)$,
to make $F^\al$ continuous the $F_n$ must have discontinuities
chosen so that the jumps in $J^\al(Z,\P)$ and $F_n$ exactly cancel.
The simplest example of this is that $F_n(z_1,\ldots,z_n)$ must jump
by $F_{n-1}(z_1,\ldots,z_{l-1},z_l+z_{l+1},z_{l+2},\ldots,z_n)$
across the real hypersurface $z_{l+1}/z_l\in(0,\iy)$ in~$(\C^\t)^n$.

We shall show that the condition that $F^\al$ be holomorphic and
continuous, plus a few extra assumptions on the symmetry and growth
of the $F_n$ and the normalization $F_1\equiv(2\pi i)^{-1}$,
actually determine the $F_n$ {\it uniquely}. Furthermore, on the
open subset of $(\C^\t)^n$ where $F_n$ is continuous it satisfies
the p.d.e.
\e
\d F_n(z_1,\ldots,z_n)=\sum_{k=1}^{n-1}
\begin{aligned}[t]
&F_k(z_1,\ldots,z_k)F_{n-k}(z_{k+1},\ldots,z_n)\,\cdot\\
&\left[\frac{\d z_{k+1}+\cdots+\d z_n}{z_{k+1}+\cdots+z_n} -\frac{\d
z_1+\cdots+\d z_k}{z_1+\cdots+z_k}\right].
\end{aligned}
\label{gf1eq3}
\e
This in turn implies that the generating functions $F^\al$ satisfy
the p.d.e.
\e
\d F^\al(Z,\P)=-\sum_{\be,\ga\in
K(\T)\sm\{0\}:\al=\be+\ga}\bar\chi(\be,\ga)F^\be(Z,\P)F^\ga(Z,\P)
\frac{\d(Z(\be))}{Z(\be)}.
\label{gf1eq4}
\e

It seems remarkable that simply requiring the $F^\al$ to be
holomorphic and continuous implies they must satisfy the p.d.e.\
\eq{gf1eq4}, which has appeared more-or-less out of nowhere. In the
Gromov--Witten case the generating function $\mathcal S$ also
satisfies a p.d.e., the WDVV equation. Note however that the WDVV
equation holds because of identities upon Gromov--Witten invariants,
but in our case \eq{gf1eq4} holds because of any identities not on
the $J^\al(Z,\P)$ for fixed $(Z,\P)$, but rather because of
identities on how the $J^\al(Z,\P)$ transform as $(Z,\P)$ changes.

Just as the WDVV equation implies the flatness of a connection
constructed using the Gromov--Witten potential, so we can interpret
\eq{gf1eq4} in terms of flat connections. Define $\L$ to be the
$\C$-Lie algebra with basis formal symbols $c^\al$ for $\al\in
K(\T)$, and Lie bracket $[c^\al,c^\be]=\bar\chi(\al,\be)
c^{\al+\be}$. Ignoring questions of convergence, define an
$\L$-valued connection matrix $\Ga$ on $\Stab(\T)$ by
\e
\Ga(Z,\P)=\sum_{\al\in K(\T)\sm\{0\}}F^\al(Z,\P)\,c^\al\ot
\frac{\d(Z(\al))}{Z(\al)}
\label{gf1eq5}
\e
Then \eq{gf1eq4} implies that $\Ga$ is {\it flat}, that is, the
curvature $R_\Ga=\d\Ga+\ha\Ga\w\Ga\equiv 0$. But we do not expect
that $\d\Ga\equiv 0$ and $\Ga\w\Ga\equiv 0$ as happens in the
Gromov--Witten case, so we do not have a 1-parameter family of flat
connections and a Frobenius manifold type structure.

All this cries out for an explanation, but I do not have one.
However, I am convinced that the explanation should be sought in
String Theory, and that underlying this is some new piece of physics
to do with Mirror Symmetry, just as the context of the derived
category $D^b(\coh(X))$ of coherent sheaves on $X$ is the core of
the Homological Mirror Symmetry programme of Kontsevich \cite{Kont}.
So I am posting this paper on the hep-th archive to bring it to the
attention of String Theorists, and I invite any physicists with
ideas on its interpretation to please let me know.

Two possible pointers towards an interpretation are discussed in
\S\ref{gf6}. Firstly, ignoring convergence issues, we show that in
the Calabi--Yau 3-fold triangulated category case the connection
$\Ga$ above induces a flat connection on $T\Stab(\T)$, which is in
fact the Levi-Civita connection of a flat holomorphic metric
$g_{\sst\C}$ on $\Stab(\T)$, provided $g_{\sst\C}$ is nondegenerate.
Secondly, again ignoring convergence issues, for $\la\in\C^\t$ and
fixed $a,b\in\Z$ define a $(0,1)$-form on $\Stab(\T)$ by
\e
\Phi_\la(Z,\P)=\ts\sum_{\al\in K(\T)\sm\{0\}}\la^a{\rm
e}^{\la^bZ(\al)}\,\overline{F^\al(Z,\P)}\,\,
\frac{\overline{\d(Z(\al))}}{\overline{Z(\al)}}.
\label{gf1eq6}
\e
Then \eq{gf1eq4} implies an equation in $(0,2)$-forms
on~$\Stab(\T)$:
\e
\bigl(\bar\pd\Phi_\la(Z,\P)\bigr)_{\bar i\bar j}= -\ha\la^{-a-2b}
(\bar\chi)^{ij}\bigl(\pd\Phi_\la(Z,\P)\bigr)_{i\bar i}
\bigl(\pd\Phi_\la(Z,\P)\bigr)_{j\bar j},
\label{gf1eq7}
\e
using complex tensor index notation, where $(\bar\chi)^{ij}$ is the
$(2,0)$ part of $\bar\chi$. This is a little similar to the {\it
holomorphic anomaly equation\/} of Bershadsky et
al.~\cite{BCOV1,BCOV2}.

Here is a brief description of the paper. Despite this introduction
we mostly work neither with Calabi--Yau 3-folds, nor with
triangulated categories. Instead, we work with abelian categories
$\A$ such as quiver representations $\modKQ$, and slope stability
conditions $(\mu,\R,\le)$ determined by a morphism $Z:K(\A)\ra\C$.
Then we can use the author's series \cite{Joyc3,Joyc4,Joyc5,Joyc6}
on invariants counting $\mu$-semistable objects in abelian
categories; the facts we need are summarized in \S\ref{gf2}. Section
\ref{gf3} studies generating functions $f^\al$ generalizing $F^\al$
in \eq{gf1eq2}, in the abelian category setting, and expressed in
terms of Lie algebras $\L$ following~\cite{Joyc3,Joyc4,Joyc5,Joyc6}.

In \S\ref{gf31} we find conditions on $F_n$ for these $f^\al$ to be
holomorphic and continuous, including some conditions from the
triangulated category case, and show that with a few extra
assumptions any such functions $F_n$ are unique. In \S\ref{gf32} we
guess a p.d.e.\ generalizing \eq{gf1eq4} for the $f^\al$ to satisfy,
deduce that it implies \eq{gf1eq3}, and use \eq{gf1eq3} to construct
a family of functions $F_n$ by induction on $n$. Then \S\ref{gf33}
shows that these $F_n$ constructed using \eq{gf1eq3} satisfy all the
conditions of \S\ref{gf31}, and so are unique. Section \ref{gf4}
discusses $\L$-valued flat connections $\Ga$ as above, and
\S\ref{gf5} the extension to triangulated categories. Finally,
\S\ref{gf6} explains how the ideas work out for Calabi--Yau 3-folds.
\medskip

\noindent{\it Acknowledgements.} I would like to thank Philip
Candelas, Calin Lazaroiu, Bal\'azs Szendr{\accent"7D o}i, Richard
Thomas, and especially Tom Bridgeland for useful conversations. I
was supported by an EPSRC Advanced Research Fellowship whilst
writing this paper.

\section{Background material}
\label{gf2}

The author has written six long, complicated papers
\cite{Joyc1,Joyc2,Joyc3,Joyc4,Joyc5,Joyc6} developing a framework
for studying stability conditions $(\tau,T,\le)$ on an abelian
category $\A$, and interesting invariants counting $\tau$-semistable
objects in $\A$, and the transformation laws of these invariants
under change of stability condition. Sections \ref{gf21}--\ref{gf22}
explain only the minimum necessary for this paper; for much more
detail, see \cite{Joyc1,Joyc2,Joyc3,Joyc4,Joyc5,Joyc6}. Section
\ref{gf23} discusses the extension to triangulated categories.

\subsection{The general set-up of \cite{Joyc3,Joyc4,Joyc5,Joyc6}}
\label{gf21}

We start with a very brief summary of selected parts of the author's
series \cite{Joyc3,Joyc4,Joyc5,Joyc6}. Here \cite[Assumptions 7.1 \&
8.1]{Joyc3} is the data we require.

\begin{ass} Let $\K$ be an algebraically closed field and $\A$
a noetherian abelian category with $\Hom(A,B)=\Ext^0(A,B)$ and
$\Ext^1(A,B)$ finite-dimensional $\K$-vector spaces for all
$A,B\in\A$, and all compositions
$\Ext^i(B,C)\t\Ext^j(A,B)\ra\Ext^{i+j}(A,C)$ bilinear for
$i,j,i+j=0$ or 1. Let $K(\A)$ be the quotient of the Grothendieck
group $K_0(\A)$ by some fixed subgroup. Suppose that if $A\in\A$
with $[A]=0$ in $K(\A)$ then~$A\cong 0$.

To define moduli stacks of objects or configurations in $\A$, we
need some {\it extra data}, to tell us about algebraic families of
objects and morphisms in $\A$, parametrized by a base scheme $U$. We
encode this extra data as a {\it stack in exact categories} $\fF_\A$
on the {\it category of\/ $\K$-schemes} $\Sch_\K$, made into a {\it
site} with the {\it \'etale topology}. The $\K,\A,K(\A),\fF_\A$ must
satisfy some complex additional conditions \cite[Assumptions 7.1 \&
8.1]{Joyc3}, which we do not give.
\label{gf2ass1}
\end{ass}

In \cite[\S 9--\S 10]{Joyc3} we define data $\A,K(\A),\fF_\A$
satisfying Assumption \ref{gf2ass1} in some large classes of
examples, including the abelian category $\coh(X)$ of coherent
sheaves on a projective $\K$-scheme $X$, and the following:

\begin{ex} A {\it quiver\/} $Q$ is a finite directed graph.
That is, $Q$ is a quadruple $(Q_0,Q_1,b,e)$, where $Q_0$ is a finite
set of {\it vertices}, $Q_1$ is a finite set of {\it arrows}, and
$b,e:Q_1\ra Q_0$ are maps giving the {\it beginning} and {\it end\/}
of each arrow.

A {\it representation} $(V,\rho)$ of $Q$ consists of
finite-dimensional $\K$-vector spaces $V_v$ for each $v\in Q_0$, and
linear maps $\rho_a:V_{b(a)}\ra V_{e(a)}$ for each $a\in Q_1$. A
{\it morphism} of representations $\phi:(V,\rho)\ra(W,\si)$ consists
of $\K$-linear maps $\phi_v:V_v\ra W_v$ for all $v\in Q_0$ with
$\phi_{e(a)}\ci\rho_a=\si_a\ci\phi_{b(a)}$ for all $a\in Q_1$. Write
$\modKQ$ for the {\it abelian category of representations} of $Q$.
It is of finite length.

Write $\N^{Q_0}$ and $\Z^{Q_0}$ for the sets of maps $Q_0\!\ra\!\N$
and $Q_0\!\ra\!\Z$, where $\N=\{0,1,2,\ldots\}\subset\Z$. Define the
{\it dimension vector} $\bdim(V,\rho)\!\in\!\N^{Q_0}\subset\Z^{Q_0}$
of $(V,\rho)\in\modKQ$ by $\bdim(V,\rho):v\mapsto\dim_\K V_v$. This
induces a surjective group homomorphism $\bdim:K_0(\modKQ)\!\ra\!
\Z^{Q_0}$. Define $K(\modKQ)$ to be the quotient of $K_0(\modKQ)$ by
the kernel of $\bdim$. Then $K(\modKQ)\cong\Z^{Q_0}$, and for
simplicity we identify $K(\modKQ)$ and $\Z^{Q_0}$, so that for
$(V,\rho)\in\modKQ$ the class $[(V,\rho)]$ in $K(\modKQ)$ is $\bdim
(V,\rho)$. As in \cite[Ex.~10.5]{Joyc3} we can define a stack in
exact categories $\fF_\modKQ$ so that $\A=\modKQ,K(\modKQ),
\fF_\modKQ$ satisfy Assumption~\ref{gf2ass1}.
\label{gf2ex1}
\end{ex}

We will need the following notation \cite[Def.~7.3]{Joyc3},
\cite[Def.~3.8]{Joyc5}:

\begin{dfn} We work in the situation of Assumption \ref{gf2ass1}.
Define
\e
C(\A)=\bigl\{[U]\in K(\A):U\in\A,\;\> U\not\cong 0\bigr\}\subset
K(\A),
\label{gf2eq1}
\e
and $\bar C(\A)=C(\A)\cup\{0\}$. That is, $C(\A)$ is the set of
classes in $K(\A)$ of nonzero objects $U\in\A$, and $\bar C(\A)$ the
set of classes of objects in $\A$. We think of $C(\A)$ as the
`positive cone' and $\bar C(\A)$ as the `closed positive cone' in
$K(\A)$. In Example \ref{gf2ex1} we have $\bar C(\A)=\N^{Q_0}$
and~$C(\A)=\N^{Q_0}\sm\{0\}$.

A set of $\A$-{\it data} is a triple $(I,\pr,\ka)$ such that
$(I,\pr)$ is a finite partially ordered set (poset) and $\ka:I\ra
C(\A)$ a map. In this paper we will be interested only in the case
when $\pr$ is a total order, so that $(I,\pr)$ is uniquely
isomorphic to $(\{1,\ldots,n\},\le)$ for $n=\md{I}$. We {\it
extend\/ $\ka$ to the set of subsets of\/} $I$ by defining
$\ka(J)=\sum_{j\in J}\ka(j)$. Then $\ka(J)\in C(\A)$ for all
$\emptyset\ne J\subseteq I$, as $C(\A)$ is closed under addition.
\label{gf2def1}
\end{dfn}

Then \cite[\S 7]{Joyc3} defines {\it moduli stacks\/} $\fObj_\A$ of
objects in $\A$, and $\fObj^\al_\A$ of objects in $\A$ with class
$\al$ in $K(\A)$, for each $\al\in\bar C(\A)$. They are Artin
$\K$-stacks, locally of finite type, with $\fObj_\A^\al$ an open and
closed $\K$-substack of $\fObj_\A$. The underlying geometric spaces
$\fObj_\A(\K),\fObj_\A^\al(\K)$ are the sets of isomorphism classes
of objects $U$ in $\A$, with $[U]=\al$ for $\fObj_\A^\al(\K)$.

In \cite[\S 4]{Joyc5} we study ({\it weak\/}) {\it stability
conditions} on $\A$, generalizing Rudakov \cite{Ruda}. The next
three definitions are taken from~\cite[Def.s 4.1--4.3, 4.6 \&
4.7]{Joyc5}.

\begin{dfn} Let Assumption \ref{gf2ass1} hold and $C(\A)$ be as in
\eq{gf2eq1}. Suppose $(T,\le)$ is a totally ordered set, and
$\tau:C(\A)\ra T$ a map. We call $(\tau,T,\le)$ a {\it stability
condition} on $\A$ if whenever $\al,\be,\ga\in C(\A)$ with
$\be=\al+\ga$ then either $\tau(\al)\!<\!\tau(\be) \!<\!\tau(\ga)$,
or $\tau(\al)\!>\!\tau(\be)\!>\!\tau(\ga)$, or
$\tau(\al)\!=\!\tau(\be)\!=\!\tau(\ga)$. We call $(\tau,T,\le)$ a
{\it weak stability condition} on $\A$ if whenever $\al,\be, \ga\in
C(\A)$ with $\be=\al+\ga$ then either $\tau(\al)\!\le\!
\tau(\be)\!\le\!\tau(\ga)$, or $\tau(\al)\!\ge\!\tau(\be)\!\ge
\!\tau(\ga)$.
\label{gf2def2}
\end{dfn}

\begin{dfn} Let $(\tau,T,\le)$ be a weak stability condition on
$\A,K(\A)$ as above. Then we say that a nonzero object $U$ in $\A$
is
\begin{itemize}
\setlength{\itemsep}{0pt}
\setlength{\parsep}{0pt}
\item[(i)] $\tau$-{\it semistable} if for all $S\subset U$ with
$S\not\cong 0,U$ we have $\tau([S])\le\tau([U/S])$;
\item[(ii)] $\tau$-{\it stable} if for all $S\subset U$ with
$S\not\cong 0,U$ we have $\tau([S])<\tau([U/S])$; and
\item[(iii)] $\tau$-{\it unstable} if it is not $\tau$-semistable.
\end{itemize}
\label{gf2def3}
\end{dfn}

\begin{dfn} Let Assumption \ref{gf2ass1} hold and $(\tau,T,\le)$
be a weak stability condition on $\A$. For $\al\in C(\A)$ define
\begin{align*}
\Oss^\al(\tau)&=\bigl\{[U]\in\fObj_\A^\al(\K):\text{$U$ is
$\tau$-semistable}\bigr\}\subset\fObj_\A(\K).
\end{align*}
Write $\dss^\al(\tau):\fObj_\A(\K)\ra\{0,1\}$ for its characteristic
function.

We call $(\tau,T,\le)$ a {\it permissible} weak stability condition
if:
\begin{itemize}
\setlength{\itemsep}{0pt}
\setlength{\parsep}{0pt}
\item[(i)] $\A$ is $\tau$-{\it artinian}, that is, there are
no chains of subobjects $\cdots\!\subset\!A_2\!\subset\!
A_1\!\subset\!U$ in $\A$ with $A_{n+1}\!\ne\!A_n$ and
$\tau([A_{n+1}])\!\ge\!\tau([A_n/A_{n+1}])$ for all $n$; and
\item[(ii)] $\Oss^\al(\tau)$ is a {\it constructible set\/} in
$\fObj_\A$ for all $\al\in C(\A)$, using the theory of constructible
sets and functions on Artin $\K$-stacks developed in~\cite{Joyc1}.
\end{itemize}
\label{gf2def4}
\end{dfn}

Examples of (weak) stability conditions on $\A=\modKQ$ and
$\A=\coh(X)$ are given in \cite[\S 4.3--\S 4.4]{Joyc5}. Most of them
are permissible. Here is~\cite[Ex.~4.14]{Joyc5}.

\begin{ex} Let Assumption \ref{gf2ass1} hold, and $c,r:K(\A)\ra\R$
be group homomorphisms with $r(\al)>0$ for all $\al\in C(\A)$.
Define $\mu:C(\A)\ra\R$ by $\mu(\al)=c(\al)/r(\al)$ for $\al\in
C(\A)$. Then $\mu$ is called a {\it slope function} on $K(\A)$, and
$(\mu,\R,\le)$ is a {\it stability condition} on~$\A$.

It will be useful later to re-express this as follows. Define the
{\it central charge} $Z:K(\A)\ra\C$ by $Z(\al)=-c(\al)+ir(\al)$. The
name will be explained in \S\ref{gf23}. Then $Z\in\Hom\bigl(K(\A),
\C\bigr)$ is a group homomorphism, and maps $C(\A)$ to the upper
half plane $H=\{x+iy:x\in\R,$ $y>0\}$ in $\C$.

For $\al\in C(\A)$, the argument $\arg\ci Z(\al)$ lies in $(0,\pi)$,
and clearly $\mu(\al)=-\cot\ci\arg\ci Z(\al)$, where $\cot$ is the
cotangent function. So $(\mu,\R,\le)$ can be recovered from $Z$.
Since $-\cot:(0,\pi)\ra\R$ is strictly increasing, it fixes orders
in $\R$. Thus $(\arg\ci Z,\R,\le)$ is an equivalent stability
condition to $(\mu,\R,\le)$, that is, $U\in\A$ is $\mu$-(semi)stable
if and only if it is $\arg\ci Z$-(semi)stable. Write
\e
\begin{split}
\Stab(\A)=\bigl\{Z\in&\Hom(K(\A),\C):\text{$Z(C(\A))\subset H$,
and the stability}\\
&\text{condition $(\mu,\R,\le)$ defined by $Z$ is
permissible}\bigr\}.
\end{split}
\label{gf2eq2}
\e
In the cases we are interested in $\Stab(\A)$ is an {\it open\/}
subset of the complex vector space $\Hom(K(\A),\C)$, and so is a
{\it complex manifold}.

Such stability conditions can be defined on all the quiver examples
of \cite[\S 10]{Joyc3}, and they are automatically {\it permissible}
by \cite[Cor.~4.13]{Joyc5}. In Example \ref{gf2ex1}, as
$K(\A)=\Z^{Q_0}$ and $C(\A)=\N^{Q_0}\sm\{0\}$ we may write $c,r$ as
\begin{equation*}
c(\al)=\ts\sum_{v\in Q_0}c_v\,(\bdim\al)(v) \quad\text{and}\quad
r(\al)=\ts\sum_{v\in Q_0}r_v\,(\bdim\al)(v),
\end{equation*}
where $c_v\in\R$ and $r_v\in(0,\iy)$ for all $v\in Q_0$.
Thus~$\Stab(\A)=H^{Q_0}\subset\C^{Q_0}$.

The usual notion of slope stability on $\A=\coh(X)$ for $X$ a smooth
projective curve is a slight generalization of the above. We take
$c([U])$ to be the {\it degree} and $r([U])$ the {\it rank\/} of
$U\in\coh(X)$. But then for $\al\in C(\A)$ coming from a torsion
sheaf $U$ we have $r(\al)=0$ and $c(\al)>0$, so we must allow $\mu$
to take values in $(-\iy,+\iy]$, with $\mu(\al)=+\iy$ if~$r(\al)=0$.
\label{gf2ex2}
\end{ex}

Here \cite[Th.~4.4]{Joyc5} is a useful property of weak stability
conditions. We call $0\!=\!A_0\!\subset\!\cdots\!\subset\!A_n\!=\!U$
in Theorem \ref{gf2thm1} the {\it Harder--Narasimhan filtration}
of~$U$.

\begin{thm} Let\/ $(\tau,T,\le)$ be a weak stability condition on an
abelian category $\A$. Suppose $\A$ is noetherian and\/
$\tau$-artinian. Then each\/ $U\in\A$ admits a unique filtration
$0\!=\!A_0\!\subset\!\cdots\!\subset\!A_n\!=\!U$ for $n\ge 0$, such
that\/ $S_k\!=\!A_k/A_{k-1}$ is $\tau$-semistable for
$k=1,\ldots,n$, and\/~$\tau([S_1])>\tau([S_2])>\cdots>\tau([S_n])$.
\label{gf2thm1}
\end{thm}

\subsection{A framework for discussing counting invariants}
\label{gf22}

Given $\A,K(\A),\fF_\A$ satisfying Assumption \ref{gf2ass1} and weak
stability conditions $(\tau,T,\le),(\ti\tau,\ti T,\le)$ on $\A$, the
final paper \cite{Joyc6} in the series was mostly concerned with
defining interesting {\it invariants} $I^\al_{\rm
ss}(\tau),J^\al(\tau), \ldots$ which `count' $\tau$-semistable
objects in class $\al$ for all $\al\in C(\A)$, and computing the
{\it transformation laws} which these invariants satisfy under
changing from $(\tau,T,\le)$ to~$(\ti\tau,\ti T,\le)$.

These different invariants all share a common structure, involving
an algebra and a Lie algebra. We will now abstract this structure
(which was not done in \cite{Joyc6}) and express the various
invariants of \cite{Joyc6} as examples of this structure. We will
need the following notation, from \cite[Def.s 4.2, 4.4 \&
5.1]{Joyc6}. In our first two definitions, $S,U(*,\tau,\ti\tau)$ are
called {\it transformation coefficients}, and are combinatorial
factors appearing in transformation laws from $(\tau,T,\le)$
to~$(\ti\tau,\ti T,\le)$.

\begin{dfn} Let Assumption \ref{gf2ass1} hold, $(\tau,T,\le),
(\ti\tau,\ti T,\le)$ be weak stability conditions on $\A$, and
$(\{1,\ldots,n\},\le,\ka)$ be $\A$-data. If for all $i=1,\ldots,n-1$
we have either
\begin{itemize}
\setlength{\itemsep}{0pt}
\setlength{\parsep}{0pt}
\item[{\rm(a)}] $\tau\ci\ka(i)\le\tau\ci\ka(i+1)$ and
$\ti\tau\ci\ka(\{1,\ldots,i\})>\ti\tau\ci\ka(\{i+1,\ldots,n\})$ or
\item[{\rm(b)}] $\tau\ci\ka(i)>\tau\ci\ka(i+1)$ and~$\ti\tau
\ci\ka(\{1,\ldots,i\})\le\ti\tau\ci\ka(\{i+1,\ldots,n\})$,
\end{itemize}
then define $S(\{1,\ldots,n\},\le,\ka,\tau,\ti\tau)=(-1)^r$, where
$r$ is the number of $i=1,\ldots,n-1$ satisfying (a). Otherwise
define~$S(\{1,\ldots,n\},\le,\ka,\tau,\ti\tau)=0$.

If $(I,\pr,\ka)$ is $\A$-data with $\pr$ a {\it total order}, there
is a unique bijection $\phi:\{1,\ldots,n\}\ra I$ with $n=\md{I}$ and
$\phi_*(\le)=\pr$, and $(\{1,\ldots,n\},\le, \ka\ci\phi)$ is
$\A$-data. Define~$S(I,\pr,\ka,\tau,\ti\tau)
=S(\{1,\ldots,n\},\le,\ka\ci\phi,\tau,\ti\tau)$.
\label{gf2def5}
\end{dfn}

\begin{dfn} Let Assumption \ref{gf2ass1} hold, $(\tau,T,\le),
(\ti\tau,\ti T,\le)$ be weak stability conditions on $\A$, and
$(\{1,\ldots,n\},\le,\ka)$ be $\A$-data. Define
\e
\begin{gathered}
U(\{1,\ldots,n\},\le,\ka,\tau,\ti\tau)=
\sum_{1\le l\le m\le n}\\
\sum_{\substack{
\text{surjective $\psi:\{1,\ldots,n\}\!\ra\!\{1,\ldots,m\}$}\\
\text{and\/ $\xi:\{1,\ldots,m\}\!\ra\!\{1,\ldots,l\}$:}\\
\text{$i\!\le\!j$ implies $\psi(i)\!\le\!\psi(j)$,
$i\!\le\!j$ implies $\xi(i)\!\le\!\xi(j)$.}\\
\text{Define $\la:\{1,\ldots,m\}\ra C(\A)$ by $\la(b)=\ka(\psi^{-1}(b))$.}\\
\text{Define $\mu:\{1,\ldots,l\}\ra C(\A)$ by $\mu(a)=\la(\xi^{-1}(a))$.}\\
\text{Then $\tau\ci\ka\equiv\tau\ci\la\ci\mu:I\ra T$ and\/
$\ti\tau\ci\mu\equiv\ti\tau(\al)$}}} \!\!\!\!\!\!\!\!
\begin{aligned}[t]
\prod_{a=1}^lS(\xi^{-1}(\{a\}),\le,\la,\tau,\ti\tau)\cdot&\\
\frac{(-1)^{l-1}}{l}\cdot\prod_{b=1}^m\frac{1}{\md{\psi^{-1}(b)}!}\,&.
\end{aligned}
\end{gathered}
\label{gf2eq3}
\e

If $(I,\pr,\ka)$ is $\A$-data with $\pr$ a {\it total order}, there
is a unique bijection $\phi:\{1,\ldots,n\}\ra I$ with $n=\md{I}$ and
$\phi_*(\le)=\pr$, and $(\{1,\ldots,n\},\le, \ka\ci\phi)$ is
$\A$-data. Define~$U(I,\pr,\ka,\tau,\ti\tau)
=U(\{1,\ldots,n\},\le,\ka\ci\phi,\tau,\ti\tau)$.
\label{gf2def6}
\end{dfn}

\begin{dfn} Let Assumption \ref{gf2ass1} hold and $(\tau,T,\le),
(\ti\tau,\ti T,\le)$ be weak stability conditions on $\A$. We say
{\it the change from $(\tau,T,\le)$ to $(\ti\tau,\ti T,\le)$ is
locally finite} if for all constructible $C\subseteq\fObj_\A(\K)$,
there are only finitely many sets of $\A$-data $(\{1,\ldots,n\},\le,
\ka)$ for which $S(\{1,\ldots,n\},\le,\ka,\tau,\ti\tau)\!\ne\!0$ and
\begin{equation*}
C\cap\bs\si(\{1,\ldots,n\})_*\bigl(\Mss(\{1,\ldots,n\},\le,\ka,
\tau)_\A\bigr)\ne\emptyset.
\end{equation*}
We say {\it the change from $(\tau,T,\le)$ to $(\ti\tau,\ti T,\le)$
is globally finite} if this holds for $C=\fObj_\A^\al(\K)$ (which is
not constructible, in general) for all $\al\in C(\A)$. Since any
constructible $C\subseteq\fObj_\A(\K)$ is contained in a finite
union of $\fObj_\A^\al(\K)$, globally finite implies locally finite.
\label{gf2def7}
\end{dfn}

The following encapsulates the structure common to most of the
invariants studied in \cite{Joyc6}, with some oversimplifications we
discuss in Remark~\ref{gf2rem1}.

\begin{ass} Let Assumption \ref{gf2ass1} hold. Suppose we are given
a $\C$-algebra $\H$ with identity 1 and multiplication $*$ (which is
associative, but not in general commutative), with a decomposition
into $\C$-vector subspaces $\H=\bigop_{\al\in\bar C(\A)}\H^\al$,
such that $1\in\H^0$ and $\H^\al*\H^\be\subseteq\H^{\al+\be}$ for
all~$\al,\be\in\bar C(\A)$.

Suppose we are given a $\C$-Lie subalgebra $\L$ of $\H$ with Lie
bracket $[f,g]=f*g-g*f$, with a decomposition into $\C$-vector
subspaces $\L=\bigop_{\al\in C(\A)}\L^\al$ such that
$\L^\al\subseteq\H^\al$ and $[\L^\al,\L^\be]\subseteq\L^{\al+\be}$
for all~$\al,\be\in C(\A)$.

Whenever $(\tau,T,\le)$ is a permissible weak stability condition on
$\A$, let there be given elements $\de^\al(\tau)\in\H^\al$ and
$\ep^\al(\tau)\in\L^\al$ for all $\al\in C(\A)$. These satisfy
\ea
\ep^\al(\tau)&=
\!\!\!\!\!\!\!\!\!\!\!\!\!\!\!\!
\sum_{\substack{\text{$\A$-data $(\{1,\ldots,n\},\le,\ka):$}\\
\text{$\ka(\{1,\ldots,n\})=\al$, $\tau\ci\ka\equiv\tau(\al)$}}}
\!\!\!\!\!\!\!
\frac{(-1)^{n-1}}{n}\,\,\de^{\ka(1)}(\tau)*\de^{\ka(2)}(\tau)*
\cdots*\de^{\ka(n)}(\tau),
\label{gf2eq4}\\
\de^\al(\tau)&= \!\!\!\!\!\!\!\!\!\!\!\!\!\!\!\!
\sum_{\substack{\text{$\A$-data $(\{1,\ldots,n\},\le,\ka):$}\\
\text{$\ka(\{1,\ldots,n\})=\al$, $\tau\ci\ka\equiv\tau(\al)$}}}
\!\!\!\! \frac{1}{n!}\,\,\ep^{\ka(1)}(\tau)*\ep^{\ka(2)}(\tau)*
\cdots*\ep^{\ka(n)}(\tau),
\label{gf2eq5}
\ea
where there are only finitely many nonzero terms in each sum.

If $(\tau,T,\le),(\ti\tau,\ti T,\le)$ are permissible weak stability
conditions on $\A$ and the change from $(\tau,T,\le)$ to
$(\ti\tau,\ti T,\le)$ is globally finite, for all $\al\in C(\A)$ we
have
\begin{gather}
\begin{gathered}
\sum_{\substack{\text{$\A$-data $(\{1,\ldots,n\},\le,\ka):$}\\
\text{$\ka(\{1,\ldots,n\})=\al$}}} \!\!\!\!\!\!\!\!\!
\begin{aligned}[t]
S(\{1,&\ldots,n\},\le,\ka,\tau,\ti\tau)\cdot\\
&\de^{\ka(1)}(\tau)*\de^{\ka(2)}(\tau)*\cdots*
\de^{\ka(n)}(\tau)=\de^\al(\ti\tau),
\end{aligned}
\end{gathered}
\label{gf2eq6}\\
\begin{gathered}
\sum_{\substack{\text{$\A$-data $(\{1,\ldots,n\},\le,\ka):$}\\
\text{$\ka(\{1,\ldots,n\})=\al$}}} \!\!\!\!\!\!\!\!\!\!\!\!\!\!\!
\begin{aligned}[t]
U(\{1,&\ldots,n\},\le,\ka,\tau,\ti\tau)\cdot\\
&\ep^{\ka(1)}(\tau)*\ep^{\ka(2)}(\tau)*\cdots*
\ep^{\ka(n)}(\tau)=\ep^\al(\ti\tau),
\end{aligned}
\end{gathered}
\label{gf2eq7}
\end{gather}
where there are only finitely many nonzero terms in each sum.

Equation \eq{gf2eq7} may be rewritten:
\e
\begin{aligned}
&\ep^\al(\ti\tau)=\\
&\sum_{\substack{\text{iso classes}\\
\text{of finite}\\ \text{sets $I$}}} \frac{1}{\md{I}!}\!
\sum_{\substack{\ka:I\ra C(\A):\\ \ka(I)=\al}}\!
\raisebox{-12pt}{\begin{Large}$\displaystyle\Biggl[$\end{Large}}
\!\!
\sum_{\substack{\text{total orders $\pr$ on $I$.}\\
\text{Write $I=\{i_1,\ldots,i_n\}$,}\\
\text{$i_1\pr i_2\pr\cdots\pr i_n$}}}\,
\begin{aligned}[t]
&U(I,\pr,\ka,\tau,\ti\tau)\cdot\\
&\ep^{\ka(i_1)}(\tau)\!*\!\cdots\!*\!\ep^{\ka(i_n)}(\tau)
\end{aligned}
\raisebox{-12pt}{\begin{Large}$\displaystyle\Biggr]$\end{Large}}.
\end{aligned}
\label{gf2eq8}
\e
The term $[\cdots]$ in \eq{gf2eq8} is a finite $\Q$-linear
combination of multiple commutators of $\ep^{\ka(i)}$ for $i\in I$,
and so it lies in the Lie algebra $\L$, not just the algebra $\H$.
Thus \eq{gf2eq7} and \eq{gf2eq8} can be regarded as identities in
$\L$ rather than~$\H$.
\label{gf2ass2}
\end{ass}

\begin{rem}{\bf(a)} $\de^\al(\tau)$ is an invariant of the moduli
space $\Oss^\al(\tau)$ of $\tau$-semistable objects in class $\al$,
which `counts' such $\tau$-semistable objects. Usually it is of the
form $\de^\al(\tau)=\Phi(\dss^\al(\tau))$, where
$\dss^\al(\tau)\in\CF(\fObj_\A)$ is the characteristic function of
$\Oss^\al(\tau)$, $\CF(\fObj_\A)$ is the vector space of
constructible functions on the Artin $\K$-stack $\fObj_\A$ as in
\cite{Joyc1}, and $\Phi:\CF(\fObj_\A)\ra\H$ is a linear map with
special multiplicative properties.
\smallskip

\noindent{\bf(b)} The $\ep^\al(\tau)$ are an alternative set of
generators to the $\de^\al(\tau)$. Here \eq{gf2eq5} is the inverse
of \eq{gf2eq4}, and given \eq{gf2eq3}--\eq{gf2eq5}, equations
\eq{gf2eq6} and \eq{gf2eq7} are equivalent. Thus, the main
nontrivial claim about the $\ep^\al(\tau)$ is that they lie in the
Lie algebra $\L$, which may be much smaller than $\H$. Roughly
speaking, the $\ep^\al(\tau)$ count $\tau$-semistable objects $S$ in
class $\al$ weighted by a rational number depending on the
factorization of $S$ into $\tau$-stables, which is 1 if $U$ is
$\tau$-stable. If $S$ is decomposable this weight is 0, so
$\ep^\al(\tau)$ counts only indecomposable $\tau$-semistables. The
Lie algebra $\L$ is the part of $\H$ `supported on indecomposables'.
\smallskip

\noindent{\bf(c)} In \cite{Joyc6} we worked with (Lie) algebras over
$\Q$, not $\C$. But here we complexify, as we shall be discussing
holomorphic functions into~$\H,\L$.
\smallskip

\noindent{\bf(d)} In parts of \cite{Joyc6}, equations
\eq{gf2eq6}--\eq{gf2eq8} are only proved under an extra assumption,
the existence of a third weak stability condition $(\hat\tau,\hat
T,\le)$ compatible with $(\tau,T,\le),(\ti\tau,\ti T,\le)$ in
certain ways. But we will not worry about this.
\smallskip

\noindent{\bf(e)} In parts of \cite{Joyc6} we relax the assumption
that $(\tau,T,\le),(\ti\tau,\ti T,\le)$ are {\it permissible\/}
(taking them instead to be $\tau$-{\it artinian}, or {\it
essentially permissible\/}), and we allow the change from
$(\tau,T,\le)$ to $(\ti\tau,\ti T,\le)$ to be {\it locally finite}
rather than globally finite. Then equations \eq{gf2eq4}--\eq{gf2eq8}
need no longer have only finitely many nonzero terms, and they are
interpreted using a notion of convergence in~$\H$.
\smallskip

\noindent{\bf(f)} The $\de^\al(\tau),\ep^\al(\tau)$ are only the
simplest of the invariants studied in \cite{Joyc6} --- we could call
them `one point invariants', as they depend on only one class
$\al\in C(\A)$. We also considered systems of `$n$ point invariants'
depending on $n$ classes $\al\in C(\A)$, which will not enter this
paper. One thing that makes the one point invariants special is that
their transformation laws \eq{gf2eq6}--\eq{gf2eq7} depend only on
other one point invariants, not on $n$ point invariants for
all~$n\ge 1$.
\label{gf2rem1}
\end{rem}

The next six examples explain how various results in \cite{Joyc6}
fit into the framework of Assumption~\ref{gf2ass2}.

\begin{ex} Let Assumption \ref{gf2ass1} hold with $\K$ of
characteristic zero. Take $\H=\CF(\fObj_\A)\ot_\Q\C$, the vector
space of $\C$-valued {\it constructible functions} on $\fObj_\A$,
and $\H^\al$ the subspace of functions supported on $\fObj_\A^\al$.
The multiplication $*$ on $\H$, studied at length in \cite{Joyc4},
has the following approximate form: for $V\in\A$, $(f*g)([V])$ is
the `integral' over all short exact sequences $0\ra U\ra V\ra W\ra
0$ in $\A$ of $f([U])g([W])$, with respect to a measure defined
using the Euler characteristic of constructible subsets of
$\K$-stacks.

The identity is $1=\de_{[0]}$, the characteristic function of
$[0]\in\fObj_\A(\K)$. The Lie subalgebra $\L$ is
$\CFi(\fObj_\A)\ot_\Q\C$, functions supported on points $[U]$ for
$U\in\A$ indecomposable, and $\de^\al(\tau)=\dss^\al(\tau)$, the
characteristic function of $\Oss^\al(\tau)$. Then
\cite{Joyc4,Joyc5,Joyc6} show Assumption \ref{gf2ass2} holds, except
that \eq{gf2eq6}--\eq{gf2eq8} are only proved under extra conditions
as in Remark \ref{gf2rem1}(d) above.

We can also replace $\H=\CF(\fObj_\A)\ot_\Q\C$ and
$\L=\CFi(\fObj_\A)\ot_\Q\C$ by the much smaller (Lie) subalgebras
$\Ht_\tau\ot_\Q\C,\Lt_\tau\ot_\Q\C$ of \cite[\S 7]{Joyc5}, since by
\cite[\S 5]{Joyc6} these are very often independent of the choice of
permissible weak stability condition $(\tau,T,\le)$ used to define
them.
\label{gf2ex3}
\end{ex}

\begin{ex} Let Assumption \ref{gf2ass1} hold. Take $\H=\SFa(\fObj_\A)
\ot_\Q\C$, the algebra of stack functions on $\fObj_\A$ with algebra
stabilizers defined in \cite[\S 5]{Joyc4}, using the theory of {\it
stack functions} from \cite{Joyc2}, a universal generalization of
constructible functions. Let $\L=\SFai(\fObj_\A)\ot_\Q\C$, the
subspace of $\H$ supported on `virtual indecomposables', and let
$\H^\al,\L^\al$ be the subspaces of $\H,\L$ supported on
$\fObj_\A^\al$. Set $\de^\al(\tau)=\bdss^\al(\tau)$, in the notation
of \cite{Joyc5}. Then \cite{Joyc4,Joyc5,Joyc6} show Assumption
\ref{gf2ass2} holds, but with \eq{gf2eq6}--\eq{gf2eq8} only proved
under extra conditions.

This also works with $\SFa(\fObj_\A)$ replaced by one of the
`twisted stack function' spaces $\oSFa(\fObj_\A,\Up,\La)$,
$\oSFa(\fObj_\A,\Up,\La^\ci)$, $\oSFa(\fObj_\A,\Th,\Om)$
of~\cite{Joyc4}.

We can also replace $\H=\SFa(\fObj_\A)\ot_\Q\C$ and
$\L=\SFai(\fObj_\A)\ot_\Q\C$ by the much smaller (Lie) subalgebras
$\bHt_\tau\ot_\Q\C,\bLt_\tau\ot_\Q\C$ of \cite[\S 8]{Joyc5}, since
by \cite[\S 5]{Joyc6} these are very often independent of the choice
of permissible weak stability condition $(\tau,T,\le)$ used to
define them.
\label{gf2ex4}
\end{ex}

\begin{ex} Let Assumption \ref{gf2ass1} hold and $\chi:K(\A)\t
K(\A)\ra\Z$ be biadditive and satisfy
\e
\dim_\K\Hom(U,V)-\dim_\K\Ext^1(U,V)=\chi\bigl([U],[V]\bigr)
\quad\text{for all $U,V\in\A$.}
\label{gf2eq9}
\e
This happens when $\A=\coh(X)$ with $X$ a smooth projective curve,
and for $\A=\modKQ$ in Example \ref{gf2ex1} with $\chi$ given by the
{\it Ringel form}
\e
\chi(\al,\be)=\ts\sum_{v\in Q_0}\al(v)\be(v)-\sum_{a\in
Q_1}\al(b(a))\be(e(a)) \;\>\text{for $\al,\be\in\Z^{Q_0}$.}
\label{gf2eq10}
\e

Define $\La=\C(z)$, the algebra of rational functions $p(z)/q(z)$
for polynomials $p,q$ with coefficients in $\C$ and $q\ne 0$, and
define a special element $\el=z^2$ in $\La$. Define $\La^\ci$ to be
the subalgebra of $p(z)/q(z)$ in $\La$ for which $z\pm 1$ do not
divide $q$. The facts we need about $\La,\La^\ci$ are that the {\it
virtual Poincar\'e polynomial\/} $P(X;z)$ of a $\K$-variety $X$
takes values in $\La^\ci\subset\La$, and $\el=P(\K;z)$ for $\K$ the
affine line, and $\el$ and $\el^k+\el^{k-1}+\cdots+1$ are invertible
in $\La^\ci$, and $\el-1$ is invertible in~$\La$.

Let $a^\al$ for $\al\in\bar C(\A)$ be formal symbols, and define
$\H=A(\A,\La,\chi)$ as in \cite[\S 6.2]{Joyc4} to be the
$\La$-module with basis $\{a^\al:\al\in\bar C(\A)\}$, with the
obvious notions of addition and multiplication by $\C$. Define a
{\it multiplication} $*$ on $\H$ by
\e
\bigl(\ts\sum_{i\in I}\la_i\,a^{\al_i}\bigr)*\bigl(\ts\sum_{j\in
J}\mu_j\,a^{\be_j}\bigr)=\ts\sum_{i\in I}\sum_{j\in
J}\la_i\mu_j\el^{-\chi(\be_j,\al_i)}\,a^{\al_i+\be_j}.
\label{gf2eq11}
\e
Then $\H$ is a $\C$-algebra, with identity $a^0$. Define
$\H^\al=\La\cdot a^\al$ for $\al\in\bar C(\A)$. Define
$\L^\al=(\el-1)^{-1}\La^\ci\cdot a^\al\subset\H^\al$ for $\al\in
C(\A)$, and $\L=\bigop_{\al\in C(\A)}\L^\al$. Then $\L$ is a Lie
subalgebra of $\H$, as~$(\el^{-\chi(\be,\al)}-
\el^{-\chi(\al,\be)})/(\el-1)\in\La^\ci$.

For $(\tau,T,\le)$ a permissible weak stability condition on $\A$
and $\al\in C(\A)$, define $\de^\al(\tau)=\Iss^\al(\tau)a^\al$,
where $\Iss^\al(\tau)$ is the {\it virtual Poincar\'e function\/} of
$\Oss^\al(\tau)$, as defined in \cite[\S 4.2]{Joyc2}, where we
regard $\Oss^\al(\tau)$ as a finite type open $\K$-substack with
affine geometric stabilizers in the Artin $\K$-stack $\fObj_\A^\al$.
Define $\ep^\al(\tau)$ by \eq{gf2eq4}. Then $\ep^\al(\tau)\in
\H^\al$, so we can write $\ep^\al(\tau)=(\el-1)^{-1}J^\al(\tau)
a^\al$ for $J^\al(\tau)\in\La$. We show in \cite[Th.~6.8]{Joyc6}
that $J^\al(\tau)\in\La^\ci$, so~$\ep^\al(\tau)\in\L^\al$.

Then \cite[\S 6.2]{Joyc4} and \cite[\S 6.2]{Joyc6} show that
Assumption \ref{gf2ass2} holds in its entirety when $\A=\modKQ$, and
with extra conditions as in Remark \ref{gf2rem1}(d) above in
general. It also holds with $\La$ replaced by other commutative
$\C$-algebras, and virtual Poincar\'e polynomials replaced by other
$\La$-valued `motivic invariants' $\Up$ of $\K$-varieties with
$\el=\Up(\K)$; for details see~\cite{Joyc2,Joyc4,Joyc6}.
\label{gf2ex5}
\end{ex}

\begin{ex} Let $\K$ be an algebraically closed field and $X$ a smooth
projective surface over $\K$ with $K_X^{-1}$ {\it numerically
effective} ({\it nef\/}). Take $\A=\coh(X)$ with data $K(\A),\fF_\A$
satisfying Assumption \ref{gf2ass1} as in \cite[Ex.~9.1]{Joyc3}.
Then there is a biadditive $\chi:K(\A)\t K(\A)\ra\Z$ such that for
all $U,V\in\A$ we have
\e
\dim_\K\Hom(U,V)-\dim_\K\Ext^1(U,V)+\dim_\K\Ext^2(U,V)=
\chi\bigl([U],[V]\bigr).
\label{gf2eq12}
\e

Define $\La,\H,*,\H^\al$ as in Example \ref{gf2ex5}, but set
$\L^\al=\H^\al$ for $\al\in C(\A)$ and $\L=\bigop_{\al\in
C(\A)}\L^\al$. Then in \cite[\S 6.4]{Joyc6}, for a class of weak
stability conditions $(\tau,T,\le)$ on $\A$ based on Gieseker
stability, we define invariants $\Iss^\al(\tau),\bar
J^\al(\tau)\in\La$ such that Assumption \ref{gf2ass2} holds with
$\de^\al(\tau)=\Iss^\al(\tau)a^\al$ and
$\ep^\al(\tau)=(\el-1)^{-1}\bar J^\al(\tau)a^\al$. But we do not
prove that $\bar J^\al(\tau)\in\La^\ci$, which is why we modify the
definitions of~$\L^\al,\L$.
\label{gf2ex6}
\end{ex}

Examples \ref{gf2ex5} and \ref{gf2ex6} illustrate the relationship
between `invariants' $\Iss^\al(\tau),\ab J^\al(\tau),\ab \bar
J^\al(\tau)$ which `count' $\tau$-semistables in class $\al$, and
our (Lie) algebra approach. In this case, the transformation laws
\eq{gf2eq6}--\eq{gf2eq7} for $\de^\al(\tau),\ep^\al(\tau)$ are
equivalent to the following laws for $\Iss^\al(\tau),J^\al(\tau)$,
from~\cite[Th.~6.8]{Joyc6}:
\ea
\begin{gathered}
\Iss^\al(\ti\tau)=
\sum_{\begin{subarray}{l}
\text{$\A$-data $(\{1,\ldots,n\},\le,\ka):$}\\
\text{$\ka(\{1,\ldots,n\})=\al$}\end{subarray}
\!\!\!\!\!\!\!\!\!\!\!\!\!\!\!\!\!\!\!\!\!\!\!\!
\!\!\!\!\!\!\!\!\!\!\!\!\!\!\!\!\! }
\begin{aligned}[t]
S(\{1,\ldots,n\},\le,\ka,\tau,\ti\tau)\cdot
\el^{-\sum_{1\le i<j\le n}\chi(\ka(j),\ka(i))}\cdot&\\
\ts\prod_{i=1}^n\Iss^{\ka(i)}(\tau)&,
\end{aligned}
\end{gathered}
\label{gf2eq13}\\
\begin{gathered}
J^\al(\ti\tau)=
\sum_{\begin{subarray}{l}
\text{$\A$-data $(\{1,\ldots,n\},\le,\ka):$}\\
\text{$\ka(\{1,\ldots,n\})=\al$}\end{subarray}
\!\!\!\!\!\!\!\!\!\!\!\!\!\!\!\!\!\!\!\!\!\!\!\!
\!\!\!\!\!\!\!\!\!\!\!\!\!\!\!\!\! }
\begin{aligned}[t]
U(\{1,\ldots,n\},\le,\ka,\tau,\ti\tau)\cdot
\el^{-\sum_{1\le i<j\le n}\chi(\ka(j),\ka(i))}\cdot&\\
(\el-1)^{1-n}\ts\prod_{i=1}^nJ^{\ka(i)}(\tau)&.
\end{aligned}
\end{gathered}
\label{gf2eq14}
\ea
Observe that \eq{gf2eq6}--\eq{gf2eq7} are simpler than
\eq{gf2eq13}--\eq{gf2eq14}, since the powers of $\el$ in
\eq{gf2eq13}--\eq{gf2eq14} are packaged in the multiplication $*$ in
$\H$. This is more pronounced in our next two examples, where the
formulae for $*$ are much more complicated, so the transformation
laws for invariants are too. One moral is that working in the
framework of Assumption \ref{gf2ass2} is simpler than working with
systems of invariants, which is why we have adopted it.

\begin{ex} Let Assumption \ref{gf2ass1} hold and $\chi:K(\A)\t
K(\A)\ra\Z$ be biadditive and satisfy \eq{gf2eq9}, and let
$\La,\La^\ci,\el$ be as in Example \ref{gf2ex5}. Consider pairs
$(I,\ka)$ with $I$ a finite set and $\ka:I\ra C(\A)$ a map. Define
an equivalence relation `$\approx$' on such $(I,\ka)$ by
$(I,\ka)\approx(I',\ka')$ if there exists a bijection $i:I\ra I'$
with $\ka'\ci i=\ka$. Write $[I,\ka]$ for the $\approx$-equivalence
class of $(I,\ka)$. Introduce formal symbols $b_{[I,\ka]}$ for all
such equivalence classes~$[I,\ka]$.

As in \cite[\S 6.3]{Joyc4}, let $\H=B(\A,\La,\chi)$ be the
$\La$-module with basis the $b_{[I,\ka]}$. Define
$\H^\al=\bigop_{[I,\ka]: \ka(I)=\al}\La\cdot b_{[I,\ka]}$. Define a
{\it multiplication} $*$ on $\H$ by
\begin{gather}
b_{[I,\ka]}*b_{[J,\la]}=
\sum_{\text{eq. classes $[K,\mu]$}}b_{[K,\mu]}\,\cdot\,
\frac{(\el-1)^{\md{K}-\md{I}-\md{J}}}{\md{\Aut(K,\mu)}}\,\cdot
\label{gf2eq15}
\\
\raisebox{-9pt}{\begin{Large}$\displaystyle\biggl[$\end{Large}}
\sum_{\substack{\text{iso.}\\ \text{classes}\\ \text{of finite}\\
\text{sets $L$}}}\!\!
\frac{(-1)^{\md{L}-\md{K}}}{\md{L}!}\!\!\!\!\!\!\!\!
\sum_{\substack{\text{$\phi:I\ra L$, $\psi:J\ra L$ and}\\
\text{$\th:L\!\ra\!K$: $\phi\!\amalg\!\psi$ surjective,}\\
\text{$\mu(k)=\ka((\th\ci\phi)^{-1}(k))+$}\\
\text{$\la((\th\ci\psi)^{-1}(k))$, $k\in K$}}}\,\prod_{k\in
K}\!(\md{\th^{-1}(k)}\!-\!1)!
\prod_{\begin{subarray}{l} i\in I,\; j\in J:\\
\phi(i)=\psi(j)\end{subarray} \!\!\!\!\!\!\!\!\!\!\!\!\!\!\!\! }
\el^{-\chi(\la(j),\ka(i))}
\raisebox{-9pt}{\begin{Large}$\displaystyle\biggr]$\end{Large}},
\nonumber
\end{gather}
extended $\La$-bilinearly. Then $\H$ is a $\C$-algebra with
identity~$b_{[\emptyset,\emptyset]}$.

For $\al\in C(\A)$ define $b^\al=b_{[\{1\},\al']}$ where
$\al'(1)=\al$, define $\L^\al=\La^\ci\cdot b^\al$ and
$\L=\bigop_{\al\in C(\A)}\L^\al$. Equation \eq{gf2eq15} yields
\e
[b^\al,b^\be]=\frac{\el^{-\chi(\be,\al)}-\el^{-\chi(\al,\be)}}{
\el-1}\,b^{\al+\be},
\label{gf2eq16}
\e
and $(\el^{-\chi(\be,\al)}-\el^{-\chi(\al,\be)})/(\el-1)\in\La^\ci$,
so $\L$ is a Lie subalgebra of~$\H$.

If $(\tau,T,\le)$ is a permissible weak stability condition we put
$\ep^\al(\tau)=J^\al(\tau)b^\al$ for the same $J^\al(\tau)\in
\La^\ci$ as in Example \ref{gf2ex5}. We then define $\de^\al(\tau)$
by \eq{gf2eq5}, giving a much more complicated answer than in
Example \ref{gf2ex5}. From \cite[\S 6.3]{Joyc4} and \cite[\S
6]{Joyc6} it follows that Assumption \ref{gf2ass2} holds in its
entirety when $\A=\modKQ$, and with extra conditions as in Remark
\ref{gf2rem1}(d) above in general.
\label{gf2ex7}
\end{ex}

\begin{ex} Let Assumption \ref{gf2ass1} hold and $\bar\chi:K(\A)\t
K(\A)\ra\Z$ be antisymmetric and biadditive and satisfy
\e
\begin{split}
&\bigl(\dim_\K\Hom(U,V)-\dim_\K\Ext^1(U,V)\bigr)-\\
&\bigl(\dim_\K\Hom(V,U)-\dim_\K\Ext^1(V,U)\bigr)=
\bar\chi\bigl([U],[V]\bigr)\quad\text{for all $U,V\in\A$.}
\end{split}
\label{gf2eq17}
\e
Note that \eq{gf2eq9} implies \eq{gf2eq17} with $\bar\chi(\al,\be)=
\chi(\al,\be)-\chi(\be,\al)$, so this holds for $\A=\modKQ$ and
$\A=\coh(X)$ for $X$ a smooth projective curve. But we also show in
\cite[\S 6.6]{Joyc4} using Serre duality that \eq{gf2eq17} holds
when $\A=\coh(X)$ for $X$ a {\it Calabi--Yau\/ $3$-fold\/}
over~$\K$.

As in Example \ref{gf2ex7}, introduce symbols $c_{[I,\ka]}$ for all
equivalence classes $[I,\ka]$, and let $\H=C(\A,\Om,\ha\bar\chi)$ be
the $\C$-vector space with basis the $c_{[I,\ka]}$. Define
$\H^\al=\bigop_{[I,\ka]:\ka(I)=\al}\C\cdot c_{[I,\ka]}$. Define a
{\it multiplication} $*$ on $\H$ by
\begin{gather}
c_{[I,\ka]}*c_{[J,\la]}=
\sum_{\text{eq. classes $[K,\mu]$}}c_{[K,\mu]}\,\cdot\,
\frac{1}{\md{\Aut(K,\mu)}}\!\!\!\!
\sum_{\substack{\text{$\eta:I\!\ra\!K$, $\ze:J\!\ra\!K$:}\\
\text{$\mu(k)=\ka(\eta^{-1}(k))+\la(\ze^{-1}(k))$}}} \nonumber
\\
\raisebox{-9pt}{\begin{Large}$\displaystyle\biggl[$\end{Large}}
\sum_{\substack{\text{simply-connected directed graphs $\Ga$:}\\
\text{vertices $I\amalg J$, edges $\mathop{\bu} \limits^{\sst
i}\ra\mathop{\bu}\limits^{\sst j}$, $i\in I$, $j\in J$,}\\
\text{conn. components $\eta^{-1}(k)\amalg\ze^{-1}(k)$, $k\in K$}}}
\,\,\,\prod_{\substack{\text{edges}\\
\text{$\mathop{\bu}\limits^{\sst i}\ra\mathop{\bu}\limits^{\sst
j}$}\\ \text{in $\Ga$}}}\ha\bar\chi(\ka(i),\la(j))
\raisebox{-9pt}{\begin{Large}$\displaystyle\biggr]$\end{Large}},
\label{gf2eq18}
\end{gather}
extended $\C$-bilinearly. Then $\H$ is a $\C$-algebra with identity
$c_{[\emptyset,\emptyset]}$. For $\al\in C(\A)$ define
$c^\al=c_{[\{1\},\al']}$ where $\al'(1)=\al$, define $\L^\al=\C\cdot
c^\al$ and $\L=\bigop_{\al\in C(\A)}\L^\al$. Equation \eq{gf2eq18}
yields $[c^\al,c^\be]= \bar\chi(\al,\be)\,c^{\al+\be}$, so $\L$ is a
Lie subalgebra.

Then \cite[\S 6.5]{Joyc6} defines invariants $J^\al(\tau)\in\Q$ for
$\al\in C(\A)$, such that if we set
$\ep^\al(\tau)=J^\al(\tau)\,c^\al$ and define $\de^\al(\tau)$ by
\eq{gf2eq5} then Assumption \ref{gf2ass2} holds, with extra
conditions as in Remark \ref{gf2rem1}(d) above. These invariants
$J^\al(\tau)$ are defined using the {\it Euler characteristic} of
constructible sets in Artin $\K$-stacks, in a rather subtle way. As
Euler characteristic and virtual Poincar\'e polynomials are related
by $\chi(X)=P(X;-1)$, these are specializations of the virtual
Poincar\'e polynomial invariants of Examples \ref{gf2ex5}
and~\ref{gf2ex7}.

Note that we cannot define the $\de^\al(\tau)$ directly, but only
reconstruct them from the $\ep^\al(\tau)$. In the notation of
Example \ref{gf2ex4}, this is because $\ep^\al(\tau)$ is defined
using a Lie algebra morphism $\Psi:\SFai(\fObj_\A)\ra\L$ which does
{\it not\/} extend to an algebra morphism
$\Psi:\SFa(\fObj_\A)\ra\H$, so we cannot define
$\de^\al(\tau)=\Psi(\bdss^\al(\tau))$ as we might hope. The above
also holds with $\C$ replaced by other commutative $\C$-algebras
$\Om$, and Euler characteristics replaced by other $\Om$-valued
`motivic invariants' $\Th$ of $\K$-varieties with $\Th(\K)=1$; for
details see~\cite{Joyc2,Joyc4,Joyc6}.

To rewrite \eq{gf2eq7} as a transformation law for the $J^\al(\tau)$
we need to compute $c^{\al_1}*\cdots* c^{\al_n}$ in $\H$. Actually
it is enough to know the projection of this to $\L$. As in \cite[\S
6.5]{Joyc6}, calculation shows this is given by:
\begin{gather}
c^{\al_1}*\cdots*c^{\al_n}=\text{ terms in $c_{[I,\ka]}$,
$\md{I}>1$, }
\label{gf2eq19}\\
+\raisebox{-6pt}{\begin{Large}$\displaystyle\biggl[$\end{Large}}
\frac{1}{2^{n-1}}\!\!\!\!\!
\sum_{\substack{\text{connected, simply-connected digraphs
$\Ga$:}\\
\text{vertices $\{1,\ldots,n\}$, edge $\mathop{\bu} \limits^{\sst
i}\ra\mathop{\bu}\limits^{\sst j}$ implies $i<j$}}} \,\,\,
\prod_{\substack{\text{edges}\\
\text{$\mathop{\bu}\limits^{\sst i}\ra\mathop{\bu}\limits^{\sst
j}$}\\ \text{in $\Ga$}}}\bar\chi(\al_i,\al_j)
\raisebox{-6pt}{\begin{Large}$\displaystyle\biggr]$\end{Large}}
c^{\al_1+\cdots+\al_n}. \nonumber
\end{gather}
Here a {\it digraph\/} is a directed graph.

Let $(\tau,T,\le), (\ti\tau,\ti T,\le)$ be weak stability conditions
on $\A$, $\Ga$ be a connected, simply-connected digraph with finite
vertex set $I$, and $\ka:I\ra C(\A)$. Define
$V(I,\Ga,\ka,\tau,\ti\tau)\in\Q$ by
\e
V(I,\Ga,\ka,\tau,\ti\tau)=\frac{1}{2^{\md{I}-1}\md{I}!}
\!\!\sum_{\substack{\text{total orders $\pr$ on $I$:}\\ \text{edge
$\mathop{\bu} \limits^{\sst i}\ra\mathop{\bu}\limits^{\sst j}$ in
$\Ga$ implies $i\pr j$}}} \!\!\!\!\! U(I,\pr,\ka,\tau,\ti\tau).
\label{gf2eq20}
\e
Then using $\ep^\al(\tau)=J^\al(\tau)\,c^\al$ and \eq{gf2eq19}, it
turns out \cite[Th.~6.28]{Joyc6} that \eq{gf2eq7} is equivalent to
\e
J^\al(\ti\tau)\!=\!\!\!\!
\sum_{\substack{\text{iso.}\\ \text{classes}\\
\text{of finite}\\ \text{sets $I$}}}\,\,
\sum_{\substack{\ka:I\ra C(\A):\\ \ka(I)=\al}}\,\,
\sum_{\begin{subarray}{l} \text{connected,}\\
\text{simply-connected}\\ \text{digraphs $\Ga$,}\\
\text{vertices $I$}\end{subarray}\!\!\!\!\!\!\!\!\!\!\!\!\!\!\!\!\!
\!\!\!\!\!\!\!\!\!\!} V(I,\Ga,\ka,\tau,\ti\tau)
\begin{aligned}[t]
&\cdot\prod\limits_{\text{edges \smash{$\mathop{\bu}\limits^{\sst
i}\ra\mathop{\bu}\limits^{\sst j}$} in
$\Ga$}\!\!\!\!\!\!\!\!\!\!\!\!\!\!\!\!\!\!\!\!\!\!\!
\!\!\!\!\!\!\!\!\!\!\!\!\!\!} \bar\chi(\ka(i),\ka(j))\\
&\cdot\prod\nolimits_{i\in I}J^{\ka(i)}(\tau).
\end{aligned}
\label{gf2eq21}
\e
\label{gf2ex8}
\end{ex}

Example \ref{gf2ex8} is the reason why the title of the paper
involves Calabi--Yau 3-folds, why we believe that the ideas of this
paper have to do with Mirror Symmetry and String Theory, and why we
want to bring them to the attention of String Theorists in
particular so that they may explain them in physical terms. In
brief, the point is this.

In \cite[\S 6.5]{Joyc6}, as the culmination of a great deal of work
in \cite{Joyc1,Joyc2,Joyc3,Joyc4,Joyc5,Joyc6}, the author defined
invariants $J^\al(\tau)\in\Q$ `counting' $\tau$-semistable sheaves
in class $\al\in K(\A)$ on a Calabi--Yau 3-fold $X$, which transform
according to a complicated transformation law \eq{gf2eq21} under
change of weak stability condition, reminiscent of Feynman diagrams.

The author expects that some related invariants which extend
Donaldson--Thomas invariants and transform according to the same law
\eq{gf2eq21} should be important in String Theory, perhaps counting
numbers of branes or BPS states. For conjectures on this see
\cite[\S 6.5]{Joyc6}. This paper will study natural ways of
combining these invariants in holomorphic generating functions; the
author expects that these generating functions, and the equations
they satisfy, should also be significant in String Theory.

\subsection{Comments on the extension to triangulated categories}
\label{gf23}

The series \cite{Joyc3,Joyc4,Joyc5,Joyc6} studied only abelian
categories, such as the coherent sheaves $\coh(X)$ on a projective
$\K$-scheme $X$. But for applications to String Theory and Mirror
Symmetry, the whole programme should be extended to {\it
triangulated categories}, such as the {\it bounded derived
category\/} $D^b(\coh(X))$ of coherent sheaves on $X$. The issues
involved in this are discussed in \cite[\S 7]{Joyc6}. For a recent
survey on derived categories of coherent sheaves on Calabi--Yau
$m$-folds, see Bridgeland~\cite{Brid2}.

One justification for this is Kontsevich's {\it Homological Mirror
Symmetry} proposal \cite{Kont}, which explains Mirror Symmetry of
Calabi--Yau 3-folds $X,\hat X$ as an equivalence between
$D^b(\coh(X))$ and the derived Fukaya category $D^b(F(\hat X))$ of
$\hat X$. This relates the complex algebraic geometry of $X$,
encoded in $D^b(\coh(X))$, to the symplectic geometry of $\hat X$,
encoded in $D^b(F(\hat X))$. Building on Kontsevich's ideas,
triangulated categories of branes have appeared in String Theory in
the work of Douglas, Aspinwall, Diaconescu, Lazaroiu and others.

The following notion of stability condition on a triangulated
category, due to Bridgeland \cite[\S 1.1]{Brid1}, will be important
in this programme. For background on triangulated categories, see
Gelfand and Manin~\cite{GeMa}.

\begin{dfn} Let $\T$ be a triangulated category, and $K(\T)$ the
quotient of its Grothendieck group $K_0(\T)$ by some fixed subgroup.
For instance, if $\T$ is of {\it finite type} over a field $\K$ we
can take $K(\T)$ to be the {\it numerical Grothendieck group}
$K^{\rm num}(\T)$ as in \cite[\S 1.3]{Brid1}, and then Bridgeland
calls the resulting stability conditions {\it numerical stability
conditions}.

A {\it stability condition} $(Z,\P)$ on $\T$ consists of a group
homomorphism $Z:K(\T)\ra\C$ called the {\it central charge}, and
full subcategories $\P(\phi)\subset\T$ for each $\phi\in\R$ of {\it
semistable objects with phase} $\phi$, satisfying:
\begin{itemize}
\setlength{\itemsep}{0pt}
\setlength{\parsep}{0pt}
\item[(a)] If $S\in\P(\phi)$ then $Z([S])=m([S]){\rm e}^{i\pi\phi}$
for some $m([S])\in(0,\iy)$;
\item[(b)] for all $t\in\R$, $\P(t+1)=\P(t)[1]$;
\item[(c)] if $t_1>t_2$ and $S_j\in\P(t_j)$ for $j=1,2$ then
$\Hom_\T(S_1,S_2)=0$; and
\item[(d)] for $0\ne U\in\T$ there is a finite sequence
$t_1>t_2>\cdots>t_n$ in $\R$ and a collection of distinguished
triangles with $S_j\in\P(t_j)$ for all~$j$:
\end{itemize}
\e
\xymatrix@C=10pt@R=15pt{ 0=A_0 \ar[rr] && A_1 \ar[rr]\ar[dl] && A_2
\ar[r] \ar[dl] & \cdots \ar[r] & A_{n-1} \ar[rr] && A_n=U. \ar[dl]\\
& S_1 \ar@{-->}[ul] && S_2 \ar@{-->}[ul] &&&& S_n \ar@{-->}[ul] }
\label{gf2eq22}
\e
\label{gf2def8}
\end{dfn}

This is the generalization to triangulated categories of the {\it
slope function} stability conditions of Example \ref{gf2ex2}. In
both cases we have a central charge homomorphism $Z:K(\A)\ra\C$ or
$Z:K(\T)\ra\C$, and semistability can be expressed in terms of
$\arg\ci Z$. In the abelian case $\arg\ci Z$ takes a unique value in
$(0,\pi)$, but in the triangulated case one has to choose a value of
$\arg\ci Z$ and lift phases from $\R/2\pi\Z$ to $\R$. This need to
choose phases is why in the abelian case $Z$ determines the
stability condition, but in the triangulated case we also need extra
data $\P$. Equation \eq{gf2eq22} is the analogue of Theorem
\ref{gf2thm1}, since both decompose an arbitrary object $U\in\A$ or
$\T$ into semistable objects $S_1,\ldots,S_n$ with phases satisfying
$\mu(S_1)>\cdots>\mu(S_n)$ or~$t_1>\cdots>t_n$.

There is also a generalized notion of stability condition on $\T$
due to Gorodentscev et al.\ \cite{GKR}, not involving a central
charge,  which is closer in spirit to Definition \ref{gf2def2}
above. But we will not use it. Here is Bridgeland's main result
\cite[Th.~1.2]{Brid1}, slightly rewritten:

\begin{thm} Let\/ $\T$ be a triangulated category and\/ $K(\T)$ as
in Definition \ref{gf2def8}. Write $\Stab(\T)$ for the set of
stability conditions $(Z,\P)$ on $\T$. Then\/ $\Stab(\T)$ has a
natural, Hausdorff topology. Let\/ $\Si$ be a connected component
of\/ $\Stab(\T)$. Then there is a complex vector subspace $V_\Si$ in
$\Hom(K(\T),\C)$ with a well-defined linear topology such that the
map $\Si\ra\Hom(K(\T),\C)$ given by $(Z,\P)\mapsto Z$ is a local
homeomorphism~$\Si\ra V_\Si$.

When $V_\Si$ is finite-dimensional, which happens automatically when
$K(\T)$ has finite rank, $\Si$ can be given the structure of a
complex manifold uniquely so that\/ $(Z,\P)\mapsto Z$ is a local
biholomorphism~$\Si\ra V_\Si$.
\label{gf2thm2}
\end{thm}

Bridgeland's stability conditions were motivated by Douglas' work on
Pi-stability, and are natural objects in String Theory. Suppose we
wish to define some kind of generating function $f^\al$ encoding
invariants `counting' $(Z,\P)$-semistable objects in class $\al$ in
$\T$. These invariants will depend on $(Z,\P)$, so the generating
function $f^\al$ should be a function on $\Stab(\T)$ (and perhaps in
other variables as well). Now Theorem \ref{gf2thm2} shows
$\Stab(\T)$ is a {\it complex manifold}, so it makes sense to
require $f^\al$ to be a {\it holomorphic function} on $\Stab(\T)$.
We can also try to make $f^\al$ {\it continuous}, despite the fact
that the invariants it encodes will change discontinuously over real
hypersurfaces in~$\Stab(\T)$.

This problem also makes sense in the abelian setting of Example
\ref{gf2ex2}, where we can try to define a generating function
$f^\al$ which is a continuous, holomorphic function on the complex
manifold $\Stab(\A)$ of \eq{gf2eq2}. In fact most of the rigorous
part of the paper is about Example \ref{gf2ex2}, but we have done it
in a way that the author expects will generalize to the triangulated
case when (if ever) the extension of \cite{Joyc3,Joyc4,Joyc5, Joyc6}
to triangulated categories has been worked out.

\section{Holomorphic generating functions}
\label{gf3}

Consider the following situation. Let Assumptions \ref{gf2ass1} and
\ref{gf2ass2} hold for $\A$, with $K(\A)$ of finite rank, and
suppose $\Stab(\A)$ in Example \ref{gf2ex2} is a nonempty open
subset of $\Hom(K(\A),\C)$, and so a complex manifold. This works
for all the quiver examples $\A=\modKQ,\nilKQ,\modKQI,\nilKQI,\modA$
of \cite[\S 10]{Joyc3}, with $\H,\L,\ldots$ chosen as in one of
Examples \ref{gf2ex3}--\ref{gf2ex8}.

Then we have a complex manifold $\Stab(\A)$ of central charges $Z$,
each of which defines a permissible stability condition
$(\mu,\R,\le)$. For this $\mu$ we have invariants
$\de^\al(\mu)\in\H^\al$ and $\ep^\al(\mu)\in\L^\al$ for all $\al\in
C(\A)$. Regarded as functions of $Z$, these $\de^\al(\mu),
\ep^\al(\mu)$ change {\it discontinuously} across real hypersurfaces
in $\Stab(\A)$ where $\arg\ci Z(\be)=\arg\ci Z(\ga)$ for $\be,\ga\in
C(\A)$ according to the transformation laws
\eq{gf2eq6}--\eq{gf2eq7}, and away from such hypersurfaces are
locally constant.

For $\al\in C(\A)$ we shall consider a generating function
$f^\al:\Stab(\A)\ra\H^\al$ of the following form, where
$(\mu,\R,\le)$ is the stability condition induced by~$Z$:
\e
f^\al(Z)=\!\!\!\!\sum_{\substack{n\ge 1,\;\al_1,\ldots,\al_n\in
C(\A):\\ \al_1+\cdots+\al_n=\al}}\!\!\!\!\!\!\!\!
F_n\bigl(Z(\al_1),\ldots,Z(\al_n)\bigr)
\ep^{\al_1}(\mu)*\ep^{\al_2}(\mu)*\cdots*\ep^{\al_n}(\mu).
\label{gf3eq1}
\e
We explain why we chose this form, and what conditions the $F_n$
must satisfy:

\begin{rem}{\bf(a)} The general form of \eq{gf3eq1} is modelled on
\eq{gf2eq4}--\eq{gf2eq7} above. The functions $F_n$ should map
$(\C^\t)^n\ra\C$, where $\C^\t=\C\sm\{0\}$. For the abelian category
case of Example \ref{gf2ex2} we have $Z(\al)\in H=\{x+iy:x\in\R,$
$y>0\}$ for $\al\in C(\A)$, so it would be enough to define $F_n$
only on $H^n$. However, for the extension to the triangulated
category case discussed in \S\ref{gf23} we must allow
$Z(\al_k)\in\C^\t$, which is why we chose the domain~$(\C^\t)^n$.
\smallskip

\noindent{\bf(b)} We require that the functions $F_n$ satisfy
\e
F_n(\la z_1,\ldots,\la z_n)=F_n(z_1,\ldots,z_n)\quad \text{for all
$\la,z_1,\ldots,z_n\in\C^\t$.}
\label{gf3eq2}
\e
The reason is easiest to explain in the triangulated category case.
Let $\T,K(\T)$ and $(Z,\P)$ be as in Definition \ref{gf2def8}, and
let $r>0$ and $\psi\in\R$. Define a new stability condition
$(Z',\P')$ on $\T$ by $Z'=r{\rm e}^{i\psi}Z$ and
$\P'(\phi)=\P(\phi-\psi/\pi)$.

This gives an action of $(0,\iy)\t\R$ on $\Stab(\T)$, which does not
change the sets of $(Z,\P)$-semistable objects, but only their
phases $\phi$. So we expect that in an appropriate extension of
Assumption \ref{gf2ass2} to the triangulated case, the invariants
$\de^\al(Z,\P),\ep^\al(Z,\P)$ `counting' $(Z,\P)$-semistable objects
in class $\al$ should be also unchanged by this action. Therefore we
can try and make $f^\al$ and each term in \eq{gf3eq1} invariant
under $Z\mapsto r{\rm e}^{i\psi}Z$, which is equivalent
to~\eq{gf3eq2}.

We can make a similar argument in the abelian case Example
\ref{gf2ex2}, but we have to restrict to $r{\rm e}^{i\psi}$ such
that $r{\rm e}^{i\psi}Z\bigl[C(\A)\bigr]\subset H$, which makes the
argument less persuasive. Requiring $f^\al$ and $F_n$ instead to be
{\it homogeneous of degree} $d\in\Z$, so that $F_n(\la
z_1,\ldots,\la z_n)=\la^d F_n(z_1,\ldots,z_n)$ for all $\la,z_k$, is
equivalent to replacing $f^\al(Z)$ by $Z(\al)^df^\al(Z)$. So we lose
nothing by restricting to~$d=0$.
\smallskip

\noindent{\bf(c)} Equation \eq{gf3eq2} implies that $F_1$ is
constant, say $F_1\equiv c$. For $\la\in\C^\t$ we may replace
$f^\al,F_n,c$ by $\la f^\al,\la F_n,\la c$ without changing whether
$f^\al$ is holomorphic or continuous, so all nonzero choices of $c$
are equivalent. We shall take
\e
F_1\equiv (2\pi i)^{-1},
\label{gf3eq3}
\e
as this simplifies formulae in \S\ref{gf32} and the rest of the
paper.

Think of \eq{gf3eq1} as saying $f^\al(Z)=c\,\ep^\al(\mu)+$`higher
order terms'. If $\ep^\al(\mu)$ is an invariant `counting'
$\mu$-semistables in class $\al$, then so is $f^\al(Z)$, to leading
order. But $\ep^\al(\mu)$ changes discontinuously with $Z$, whereas
$f^\al(Z)$ includes higher order correction terms which smooth out
these changes and make $f^\al$ continuous.
\smallskip

\noindent{\bf(d)} Following equations \eq{gf2eq7} and \eq{gf2eq8},
we may rewrite \eq{gf3eq1} as
\e
\begin{aligned}
&f^\al(Z)=\\
&\sum_{\substack{\text{iso classes}\\
\text{of finite}\\ \text{sets $I$}}} \frac{1}{\md{I}!}\!
\sum_{\substack{\ka:I\ra C(\A):\\ \ka(I)=\al}}\!
\raisebox{-12pt}{\begin{Large}$\displaystyle\Biggl[$\end{Large}}
\!\!
\sum_{\substack{\text{total orders $\pr$ on $I$.}\\
\text{Write $I=\{i_1,\ldots,i_n\}$,}\\
\text{$i_1\pr i_2\pr\cdots\pr i_n$}}}\!\!\!\!\!\!\!
\begin{aligned}[t]
F_{\md{I}}&(Z\ci\ka(i_1),\ldots,Z\ci\ka(i_n))\cdot\\
&\ep^{\ka(i_1)}(\mu)\!*\!\cdots\!*\!\ep^{\ka(i_n)}(\mu)
\end{aligned}
\raisebox{-12pt}{\begin{Large}$\displaystyle\Biggr]$\end{Large}}.
\end{aligned}
\label{gf3eq4}
\e
As for \eq{gf2eq8}, we shall require the functions $F_n$ to have the
property that the term $[\cdots]$ in \eq{gf3eq4} is a finite
$\C$-linear combination of multiple commutators of $\ep^{\ka(i)}$
for $i\in I$, and so it lies in the Lie algebra $\L$, not just the
algebra $\H$. Thus \eq{gf3eq1} and \eq{gf3eq4} make sense in $\L$,
and $f^\al$ actually maps~$\Stab(\A)\ra\L^\al$.

This is why we choose to write \eq{gf3eq1} in terms of the
$\ep^\al(\mu)$ rather than the $\de^\al(\mu)$. By substituting
\eq{gf2eq4} into \eq{gf3eq1} we get another equation of the same
form for $f^\al$, but with $\de^{\al_i}(\mu)$ instead of
$\ep^{\al_i}(\mu)$, and different functions $F_n$. But using the
$\ep^{\al_i}(\mu)$ means we can work in $\L$ rather than $\H$, which
is a great simplification if $\L$ is much smaller than $\H$. This
happens in Example \ref{gf2ex8}, our motivating Calabi--Yau 3-fold
example, when $\L^\al=\C\cdot c^\al$ so $f^\al$ is really just a
holomorphic function, but $\H^\al$ is in general
infinite-dimensional.

Now for $\md{I}>1$, if a $\C$-linear combination of products of
$\ep^{\ka(i)}(\mu)$ for $i\in I$ is a sum of multiple commutators,
it is easy to see that the sum of the coefficients of the products
must be zero in $\C$. Thus, a necessary condition for $[\cdots]$ in
\eq{gf3eq4} to be a linear combination of multiple commutators is
that
\e
\ts\sum_{\si\in S_n}F_n(z_{\si(1)},\ldots,z_{\si(n)})=0
\quad\text{for all $n>1$ and $(z_1,\ldots,z_n)\in(\C^\t)^n$,}
\label{gf3eq5}
\e
where $S_n$ is the {\it symmetric group\/} of permutations $\si$
of~$\{1,\ldots,n\}$.
\smallskip

\noindent{\bf(e)} We require that $f^\al$ be a {\it continuous\/}
and {\it holomorphic\/} function on $\Stab(\A)$. These translate to
conditions on the functions $F_n$. In \S\ref{gf31} we will compute
the conditions on $F_n$ for $f^\al$ to be continuous; it turns out
that across real hypersurfaces $\arg z_l=\arg z_{l+1}$, $F_n$ must
jump by expressions in $F_k$ for $k<n$. For $f^\al$ to be
holomorphic, it is enough that the $F_n$ be holomorphic wherever
they are continuous.

Thus, $F_n$ is a branch of a multivalued holomorphic function,
except along $\arg z_l=\arg z_{l+1}$ where it jumps discontinuously
from one branch to another; but the discontinuities in
$\ep^\al(\mu)$ and $F_n(\cdots)$ cancel out to make $f^\al$
continuous. A simple comparison is a branch of $\log z$ on $\C^\t$,
cut along~$(0,\iy)$.
\smallskip

\noindent{\bf(f)} We shall ensure uniqueness of the $F_n$ by
imposing a {\it growth condition\/}:
\e
\bmd{F_n(z_1,\ldots,z_n)}=o\bigl(\md{z_k}^{-1}\bigr) \quad\text{as
$z_k\ra 0$ with $z_l$ fixed, $l\ne k$, for all $k$.}
\label{gf3eq6}
\e
This may assist the convergence of \eq{gf3eq1} in situations when
the sum is infinite.
\label{gf3rem1}
\end{rem}

For the moment we impose the following extra condition. It implies
there are only finitely many possibilities for $n$ and
$\al_1,\ldots,\al_n$ in \eq{gf3eq1}, and so avoids problems with
infinite sums and convergence. It holds for all the quiver examples
$\A=\modKQ,\ldots$ of \cite[\S 10]{Joyc3}, but not for $\A=\coh(X)$
when~$\dim X>0$.

\begin{ass} In the situation of Assumption \ref{gf2ass1}, for each
$\al\in C(\A)$ there are only finitely pairs $\be,\ga\in C(\A)$
with~$\al=\be+\ga$.
\label{gf3ass}
\end{ass}

In the rest of the section we construct functions $F_n$ satisfying
the requirements of Remark \ref{gf3rem1}, and show that they are
{\it unique}, and satisfy interesting {\it partial differential
equations}.

\subsection{Conditions on the functions $F_n$ for $f^\al$ to be
continuous}
\label{gf31}

Let Assumptions \ref{gf2ass1}, \ref{gf2ass2} and \ref{gf3ass} hold
for $\A$, with $K(\A)$ of finite rank, and suppose $\Stab(\A)$ in
Example \ref{gf2ex2} is a nonempty open subset of $\Hom(K(\A),\C)$,
and so a complex manifold. Let $Z,\ti Z\in\Stab(\A)$, with
associated stability conditions $(\mu,\R,\le)$ and
$(\ti\mu,\R,\le)$. We think of $Z$ as varying in $\Stab(\A)$ and
$\ti Z$ as a fixed base point.

We need some notation for the coefficients
$S,U(\{1,\ldots,n\},\le,\ka,\mu,\ti\mu)$ of \S\ref{gf21}. They
depend on the $2n$ complex numbers $Z\ci\ka(k),\ti Z\ci\ka(k)$ for
$k=1,\ldots,n$ in $H=\{x+iy:x\in\R,$ $y>0\}$, and the definition
makes sense for any $2n$ elements of $H$. Thus there are unique
functions $s_n,u_n:H^{2n}\ra\Q$ written $s_n,u_n(z_1,\ldots,z_n;\ti
z_1,\ldots,\ti z_n)$ such that
\e
\begin{split}
S(\{1,\ldots,n\},\le,\ka,\mu,\ti\mu)&\!=\!s_n(Z\!\ci\!\ka(1),\ldots,
Z\!\ci\!\ka(n);\ti Z\!\ci\!\ka(1),\ldots,\ti Z\!\ci\!\ka(n)),\\
U(\{1,\ldots,n\},\le,\ka,\mu,\ti\mu)&\!=\!u_n(Z\!\ci\!\ka(1),\ldots,
Z\!\ci\!\ka(n);\ti Z\!\ci\!\ka(1),\ldots,\ti Z\!\ci\!\ka(n)).
\end{split}
\label{gf3eq7}
\e

Then using \eq{gf2eq7} with $\al_i,\ti\mu,\mu$ in place of
$\al,\tau,\ti\tau$ respectively to express $\ep^{\al_i}(\mu)$ in
\eq{gf3eq1} in terms of $\ep^{\ka(j)}(\ti\mu)$, using \eq{gf3eq7}
and rewriting, we find that
\begin{gather}
f^\al(Z)=\sum_{\substack{n\ge 1,\;\al_1,\ldots,\al_n\in C(\A):\\
\al_1+\cdots+\al_n=\al}}\ep^{\al_1}(\ti\mu)*\ep^{\ka(2)}(\ti\mu)*\cdots*
\ep^{\al_n}(\ti\mu)\cdot
\label{gf3eq8}\\
\raisebox{-12pt}{\begin{Large}$\displaystyle\Biggl[$\end{Large}}
\!\sum_{\begin{subarray}{l} m=1,\ldots,n,\\ 0=a_0<a_1<\\
\cdots<a_m=n\end{subarray}}\!\!
\begin{aligned}[t]
&F_m\bigl(Z(\al_{a_0+1}\!+\!\cdots\!+\!\al_{a_1}),\ldots,
Z(\al_{a_{m-1}+1}\!+\!\cdots\!+\!\al_{a_m})\bigr)\\
&\ts\prod\limits_{k=1}^m\!\!u_{a_k-a_{k-1}}\bigl(\ti
Z(\al_{a_{k-1}\!+\!1}),\ldots, \ti
Z(\al_{a_k});Z(\al_{a_{k-1}\!+\!1}),\ldots, Z(\al_{a_k})\bigr)
\end{aligned}
\raisebox{-12pt}{\begin{Large}$\displaystyle\Biggr]$\end{Large}}.
\nonumber
\end{gather}
We get this by decomposing $\al=\be_1+\cdots+\be_m$ in \eq{gf3eq1},
and then decomposing $\be_k=\al_{a_{k-1}+1}+\cdots+\al_{a_k}$ as
part of an expression \eq{gf2eq7} for $\ep^{\be_k}(\mu)$,
for~$k=1,\ldots,m$.

We rewrite the bottom line $[\cdots]$ of \eq{gf3eq8} using a
function $G_n:H^{2n}\ra\C$ by
\begin{gather}
f^\al(Z)=\sum_{\substack{n\ge 1,\;\al_1,\ldots,\al_n\in C(\A):\\
\al_1+\cdots+\al_n=\al}}
\begin{aligned}[t]
&G_n\bigl(Z(\al_1),\ldots,Z(\al_n);\ti Z(\al_1),\ldots,\ti
Z(\al_n)\bigr)\cdot\\
&\ep^{\al_1}(\ti\mu)*\ep^{\al_2}(\ti\mu)*\cdots*\ep^{\al_n}(\ti\mu),
\end{aligned}
\label{gf3eq9}\\
\begin{aligned}
&\text{where}\quad G_n(z_1,\ldots,z_n;\ti z_1,\ldots,\ti z_n)=\\
&\sum_{\substack{m=1,\ldots,n,\\
0=a_0<a_1<\cdots<a_m=n}}
\begin{aligned}[t]
&F_m(z_{a_0+1}+\cdots\!+z_{a_1},\ldots,
z_{a_{m-1}+1}+\cdots+z_{a_m})\\
&\ts\prod_{k=1}^m\!\!u_{a_k-a_{k-1}}(\ti z_{a_{k-1}\!+\!1},
\ldots,\ti z_{a_k};z_{a_{k-1}\!+\!1},\ldots,z_{a_k}).
\end{aligned}
\end{aligned}
\label{gf3eq10}
\end{gather}
In \eq{gf3eq9}, the terms $\ep^{\al_1}(\ti\mu)*\cdots
*\ep^{\al_n}(\ti\mu)$ and $\ti Z(\al_1),\ldots,\ti Z(\al_n)$ are
constants independent of $Z$. Thus it is clear that $f^\al$ is
continuous, or holomorphic, provided the function
$(z_1,\ldots,z_n)\mapsto G_n(z_1,\ldots,z_n;\ti z_1, \ldots,\ti
z_n)$ is continuous, or holomorphic, for each fixed~$(\ti
z_1,\ldots,\ti z_n)\in H^n$.

We can now substitute \eq{gf2eq7} with $\al_i,\mu,\ti\mu$ in place
of $\al,\tau,\ti\tau$ respectively to express $\ep^{\al_i}(\ti\mu)$
in \eq{gf3eq9} in terms of $\ep^{\ka(j)}(\mu)$. Rewriting gives an
expression for $f^\al(Z)$ as a linear combination of
$\ep^{\al_1}(\mu)*\cdots*\ep^{\al_n}(\mu)$, as in \eq{gf3eq1}. In
fact the coefficients of $\ep^{\al_1}(\mu)*\cdots*\ep^{\al_n}(\mu)$
in the two expressions must agree; we can prove this either by using
\cite[Ex.~7.10]{Joyc5}, in which the $\ep^{\al_1}(\mu)*\cdots*
\ep^{\al_n}(\mu)$ are linearly independent in $\H$ for all
$\al_1,\ldots,\al_n$ satisfying some conditions, or by using
combinatorial properties of the coefficients $U(\cdots)$
from~\cite[Th.~4.8]{Joyc6}.

Equating the two writes $F_n\bigl(Z(\al_1),\ldots,Z(\al_n)\bigr)$ in
terms of the functions $G_m$ and $u_k$. Since $Z(\al_k),\ti
Z(\al_k)$ can take arbitrary values in $H$, we deduce an expression
for $F_n(z_1,\ldots,z_n)$ when $z_k,\ti z_k\in H$, the inverse
of~\eq{gf3eq10}:
\e
\begin{aligned}
&F_n(z_1,\ldots,z_n)=\\
&\sum_{\substack{m=1,\ldots,n,\\
0=a_0<a_1<\cdots<a_m=n}}\!\!\!\!
\begin{aligned}[t]
G_m(&z_{a_0+1}+\cdots\!+z_{a_1},\ldots,
z_{a_{m-1}+1}+\cdots+z_{a_m};\\
&\ti z_{a_0+1}+\cdots\!+\ti
z_{a_1},\ldots,\ti z_{a_{m-1}+1}+\cdots+\ti z_{a_m})\cdot\\
&\ts\prod_{k=1}^mu_{a_k-a_{k-1}}(z_{a_{k-1}\!+\!1},
\ldots,z_{a_k};\ti z_{a_{k-1}\!+\!1},\ldots,\ti z_{a_k}).
\end{aligned}
\end{aligned}
\label{gf3eq11}
\e
Note that although we want $F_n$ to map $(\C^\t)^n\ra\C$, for the
moment \eq{gf3eq11} is defined only when $z_k,\ti z_k\in H$, since
we have so far defined $u_n$ and $G_n$ only on $H^{2n}$, not on
$(\C^\t)^{2n}$. Note too that \eq{gf3eq11} holds for arbitrary~$\ti
z_1,\ldots,\ti z_n\in H$.

Here are some conclusions so far.

\begin{prop} Suppose Assumptions \ref{gf2ass1}, \ref{gf2ass2} and
\ref{gf3ass} hold for\/ $\A$ with\/ $K(\A)$ of finite rank and\/
$\Stab(\A)$ is a nonempty open subset of\/ $\Hom(K(\A),\C)$, and let
some functions\/ $F_n:(\C^\t)^n\ra\C$ be given. Then a sufficient
condition for the function\/ $f^\al$ of\/ \eq{gf3eq1} to be
continuous, or holomorphic, is that for fixed\/ $(\ti z_1,\ldots,\ti
z_n)\in H^n$ the function\/ $(z_1,\ldots,z_n)\mapsto
G_n(z_1,\ldots,z_n;\ti z_1,\ldots,\ti z_n)$ is continuous, or
holomorphic. This holds for some\/ $(\ti z_1,\ldots,\ti z_n)\in H^n$
if and only if it holds for all\/~$(\ti z_1,\ldots,\ti z_n)$.

This condition is also necessary, for all values of\/ $n$ occurring
in\/ \eq{gf3eq1}, if the terms $\ep^{\al_1}(\mu)*\ep^{\al_2}(\mu)
*\cdots*\ep^{\al_n}(\mu)$ occurring in \eq{gf3eq1} are  linearly
independent in $\H$. This happens in the examples of\/
{\rm\cite[Ex.~7.10]{Joyc5}}, for arbitrarily large~$n$.
\label{gf3prop1}
\end{prop}

To go further we must understand the functions
$s_n,u_n(z_1,\ldots,z_n;\ti z_1,\ldots,\ti z_n)$ better. From
Example \ref{gf2ex2} and Definitions \ref{gf2def5} and
\ref{gf2def6}, we see that these depend on whether the inequalities
$\arg(z_a+\cdots+z_b)>\arg(z_{b+1}+\cdots+z_c)$ and $\arg(\ti
z_a+\cdots+\ti z_b)>\arg(\ti z_{b+1}+\cdots+\ti z_c)$ hold for each
choice of $1\le a\le b<c\le n$, choosing $\arg(\cdots)$ uniquely in
$(0,\pi)$ as `$\cdots$' lies in~$H$.

For each $(\ti z_1,\ldots,\ti z_n)\in H^n$, define
\e
\begin{split}
N_{(\ti z_1,\ldots,\ti z_n)}\!=\!\bigl\{(z_1,\ldots,z_n)\!\in\!
H^n\!:\!\text{if $\arg(\ti z_a\!+\!\cdots\!+\!\ti z_b)\!>\!
\arg(\ti z_{b+1}\!+\!\cdots\!+\!\ti z_c)$}&\\
\text{then
$\arg(z_a\!+\!\cdots\!+\!z_b)\!>\!\arg(z_{b+1}\!+\!\cdots\!+\!z_c)$,
for all $1\!\le\!a\!\le\!b\!<\!c\!\le\!n$}\bigr\}.&
\end{split}
\label{gf3eq12}
\e
Then $N_{(\ti z_1,\ldots,\ti z_n)}$ is an {\it open\/} subset of
$H^n$, as it is defined by strict inequalities, and $(\ti
z_1,\ldots,\ti z_n)\in N_{(\ti z_1,\ldots,\ti z_n)}$. As in
\cite[Def.~4.10]{Joyc5} we say that $(\ti\mu,\R,\le)$ {\it
dominates} $(\mu,\R,\le)$ if $\ti\mu(\al)>\ti\mu(\be)$ implies
$\mu(\al)>\mu(\be)$ for all $\al,\be\in C(\A)$. From \cite[\S
5.2]{Joyc6}, this implies that
\e
S(\{1,\ldots,n\},\le,\ka,\mu,\ti\mu)=
\begin{cases} 1, & \mu\ci\ka(1)\!>\!\cdots\!>\!\mu\ci\ka(n),
\;\ti\mu\ci\ka\!\equiv\!\ti\mu(\al), \\
0, &\text{otherwise,} \end{cases}
\label{gf3eq13}
\e
for all $\A$-data $(\{1,\ldots,n\},\le,\ka)$
with~$\ka(\{1,\ldots,n\})=\al$.

If $(z_1,\ldots,z_n)\in N_{(\ti z_1,\ldots,\ti z_n)}$ then the same
argument shows that
\e
s_n(z_1,\ldots,z_n;\ti z_1,\ldots,\ti z_n)=
\begin{cases} 1, & {\begin{subarray}{l}
\ts\text{$\arg(z_1)\!>\!\cdots\!>\!\arg(z_n)$ and}\\
\ts\text{$\arg(\ti z_k)\!=\!\arg(\ti z_1+\cdots+\ti z_n)$, all $k$,}
\end{subarray}} \\
0, &\text{otherwise,} \end{cases}
\label{gf3eq14}
\e
since the conditions in \eq{gf3eq12} play the same role as
$\ti\mu(\al)>\ti\mu(\be)$ implies $\mu(\al)>\mu(\be)$ does in
\eq{gf3eq13}. From \eq{gf2eq3} we deduce that if
$(z_1,\ldots,z_n)\in N_{(\ti z_1,\ldots,\ti z_n)}$ then
\ea
&u_n(z_1,\ldots,z_n;\ti z_1,\ldots,\ti z_n)=
\nonumber\\
&\sum_{1\le l\le m\le n} \!\!\!\!\!\!\!\!\!\!\!\!\!\!
\sum_{\substack{
\text{surjective $\psi:\{1,\ldots,n\}\!\ra\!\{1,\ldots,m\}$}\\
\text{and\/ $\xi:\{1,\ldots,m\}\!\ra\!\{1,\ldots,l\}$:}\\
\text{$a\!\le\!b$ implies $\psi(a)\!\le\!\psi(b)$,
$a\!\le\!b$ implies $\xi(a)\!\le\!\xi(b)$,}\\
\text{$\arg(\ti z_a)\!=\!\arg(\ti z_1+\cdots+\ti z_n)$, all $a$,}\\
\text{$\psi(a)=\psi(b)$ implies $\arg(z_a)=\arg(z_b)$,}\\
\text{$\psi(a)\!<\!\psi(b)$ and
$\xi\!\ci\!\psi(a)\!=\!\xi\!\ci\!\psi(b)$ imply
$\arg(z_a)\!>\!\arg(z_b)$}}}
\!\!\!\!\!\!\!\!\!\!\!\!\!\!\!\!\!\!\!\!\!\!\!\!
\frac{(-1)^{l-1}}{l}\cdot\prod_{c=1}^m\frac{1}{\md{\psi^{-1}(c)}!}\,,
\label{gf3eq15}\\
&\text{and $u_n(z_1,\ldots,z_n;\ti z_1,\ldots,\ti z_n)=0$ if
$\arg(\ti z_k)\ne\arg(\ti z_1+\cdots+\ti z_n)$, some $k$.} \nonumber
\ea

We have been working with $z_k,\ti z_k\in H$, since $s_n,u_n,G_n$
are, so far, defined only on $H^{2n}$. We shall now restate the
conditions of Proposition \ref{gf3prop1} for $f^\al$ to be
continuous, or holomorphic, in a way which makes sense for~$z_k,\ti
z_k\in\C^\t$.

\begin{cond} Let some functions $F_n:(\C^\t)^n\ra\C$ be given for
$n\ge 1$. For all $n\ge 1$ and $(\ti z_1,\ldots,\ti z_n)\in
(\C^\t)^n$ there should exist an open neighbourhood $N_{(\ti
z_1,\ldots,\ti z_n)}$ of $(\ti z_1,\ldots,\ti z_n)$ in $(\C^\t)^n$,
such that if $(z_1,\ldots,z_n)\in N_{(\ti z_1,\ldots,\ti z_n)}$ then
$\Re(z_k\ti z_k^{-1})>0$ for $k=1,\ldots,n$. For $(z_1,\ldots,z_n)
\in N_{(\ti z_1,\ldots,\ti z_n)}$ we must have
\begin{gather}
F_n(z_1,\ldots,z_n)=\!\!\!\!\!\!\!\!\!\!\!\!\!\!\!\!\!\!\!\!\!\!\!\!\!
\sum_{\begin{subarray}{l}m=1,\ldots,n,\; 0=a_0<a_1<\cdots<a_m=n\\
\text{and $c_1,\ldots,c_m\in[0,2\pi):$}\\
\ti z_a\in {\rm e}^{ic_k}(0,\iy),\; a_{k-1}<a\le a_k,\;
k=1,\ldots,m\end{subarray}}\!\!\!\!\!\!\!\!\!\!\!\!\!\!\!\!\!\!
\begin{aligned}[t]
G_m(&z_{a_0+1}\!+\!\cdots\!+\!z_{a_1},\ldots,
z_{a_{m-1}+1}\!+\!\cdots\!+\!z_{a_m};\\
&\ti z_{a_0+1}\!+\!\cdots\!+\!\ti z_{a_1},\ldots,\ti
z_{a_{m-1}+1}\!+\!\cdots\!+\!\ti z_{a_m}) \cdot
\end{aligned}
\nonumber\\
\prod_{k=1}^m\,\,
\sum_{\begin{subarray}{l}
\text{$1\le l_k\le m_k\le a_k-a_{k-1}$}\\
\text{surjective $\psi_k:\{a_{k-1}+1,\ldots,a_k\}\!\ra\!\{1,\ldots,m_k\}$}\\
\text{and\/ $\xi_k:\{1,\ldots,m_k\}\!\ra\!\{1,\ldots,l_k\}$:}\\
\text{$a\!\le\!b$ implies $\psi_k(a)\!\le\!\psi_k(b)$,
$a\!\le\!b$ implies $\xi_k(a)\!\le\!\xi_k(b)$,}\\
\text{$\psi_k(a)=\psi_k(b)$ implies $\arg(z_a)=\arg(z_b)$,}\\
\text{$\psi_k(a)\!<\!\psi_k(b)$ and
$\xi_k\!\ci\!\psi_k(a)\!=\!\xi_k\!\ci\!\psi_k(b)$ imply
$\arg(z_a)\!>\!\arg(z_b)$,}\\
\text{taking $\arg(z_a),\arg(z_b)\in(c_k-\pi/2,c_k+\pi/2)$}
\end{subarray}} \!\!\!\!\!\!\!\!\!\!\!\!\!\!\!\!\!\!\!\!\!\!\!
\!\!\!\!\!\!\!\!\!\!\!\!\!\!\!\!\!\!\!\!\!\!\!\!\!
\frac{(-1)^{l_k-1}}{l_k}\cdot\prod_{c=1}^{m_k}
\frac{1}{\md{\psi_k^{-1}(c)}!}\,,
\label{gf3eq16}
\end{gather}
where $G_m(\cdots)$ are some functions defined on the subsets of
$(\C^\t)^{2m}$ required by \eq{gf3eq16}, such that the maps
$(z_1,\ldots,z_m)\mapsto G_m(z_1,\ldots,z_m;\ti z_1,\ldots,\ti z_m)$
are continuous (for $f^\al$ to be continuous), and holomorphic (for
$f^\al$ to be holomorphic), in their domains.
\label{gf3cond}
\end{cond}

Here are some remarks on this:
\begin{itemize}
\setlength{\itemsep}{0pt}
\setlength{\parsep}{0pt}
\item From \eq{gf3eq15}, for $(z_1,\ldots,z_n)$ in an open neighbourhood
of $(\ti z_1,\ldots,\ti z_n)$ we see that the term $u_{a_k-a_{k-1}}
(z_{a_{k-1}\!+\!1},\ldots,z_{a_k};\ti z_{a_{k-1}\!+\!1},\ldots,\ti
z_{a_k})$ in \eq{gf3eq11} is zero unless $\arg(\ti z_a)=c_k$ for all
$a_{k-1}<a\le a_k$ and some $c_k$. We have put this in as a
restriction in the first line of \eq{gf3eq16}, expressing it as $\ti
z_a\in {\rm e}^{ic_k}(0,\iy)$ rather than $\arg(\ti z_a)=c_k$
because of the multivalued nature of~$\arg$.
\item We put in an extra condition $\Re(z_k\ti z_k^{-1})>0$ for all
$k$ when $(z_1,\ldots,z_n)\in N_{(\ti z_1,\ldots,\ti z_n)}$. The
main point of this is in \eq{gf3eq16} we have that $\Re\bigl({\rm
e}^{-ic_k}z_a\bigr)>0$ for $a_{k-1}<a\le a_k$, so $\Re\bigl({\rm
e}^{-ic_k}(z_{a_{k-1}+1}+\cdots+z_{a_k})\bigr)>0$, and thus the
argument $z_{a_{k-1}+1}+\cdots+z_{a_k}$ in $G_m(\cdots)$ in
\eq{gf3eq16} is nonzero. That is, \eq{gf3eq16} only needs $G_m$ to
be defined on a subset of $(\C^\t)^{2m}$.

We also use this condition in the second line, as when
$a_{k-1}<a,b\le a_k$ we can choose $\arg(z_a),\arg(z_b)$ uniquely
in~$(c_k-\pi/2,c_k+\pi/2)$.
\item When restricted to $H^n$, Condition \ref{gf3cond} is
equivalent to the conditions of Proposition \ref{gf3prop1} for
$f^\al$ to be continuous, or holomorphic, as the arguments above
show. But Condition \ref{gf3cond} also makes sense on $(\C^\t)^n$,
where we want $F_n$ to be defined for the extension to the
triangulated category case. Calculations by the author indicate that
Condition \ref{gf3cond} is the correct extension to the triangulated
case.
\item The point of restricting to neighbourhoods $N_{(\ti z_1,\ldots,
\ti z_n)}$ is partly because there we can use the formula
\eq{gf3eq15}, but mostly because we do not have a meaningful
extension of $u_n$ from $H^{2n}$ to all of $(\C^\t)^{2n}$, so that
\eq{gf3eq10} and \eq{gf3eq11} do not make sense. But for any fixed
$(\ti z_1,\ldots,\ti z_n)$ we can use \eq{gf3eq15} to define
$u_n(z_1,\ldots,z_n;\ti z_1,\ldots,\ti z_n)$ for $(z_1,\ldots,z_n)$
sufficiently close to $(\ti z_1,\ldots,\ti z_n)$, and this is the
basis of Condition~\ref{gf3cond}.
\end{itemize}

Now suppose that $(\ti z_1,\ldots,\ti z_n)\in(\C^\t)^n$ with $\ti
z_{k+1}/\ti z_k\notin(0,\iy)$ for all $1\le k<n$, and let
$(z_1,\ldots,z_n)\in N_{(\ti z_1,\ldots,\ti z_n)}$. Then in the
first sum in \eq{gf3eq16} we cannot have $a_{k-1}\le a_k-2$ for any
$k$, as then $\ti z_{a_k},\ti z_{a_k-1}\in {\rm e}^{ic_k}(0,\iy)$,
contradicting $\ti z_{a_k}/\ti z_{a_k-1}\notin(0,\iy)$. Thus the
only term in the first sum is $m=n$ and $a_k=k$ for $0\le k\le n$,
so the only terms in the second line are $l_k=m_k=1$ and
$\{a_k\}\,\smash{{\buildrel\psi_k\over\longra}}\,\{1\}
\smash{{\buildrel\xi_k\over\longra}}\,\{1\}$, and \eq{gf3eq11}
reduces to
\e
\begin{gathered}
F_n(z_1,\ldots,z_n)=G_n(z_1,\ldots,z_n;\ti z_1,\ldots,\ti z_n)\\
\text{if $\ti z_{k+1}/\ti z_k\notin(0,\iy)$ for all $k$ and
$(z_1,\ldots,z_n)\in N_{(\ti z_1,\ldots,\ti z_n)}$.}
\end{gathered}
\label{gf3eq17}
\e
Thus Condition \ref{gf3cond} requires $F_n$ to be continuous, or
holomorphic, on $N_{(\ti z_1,\ldots,\ti z_n)}$, an open
neighbourhood of $(\ti z_1,\ldots,\ti z_n)$. So we deduce:

\begin{prop} Condition \ref{gf3cond} implies that
the function\/ $F_n$ must be continuous, and holomorphic, on the set
\e
\bigl\{(z_1,\ldots,z_n)\in(\C^\t)^n:\text{$z_{k+1}/z_k\notin
(0,\iy)$ for all $1\le k<n$}\bigr\}.
\label{gf3eq18}
\e
\label{gf3prop2}
\end{prop}

Similarly, suppose $(\ti z_1,\ldots,\ti z_n)\in(\C^\t)^n$ with $\ti
z_{l+1}/\ti z_l\in(0,\iy)$ for some $1\le l<n$, and $\ti z_{k+1}/\ti
z_k\notin(0,\iy)$ for all $1\le k<n$, $k\ne l$. Then in the first
sum there are two terms, $m=n$ and $a_k=k$ for $0\le k\le n$ as
before, and $m=n-1$ and $a_k=k$ for $0\le k<l$, $a_k=k+1$ for $l\le
k<n$. Rewriting $\arg(z_l)>\arg(z_{l+1})$ as $\Im(z_{l+1}/z_l)<0$,
and so on, we find \eq{gf3eq16} reduces to
\begin{gather*}
F_n(z_1,\ldots,z_n)=G_n(z_1,\ldots,z_n;\ti z_1,\ldots,\ti z_n)\\
\begin{aligned}
+G_{n-1}(&z_1,\ldots,z_{l-1},z_l+z_{l+1},z_{l+2},\ldots,z_n;\\
&\ti z_1,\ldots,\ti z_{l-1},\ti z_l+\ti z_{l+1},\ti
z_{l+2},\ldots,\ti z_n)
\end{aligned}
\,\cdot
\begin{cases}
\phantom{-}\ha, & \Im(z_{l+1}/z_l)<0,\\
\phantom{-}0, & \Im(z_{l+1}/z_l)=0,\\
-\ha, & \Im(z_{l+1}/z_l)>0.
\end{cases}
\end{gather*}
By \eq{gf3eq17} this $G_{n-1}(\cdots)$ is $F_{n-1}(z_1,\ldots,
z_{l-1},z_l+z_{l+1},z_{l+2},\ldots,z_n)$, giving:

\begin{prop} Condition \ref{gf3cond} implies that if\/ $(\ti z_1,
\ldots,\ti z_n)\in(\C^\t)^n$ with $\ti z_{l+1}/\ti z_l\in(0,\iy)$
for some $1\le l<n$, and\/ $\ti z_{k+1}/\ti z_k\notin(0,\iy)$ for
all\/ $1\le k<n$ with\/ $k\ne l$, then for $(z_1,\ldots,z_n)$ in an
open neighbourhood of\/ $(\ti z_1,\ldots,\ti z_n)$ in $(\C^\t)^n$,
the following function is continuous, and holomorphic:
\e
\begin{split}
&F_n(z_1,\ldots,z_n)-\\[-10pt]
&F_{n-1}(z_1,\ldots,z_{l-1},z_l+z_{l+1},z_{l+2},\ldots,z_n)\cdot
\begin{cases}
\phantom{-}\ha, & \Im(z_{l+1}/z_l)<0,\\
\phantom{-}0, & \Im(z_{l+1}/z_l)=0,\\
-\ha, & \Im(z_{l+1}/z_l)>0.
\end{cases}
\end{split}
\label{gf3eq19}
\e
\label{gf3prop3}
\end{prop}

To summarize: away from the real hypersurfaces
$z_{l+1}/z_l\in(0,\iy)$ in $(\C^\t)^n$ for $1\le l<n$, the $F_n$
must be continuous and holomorphic. As we cross the hypersurface
$z_{l+1}/z_l\in(0,\iy)$ at a generic point, $F_n(z_1,\ldots,z_n)$
jumps by $F_{n-1}(z_1,\ldots,z_{l-1},z_l+z_{l+1},z_{l+2},\ldots,
z_n)$, with the value on the hypersurface being the average of the
limiting values from either side.

Where several of the hypersurfaces $z_{l+1}/z_l\in(0,\iy)$
intersect, $F_n(z_1,\ldots,z_n)$ satisfies a more complicated
condition. Roughly speaking, this says that several different
sectors of \eq{gf3eq18} come together where the hypersurfaces
intersect, and on the intersection $F_n$ should be a weighted
average of the limiting values in each of these sectors. We now show
these conditions determine the functions $F_n$ uniquely, provided
they exist at all.

\begin{thm} There exists at most one family of functions
$F_n:(\C^\t)^n\ra\C$ satisfying Condition \ref{gf3cond} and
equations \eq{gf3eq2}, \eq{gf3eq3}, \eq{gf3eq5}, \eq{gf3eq6} of
Remark~\ref{gf3rem1}.
\label{gf3thm1}
\end{thm}

\begin{proof} Suppose $F_n$ and $F_n'$ for $n\ge 1$ are two families of
functions satisfying all the conditions, using functions $G_m,G_m'$
respectively in Condition \ref{gf3cond}. We shall show that
$F_n\equiv F_n'$ for all $n$, by induction on $n$. We have
$F_1\equiv F_1'\equiv(2\pi i)^{-1}$ by \eq{gf3eq3}. So let $n\ge 2$,
and suppose by induction that $F_k\equiv F_k'$ for all $k<n$. By
Condition \ref{gf3cond} and induction on $k$ this implies that
$G_k=G_k'$ for $k<n$. So taking the difference of \eq{gf3eq11} for
$F_n$ and $F_n'$ gives
\begin{align*}
f(z_1,\ldots,z_n)&=F_n(z_1,\ldots,z_n)-F_n'(z_1,\ldots,z_n)\\
&=G_n(z_1,\ldots,z_n;\ti z_1,\ldots,\ti z_n)-G_n'(z_1,\ldots,z_n;
\ti z_1,\ldots,\ti z_n)
\end{align*}
in an open neighbourhood of~$(\ti z_1,\ldots,\ti z_n)$.

As $G_n,G_n'$ are continuous and holomorphic in the $z_k$, we see
$f:(\C^\t)^n\ra\C$ is holomorphic. By \eq{gf3eq2}, $f$ is the
pullback of a holomorphic function $\ti
f:\bigl\{[z_1,\ldots,z_n]\in\CP^{n-1}:z_k\ne 0,$
$k=1,\ldots,n\bigr\}\ra\C$. Taking the difference of \eq{gf3eq6} for
$F_n,F_n'$ gives $\md{\ti f}=o(\md{z_k}^{-1})$ near points in just
one hypersurface $z_k=0$ in $\CP^{n-1}$. So by standard results in
complex analysis, $\ti f$ extends holomorphically over these parts
of $\CP^{n-1}$, and so is defined except on intersections of two or
more hypersurfaces $z_k=0$ in $\CP^{n-1}$. By Hartog's theorem $\ti
f$ extends holomorphically to all of $\CP^{n-1}$, and so is
constant. Since $n>1$, equation \eq{gf3eq5} gives $\sum_{\si\in
S_n}\ti f\bigl([z_{\si(1)},\ldots,z_{\si(n)}]\bigr)=0$ for
$z_k\in\C^\t$, forcing $\ti f\equiv 0$. Hence $f\equiv 0$ and
$F_n\equiv F_n'$. The theorem follows by induction.
\end{proof}

Note that we actually prove slightly more than the theorem says: any
functions $F_1,\ldots,F_n$ satisfying the conditions up to $n$ are
unique.

\subsection{Partial differential equations satisfied by $f^\al$ and
$F_n$}
\label{gf32}

We are trying to construct a family of holomorphic generating
functions $f^\al:\Stab(\A)\ra\L^\al$ for $\al\in C(\A)$. Clearly, it
would be interesting if this family satisfied some nontrivial
partial differential equations. We are now going to {\it guess\/} a
p.d.e.\ for the $f^\al$ to satisfy, and deduce a p.d.e.\ for the
$F_n$. We will then use this p.d.e.\ to {\it construct\/} the
functions $F_n$ that we want by induction on $n$. In \S\ref{gf33} we
will prove they satisfy Remark \ref{gf3rem1} and Condition
\ref{gf3cond}. Theorem \ref{gf3thm1} then shows these $F_n$ are
unique.

This means that the p.d.e.s that we shall guess for the $f^\al$ and
$F_n$ are actually {\it implied\/} by the general assumptions Remark
\ref{gf3rem1} and Condition \ref{gf3cond}, which seems very
surprising, as these imposed no differential equations other than
being holomorphic. One possible conclusion is that our p.d.e.s are
not simply something the author made up, but are really present in
the underlying geometry and combinatorics, and have some meaning of
their own.

To guess the p.d.e.\ we start by determining the function $F_2$.
Equation \eq{gf3eq2} implies we may write $F_2(z_1,z_2)=f(z_2/z_1)$
for some $f:\C^\t\ra\C$, and then Propositions \ref{gf3prop2} and
\ref{gf3prop3} and \eq{gf3eq3} imply that $f$ is holomorphic in
$\C\sm[0,\iy)$ with the following continuous over~$(0,\iy)$:
\begin{equation*}
f(z)-\frac{1}{2\pi i}\cdot
\begin{cases}
\phantom{-}\ha, & \Im(z)<0,\\
\phantom{-}0, & \Im(z)=0,\\
-\ha, & \Im(z)>0,
\end{cases}
\end{equation*}

Since $\log z$ cut along $(0,\iy)$ jumps by $2\pi i$ across
$(0,\iy)$, the obvious answer is $f(z)=(2\pi i)^{-2}\log z+C$ for
some constant $C$, where we define $\log z$ on $\C\sm[0,\iy)$ such
that $\Im\log z\in(0,2\pi)$. But equation \eq{gf3eq5} reduces to
$f(z)+f(z^{-1})\equiv 0$, which holds provided $C=-\pi i/(2\pi
i)^2$. This suggests that
\e
F_2(z_1,z_2)=\begin{cases} \frac{1}{(2\pi
i)^{2^{\vphantom{l}}}}\bigl(\log
(z_2/z_1)-\pi i\bigr), & z_2/z_1\notin (0,\iy),\\
\frac{1}{(2\pi i)^{2^{\vphantom{l}}}}\log(z_2/z_1), & z_2/z_1\in
(0,\iy),
\end{cases}
\label{gf3eq20}
\e
where $\log z$ is defined so that $\Im\log z\in[0,2\pi)$. It is now
easy to check that these $F_1,F_2$ satisfy Condition \ref{gf3cond}
and \eq{gf3eq2}, \eq{gf3eq3}, \eq{gf3eq5}, \eq{gf3eq6} up to $n=2$,
so Theorem \ref{gf3thm1} shows \eq{gf3eq20} is the {\it unique}
function $F_2$ which does this.

Let us now consider a simple situation in which we are interested
only in classes $\be,\ga,\be+\ga$ in $C(\A)$, and $\be,\ga$ cannot
be written as $\de+\ep$ for $\de,\ep\in C(\A)$, and the only ways to
write $\be+\ga=\de+\ep$ for $\de,\ep\in C(\A)$ are $\de,\ep=\be,\ga$
or $\de,\ep=\ga,\be$. In this case, from \eq{gf3eq1}, \eq{gf3eq3}
and \eq{gf3eq20} we see that when $Z\in\Stab(\A)$ with
$Z(\ga)/Z(\be)\notin(0,\iy)$ we have
\begin{gather*}
\ts f^\be(Z)=\frac{1}{2\pi i}\,\ep^\be(\mu),\qquad
f^\ga(Z)=\frac{1}{2\pi
i}\,\ep^\ga(\mu),\\
\ts f^{\be+\ga}(Z)\!=\!\frac{1}{2\pi
i}\ep^{\be+\ga}(\mu)\!+\!\frac{1}{(2\pi i)^2}
\left(\log\left(\frac{Z(\ga)}{Z(\be)}\right)\!-\!\pi i\right)\bigl(
\ep^\be(\mu)\!*\!\ep^\ga(\mu)\!-\!\ep^\ga(\mu)\!*\!\ep^\be(\mu)\bigr).
\end{gather*}
These satisfy the p.d.e.\ on $\Stab(\A)$, at least for
$Z(\ga)/Z(\be)\notin(0,\iy)$:
\e
\d f^{\be+\ga}(Z)=\bigl(f^\be(Z)*f^\ga(Z)-f^\ga(Z)*f^\be(Z)\bigr)\ot
\ts\left(\frac{\d(Z(\ga))}{Z(\ga)}-\frac{\d(Z(\be))}{Z(\be)}\right).
\label{gf3eq21}
\e

Here $f^{\be+\ga}$ is an $\L$-valued function on $\Stab(\A)$, so $\d
f^{\be+\ga}$ is an $\L$-valued 1-form, that is, a section of
$\L\ot_\C T^*_{\sst\mathbb C}\Stab(\A)$. Also $Z(\ga),Z(\be)$ are
complex functions on $\Stab(\A)$, so $\d(Z(\ga))/Z(\ga)$,
$\d(Z(\be))/Z(\be)$ are complex 1-forms on $\Stab(\A)$, and
tensoring over $\C$ with $f^\be(Z)*f^\ga(Z)-f^\ga(Z)*f^\be(Z)$ also
gives an $\L$-valued 1-form on $\Stab(\A)$. Note that $\ep^\be(\mu),
\ep^\ga(\mu),\ep^{\be+\ga}(\mu)$ are locally constant in $Z$ away
from $Z(\ga)/Z(\be)\in(0,\iy)$, so there are no terms from
differentiating them. Also, by construction $f^\be,f^\ga,
f^{\be+\ga}$ are continuous and holomorphic over the hypersurface
$Z(\ga)/Z(\be)\in(0,\iy)$, so by continuity \eq{gf3eq21} holds there
too.

We now guess that the generating functions $f^\al$ of \eq{gf3eq1}
should satisfy the p.d.e., for all~$\al\in C(\A)$:
\e
\begin{split}
\d f^\al(Z) &=\sum_{\be,\ga\in C(\A):\al=\be+\ga}
\bigl(f^\be(Z)*f^\ga(Z)\bigr)\ot \left(\frac{\d(Z(\ga))}{Z(\ga)}
-\frac{\d(Z(\be))}{Z(\be)}\right)\\
&=\sum_{\be,\ga\in C(\A):\al=\be+\ga}
\bigl(\ha[f^\be(Z),f^\ga(Z)]\bigr)\ot
\left(\frac{\d(Z(\ga))}{Z(\ga)}-\frac{\d(Z(\be))}{Z(\be)}\right)\\
&=-\sum_{\be,\ga\in C(\A):\al=\be+\ga}\bigl([f^\be(Z),f^\ga(Z)]
\bigr)\ot\frac{\d(Z(\be))}{Z(\be)},
\end{split}
\label{gf3eq22}
\e
where the three lines are equivalent, noting that we may exchange
$\be,\ga$. In the simple case above this reduces to \eq{gf3eq21}
when~$\al=\be+\ga$.

We can now explain our choice of constant $F_1\equiv (2\pi i)^{-1}$
in \eq{gf3eq3}. The $2\pi i$ comes from the jumping of $\log z$ over
$(0,\iy)$ as above. If we had instead set $F_1\equiv c$ for some
$c\in\C$, then $f^\al,F_n$ and the right hand side of \eq{gf3eq20}
would be multiplied by $2\pi i\,c$, and the right hand sides of
\eq{gf3eq21}--\eq{gf3eq22}, and \eq{gf4eq1} below, would be
multiplied by $(2\pi i\, c)^{-1}$. We picked $c=(2\pi i)^{-1}$ to
eliminate constant factors in the p.d.e.\ \eq{gf3eq22} and flat
connection \eq{gf4eq1}, and so simplify our equations.

For \eq{gf3eq22} to hold, it is clearly necessary that the right
hand side should be closed. We check this by applying `$\d$' to it
and using \eq{gf3eq22}, giving:
\ea
\d\raisebox{-3pt}{$\displaystyle\biggl[$}&\sum_{\be,\ga\in
C(\A):\al=\be+\ga\!\!\!\!\!\!\!\!\!\!}
\bigl(f^\be(Z)*f^\ga(Z)\bigr)\ot
\ts\Bigl(\frac{\d(Z(\ga))}{Z(\ga)}-\frac{\d(Z(\be))}{Z(\be)}\Bigr)
\raisebox{-3pt}{$\displaystyle\biggr]$}
\nonumber\\
\begin{split}
&=\sum_{\ep,\de\in C(\A):\al=\ep+\de\!\!\!\!\!\!\!\!\!\!} \bigl(\d
f^\ep(Z)*f^\de(Z)\bigr)\w
\ts\Bigl(\frac{\d(Z(\de))}{Z(\de)}-\frac{\d(Z(\ep))}{Z(\ep)}\Bigr)
\\
&\qquad+\sum_{\be,\ep\in C(\A):\al=\be+\ep\!\!\!\!\!\!\!\!\!\!}
\bigl(f^\be(Z)*\d f^\ep(Z)\bigr)\w\ts\Bigl(\frac{\d(Z(\ep))}{Z(\ep)}
-\frac{\d(Z(\be))}{Z(\be)}\Bigr)
\end{split}
\label{gf3eq23}
\\
&=\!\sum_{\be,\ga,\de\in C(\A):\al=\be+\ga+\de \!\!\!\!\!\!\!\!
\!\!\!\!\!\!\!\!\!\!\!\!\!\!\!\!\!\!\!\!\!\!\!\!\!\!\!\!\!\!\!\! }
\!\bigl(f^\be(Z)\!*\!f^\ga(Z)\!*\!f^\de(Z)\bigr)\!\ot\!\ts\Bigl[
\Bigl(\frac{\d(Z(\ga))}{Z(\ga)}\!-\!\frac{\d(Z(\be))}{Z(\be)}\Bigr)\!\w\!
\Bigl(\frac{\d(Z(\de))}{Z(\de)}\!-\!\frac{\d(Z(\be+\ga))}{Z(\be+\ga)}\Bigr)
\nonumber\\
&\qquad\qquad\qquad\qquad\quad \ts+\Bigl(\frac{\d(Z(\de))}{Z(\de)}
-\frac{\d(Z(\ga))}{Z(\ga)}\Bigr)\w
\Bigl(\frac{\d(Z(\ga+\de))}{Z(\ga+\de)}-\frac{\d(Z(\be))}{Z(\be)}\Bigr)
\Bigr]=0.
\nonumber
\ea
Here expanding the first line gives $\sum_{\be,\ga}\d
f^\be*f^\ga\w(\cdots)+\sum_{\be,\ga}f^\be*\d f^\ga\w(\cdots)$, as
the final 1-form is closed. These two terms appear in the second and
third lines, with $\be,\ga$ relabelled as $\ep,\de$ in the second
line and $\be,\ep$ in the third. The fourth and fifth lines
substitute \eq{gf3eq22} into the second and third lines, with $\ep$
in place of $\al$ for the second line and $\ep,\ga,\de$ in place of
$\al,\be,\ga$ for the third line. The final step holds as the 2-form
$[\cdots]$ on the fourth and fifth lines is zero.

Thus equation \eq{gf3eq22} has the attractive property that {\it it
implies its own consistency condition\/}; that is, the condition for
\eq{gf3eq22} to be locally solvable for $f^\al$ is equation
\eq{gf3eq22} for $\be,\ga$. We express \eq{gf3eq22} in terms of the
functions $F_n$ and~$G_n$.

\begin{prop} Suppose Assumptions \ref{gf2ass1}, \ref{gf2ass2} and
\ref{gf3ass} hold for\/ $\A$ with\/ $K(\A)$ of finite rank and\/
$\Stab(\A)$ is a nonempty open subset of\/ $\Hom(K(\A),\C)$, and let
some functions\/ $F_n:(\C^\t)^n\ra\C$ be given. Then a sufficient
condition for the functions\/ $f^\al$ of\/ \eq{gf3eq1} to satisfy\/
\eq{gf3eq22}, is that for fixed\/ $(\ti z_1,\ldots,\ti z_n)\in H^n$
the functions $(z_1,\ldots,z_n)\mapsto G_n(z_1,\ldots,z_n;\ti
z_1,\ldots,\ti z_n)$ of\/ \S\ref{gf31} should satisfy
\e
\begin{split}
\d G_n(z_1,\ldots,z_n;\ti z_1,\ldots,\ti z_n)=\sum_{k=1}^{n-1}
&G_k(z_1,\ldots,z_k;\ti z_1,\ldots,\ti z_k)\,\cdot\\
G_{n-k}(z_{k+1},\ldots,z_n;\ti z_{k+1},\ldots,\ti z_n)
&\left(\frac{\d z_{k+1}\!+\!\cdots\!+\!\d
z_n}{z_{k+1}\!+\!\cdots\!+\!z_n}\!-\!\frac{\d z_1\!+\!\cdots\!+\!\d
z_k}{z_1\!+\!\cdots\!+\!z_k}\right)
\end{split}
\label{gf3eq24}
\e
in $H^n$. This holds for some\/ $(\ti z_1,\ldots,\ti z_n)$ in $H^n$
if and only if it holds for all\/ $(\ti z_1,\ldots,\ti z_n)$
in~$H^n$.

This condition is also necessary, for all values of\/ $n$ occurring
in\/ \eq{gf3eq1}, if the terms $\ep^{\al_1}(\mu)*\ep^{\al_2}(\mu)
*\cdots*\ep^{\al_n}(\mu)$ occurring in \eq{gf3eq1} are  linearly
independent in $\H$. This happens in the examples of\/
{\rm\cite[Ex.~7.10]{Joyc5}}, for arbitrarily large~$n$.

Now suppose Condition \ref{gf3cond} holds. Then equation
\eq{gf3eq24} holds for $(z_1,\ldots,z_n)$ in $N_{(\ti z_1,\ldots,\ti
z_n)}$ for all\/ $n\ge 1$ and all fixed\/ $(\ti z_1,\ldots,\ti
z_n)\in(\C^\t)^n$ if and only if the following p.d.e.\ holds on the
domain \eq{gf3eq18} for all\/ $n\ge 1$:
\e
\d F_n(z_1,\ldots,z_n)=\sum_{k=1}^{n-1}
\begin{aligned}[t]
&F_k(z_1,\ldots,z_k)F_{n-k}(z_{k+1},\ldots,z_n)\,\cdot\\
&\left[\frac{\d z_{k+1}+\cdots+\d z_n}{z_{k+1}+\cdots+z_n} -\frac{\d
z_1+\cdots+\d z_k}{z_1+\cdots+z_k}\right].
\end{aligned}
\label{gf3eq25}
\e
\label{gf3prop4}
\end{prop}

\begin{proof} For the first part, substitute \eq{gf3eq9} in for
$f^\al,f^\be$ and $f^\ga$ in the top line of \eq{gf3eq22}. Then both
sides can be rewritten as a sum over $\al_1,\ldots,\al_n\in C(\A)$
with $\al_1+\cdots+\al_n=\al$ of $\ep^{\al_1}(\ti\mu)*\cdots*
\ep^{\al_n}(\ti\mu)$ tensored with complex 1-forms.
Equating the complex 1-form coefficients of $\ep^{\al_1}(\ti\mu)*
\cdots*\ep^{\al_n}(\ti\mu)$ gives \eq{gf3eq24}, evaluated at
$z_a=Z(\al_a)$ and $\ti z_a=\ti Z(\al_a)$, where the $G_k$ term
comes from $f^\be$ in \eq{gf3eq22} with $\be=\al_1+\cdots+\al_k$,
and the $G_{n-k}$ term comes from $f^\ga$ in \eq{gf3eq22} with
$\ga=\al_{k+1}+\cdots+\al_n$. The first two paragraphs follow, by
the same arguments used to prove Proposition~\ref{gf3prop1}.

For the final part, let Condition \ref{gf3cond} hold. If $(\ti
z_1,\ldots,\ti z_n)\in(\C^\t)^n$ with $\ti z_{k+1}/\ti
z_k\notin(0,\iy)$ for all $1\le k<n$ and $(z_1,\ldots,z_n)\in
N_{(\ti z_1,\ldots,\ti z_n)}$, then the proof of Proposition
\ref{gf3prop2} shows that
$F_n(z_1,\ldots,z_n)\!=\!G_n(z_1,\ldots,z_n;\ti z_1,\ldots,\ti
z_n)$, $F_k(z_1,\ldots,z_k)\!=\!G_k(z_1,\ldots,z_k;\ti
z_1,\ldots,\ti z_k)$ and $F_{n-k}(z_{k+1},\ldots,z_n)\ab\!=\!\ab
G_{n-k}\ab(z_{k+1},\ab\ldots,z_n;\ti z_{k+1},\ldots,\ti z_n)$. Thus
\eq{gf3eq24} is equivalent to \eq{gf3eq25} in $N_{(\ti
z_1,\ldots,\ti z_n)}$, so \eq{gf3eq24} implies \eq{gf3eq25} in the
domain~\eq{gf3eq18}.

For the reverse implication, suppose Condition \ref{gf3cond} holds
and \eq{gf3eq25} holds in \eq{gf3eq18}. Then the argument above
shows \eq{gf3eq24} holds for $(z_1,\ldots,z_n),(\ti z_1,\ldots,\ti
z_n)$ in \eq{gf3eq18} with $(z_1,\ldots,z_n)\in N_{(\ti
z_1,\ldots,\ti z_n)}$. But whether \eq{gf3eq24} holds or not is
unaffected by small changes in $(\ti z_1,\ldots,\ti z_n)$, and
\eq{gf3eq18} is dense in $(\C^\t)^n$. Thus, \eq{gf3eq24} holds for
all $(\ti z_1,\ldots,\ti z_n)\in(\C^\t)^n$ and $(z_1,\ldots,z_n)\in
N_{(\ti z_1,\ldots,\ti z_n)}$ with $(z_1,\ldots,z_n)$ in
\eq{gf3eq18}. Now the functions $(z_1,\ldots,z_m)\mapsto
G_m(z_1,\ldots,z_m;\ti z_1,\ldots,\ti z_m)$ are continuous and
holomorphic by Condition \ref{gf3cond}, so as \eq{gf3eq18} is open
and dense we see that \eq{gf3eq24} must hold for all
$(z_1,\ldots,z_n)\in N_{(\ti z_1,\ldots,\ti z_n)}$, by continuity.
\end{proof}

The proof of Proposition \ref{gf3prop4} conceals a subtlety. One
might think that for generic $Z\in\Stab(\A)$, all terms
$(Z(\al_1),\ldots,Z(\al_n))$ occurring in \eq{gf3eq1} will lie in
the open dense domain \eq{gf3eq18}, so that \eq{gf3eq25} on
\eq{gf3eq18} implies \eq{gf3eq22} for generic $Z$ in the obvious
way, and so \eq{gf3eq22} must hold for all $Z$ by continuity.
However, this is false. For example, if $\al_1=\al_2$ then
$Z(\al_2)/Z(\al_1)\equiv 1\in(0,\iy)$, so
$(Z(\al_1),\ldots,Z(\al_n))$ does not lie in \eq{gf3eq18} for any
$Z\in\Stab(\A)$. So assuming \eq{gf3eq25} on \eq{gf3eq18} apparently
tells us nothing about how such terms contribute to~\eq{gf3eq22}.

Because of this, for \eq{gf3eq22} to hold when $f^\al$ in
\eq{gf3eq1} includes terms with dependencies such as $\al_1=\al_2$,
we need $F_n$ to satisfy not just \eq{gf3eq25} on \eq{gf3eq18}, but
other more complicated conditions on the real hypersurfaces
$z_{k+1}/z_k\in(0,\iy)$ as well. The point of the proof is that
these other conditions are implied by \eq{gf3eq25} on \eq{gf3eq18}
and Condition \ref{gf3cond}, as we can express the conditions in
terms of the $G_n(z_1,\ldots,z_n;\ti z_1,\ldots,\ti z_n)$ and use
the fact that they are continuous and holomorphic in
$(z_1,\ldots,z_n)$ over the hypersurfaces~$z_{k+1}/z_k\in(0,\iy)$.

Equations \eq{gf3eq24} and \eq{gf3eq25} apparently have poles on the
hypersurfaces $z_1+\cdots+z_k=0$ and $z_{k+1}+\cdots+z_n=0$. So we
would expect solutions $G_n,F_n$ to have log-type singularities
along these hypersurfaces; in particular, this suggests that there
should not be single-valued solutions on the domain \eq{gf3eq18}. In
fact this is false, and single-valued, nonsingular solutions can
exist across these hypersurfaces. The next proposition explains why.

\begin{prop} For $n\ge 2$ the following is a nonempty, connected set
in~$\C^n$:
\e
\begin{split}
\bigl\{(z_1,\ldots,z_n)\in(\C^\t)^n:\,&\text{$z_{k+1}/z_k\notin
(0,\iy)$ for $k=1,\ldots,n-1$}\\
&\text{and\/ $z_1+\cdots+z_n=0$}\bigr\}.
\end{split}
\label{gf3eq26}
\e
If\/ $F_n$ satisfies \eq{gf3eq25} on the domain \eq{gf3eq18} then
$F_n\equiv C_n$ on \eq{gf3eq26} for some $C_n\in\C$. If\/ $F_n$ also
satisfies \eq{gf3eq5} as in Remark \ref{gf3rem1} then $C_n=0$. In
this case we have\/ $F_n(z_1,\ldots,z_n)=(z_1+\cdots+z_n)
H_n(z_1,\ldots,z_n)$ for a holomorphic function $H_n$ defined on the
domain \eq{gf3eq18}, including where $z_1+\cdots+z_n\equiv 0$. Using
these we rewrite \eq{gf3eq25} as
\e
\begin{aligned}
\d& F_n(z_1,\ldots,z_n)=\ts\sum_{k=1}^{n-1}
H_k(z_1,\ldots,z_k)H_{n-k}(z_{k+1},\ldots,z_n)\,\cdot\\
&\bigl((z_1\!+\!\cdots\!+\!z_k)(\d z_{k+1}\!+\!\cdots\!+\!\d z_n)
\!-\!(z_{k+1}\!+\!\cdots\!+\!z_n)(\d z_1\!+\!\cdots\!+\!\d
z_k)\bigr).
\end{aligned}
\label{gf3eq27}
\e
Note that \eq{gf3eq27} has no poles on~\eq{gf3eq18}.
\label{gf3prop5}
\end{prop}

\begin{proof} Let $(z_1,\ldots,z_n),(z_1',\ldots,z_n')$ lie in
\eq{gf3eq26}. We shall construct a path between them in
\eq{gf3eq26}, showing \eq{gf3eq26} is connected. It is easy to see
that
\ea
\begin{split}
\bigl\{w\in\C:&\text{$(z_1,\ldots,z_{n-2},w,z_{n-1}+z_n-w)$ lies in
\eq{gf3eq26}}\bigr\}=\\
&\C\sm\bigl(\{x(z_{n-1}+z_n):x\in[0,1]\}\cup
\{xz_{n-2}:x\in[0,\iy)\}\bigr),
\end{split}
\label{gf3eq28}\\
\begin{split}
\bigl\{w\in\C:&\text{$(z_1',\ldots,z_{n-2}',w,z_{n-1}'+z_n'-w)$ lies
in \eq{gf3eq26}}\bigr\}=\\
&\C\sm\bigl(\{x(z_{n-1}'+z_n'):x\in[0,1]\}\cup
\{xz_{n-2}':x\in[0,\iy)\}\bigr),
\end{split}
\label{gf3eq29}
\ea
which are both {\it connected\/} subsets of $\C$, containing
$z_{n-1}$ and $z_{n-1}'$ respectively. Choose some $w_0$ in both
\eq{gf3eq28} and \eq{gf3eq29} with $\md{w_0}\gg\md{z_k},\md{z_k'}$
for all $k=1,\ldots,n$. Choose paths between $z_{n-1}$ and $w_0$ in
\eq{gf3eq28}, and between $z_{n-1}'$ and $w_0$ in \eq{gf3eq29}.
These induce paths in \eq{gf3eq26} between $(z_1,\ldots,z_n)$ and
$(z_1,\ldots,z_{n-2},w_0,z_{n-1}+z_n-w_0)$, and between
$(z_1',\ldots,z_n')$ and~$(z_1',\ldots,z_{n-2}',w_0,
z_{n-1}'+z_n'-w_0)$.

It remains to find a path in \eq{gf3eq26} between $(z_1,\ldots,
z_{n-2},w_0,z_{n-1}+z_n-w_0)$ and $(z_1',\ldots,z_{n-2}',w_0,
z_{n-1}'+z_n'-w_0)$. To do this, choose a path
$(x_1(t),\ldots,x_{n-2}(t))$ for $t\in[0,1]$ between
$(z_1,\ldots,z_{n-2})$ and $(z_1',\ldots,z_{n-2}')$ in
\begin{equation*}
\begin{split}
\bigl\{(y_1,\ldots,y_{n-2})\in(\C^\t)^{n-2}:\,
&\text{$y_{k+1}/y_k\notin(0,\iy)$ for $k=1,\ldots,n-3$}\\
&\text{and $y_{n-2}/w_0\notin(0,\iy)$}\bigr\},
\end{split}
\end{equation*}
which is possible as using $y_{k+1}/y_k$, $y_{n-2}/w_0$ as
coordinates we see this domain is homeomorphic to
$(\C\sm[0,\iy))^{n-2}$, and thus is connected. Making $w_0$ larger
if necessary, we can also assume that $\md{w_0}\gg\md{x_k(t)}$ for
all $k=1,\ldots,n-2$ and $t\in[0,1]$. Then it is easy to see that
the path $(x_1(t),\ldots,x_{n-2}(t),w_0,-x_1(t)-\cdots-x_{n-2}(t)
-w_0)$ for $t\in[0,1]$ links $(z_1,\ldots,z_{n-2},w_0,z_{n-1}+
z_n-w_0)$ and $(z_1',\ldots,z_{n-2}',w_0,z_{n-1}'+z_n'-w_0)$ in
\eq{gf3eq26}. Therefore \eq{gf3eq26} is connected.

For the second part, observe that on the hypersurface
$z_1+\cdots+z_n=0$ we have $z_{k+1}+\cdots+z_n\equiv
-(z_1+\cdots+z_k)$, so the 1-form $[\cdots]$ in \eq{gf3eq25}
restricts to zero on $z_1+\cdots+z_n=0$. Thus the restriction of $\d
F_n$ to \eq{gf3eq26} is zero, so $F_n\equiv C_n$ on \eq{gf3eq26} for
some $C_n\in\C$, as \eq{gf3eq26} is connected. For generic
$(z_1,\ldots,z_n)$ in \eq{gf3eq26} all the permutations
$(z_{\si(1)},\ldots,z_{\si(n)})$ for $\si\in S_n$ lie in
\eq{gf3eq26} as well, so \eq{gf3eq5} becomes $n!C_n=0$, giving
$C_n=0$. Thus, the holomorphic function $F_n$ is zero along the
nonsingular hypersurface $z_1+\cdots+z_n=0$ in its domain
\eq{gf3eq18}. Properties of holomorphic functions imply that
$F_n(z_1,\ldots,z_n)=(z_1+\cdots+z_n) H_n(z_1,\ldots,z_n)$ for a
unique holomorphic function $H_n$ on \eq{gf3eq18}. Equation
\eq{gf3eq27} is immediate.
\end{proof}

Suppose we are given some holomorphic functions $F_n$ on the domains
\eq{gf3eq18} satisfying \eq{gf3eq25}. Analytically continue the
$F_n$ to {\it multivalued}, {\it singular} holomorphic functions
$\ti F_n$ on $(\C^\t)^n$, still satisfying \eq{gf3eq25}. The
argument above shows that $\ti F_n$ is locally constant along
$z_1+\cdots+z_n=0$, but it can take a different value on each sheet.
So \eq{gf3eq25} can have genuine poles along $z_1+\cdots+z_k=0$ and
$z_{k+1}+\cdots+z_n=0$ when $\ti F_k,\ti F_{n-k}$ are nonzero
constants.

Thus $\ti F_n$ will have log-like singularities along
$z_1+\cdots+z_k=0$ and $z_{k+1}+\cdots+z_n=0$, and more generally
singularities along $z_a+\cdots+z_b=0$ for $1\le a\le b\le n$ with
$(a,b)\ne(1,n)$. One moral is that our functions $F_n$ manage to be
single-valued and nonsingular on \eq{gf3eq18} for very special
reasons, and their analytic continuations have much worse
singularities and branching behaviour.

We now construct functions $F_n$ for $n\ge 1$ satisfying
\eq{gf3eq25}, by induction on~$n$.

\begin{prop} There exists a unique series of holomorphic functions
$F_n$ for $n\ge 1$ defined on the domain \eq{gf3eq18} with\/
$F_1\equiv(2\pi i)^{-1},$ such that\/ $F_n$ satisfies \eq{gf3eq25}
and is zero on \eq{gf3eq26}. Also $F_n(z_1,\ldots,z_n)\equiv
(z_1+\cdots+z_n)H_n(z_1,\ldots,z_n)$ for a unique holomorphic
function $H_n$ defined on \eq{gf3eq18}, and\/ \eq{gf3eq27} holds.
\label{gf3prop6}
\end{prop}

\begin{proof} Suppose by induction that for some $m\ge 2$ we have
constructed unique holomorphic functions $F_n,H_n$ on the domains
\eq{gf3eq18} for $n=1,\ldots,m-1$ satisfying all the conditions of
the proposition for $n<m$. For $m=2$ this is trivial, as we must
have $F_1(z)=(2\pi i)^{-1}$ and $H_1(z)=(2\pi i z)^{-1}$. We will
construct $F_m,H_m$, and show they are unique.

Equations \eq{gf3eq25} and \eq{gf3eq27} for $n=m$ give equivalent
expressions for $\d F_m$ on \eq{gf3eq18}, with \eq{gf3eq27} being
manifestly holomorphic on all of \eq{gf3eq18}. Write $\al_m$ for the
right hand side of \eq{gf3eq25} or \eq{gf3eq27}, so that $\al_m$ is
a holomorphic 1-form on \eq{gf3eq18}, and we need to construct $F_m$
with $\d F_m=\al_m$. Following \eq{gf3eq23}, we can compute
$\d\al_m$ by applying $\d$ to the r.h.s.\ of \eq{gf3eq25} for $n=m$,
and using \eq{gf3eq25} for $n<m$ (which holds by induction) to
substitute in for $\d F_k$ and $\d F_{n-k}$. Then everything cancels
giving $\d\al_m=0$, so $\al_m$ is a {\it closed\/} 1-form.

Although \eq{gf3eq18} is not simply-connected, it is the pullback to
$\C^m\sm\{0\}$ of
\e
\bigl\{[z_1,\ldots,z_m]\in\CP^{m-1}:\text{$z_k\ne 0$ and
$z_{k+1}/z_k\notin(0,\iy)$ for all $k$}\bigr\},
\label{gf3eq30}
\e
which is homeomorphic to $\bigl(\C\sm[0,\iy)\bigr)^{m-1}$, and so is
simply-connected. Now $\al_m$ is the pull-back of a 1-form on
\eq{gf3eq30}, which is closed as $\al_m$ is closed, and so is {\it
exact\/} as \eq{gf3eq30} is simply-connected.

Therefore $\al_m$ is an exact holomorphic 1-form on \eq{gf3eq18}, so
there exists a holomorphic function $F_m$ on \eq{gf3eq18} with $\d
F_m=\al_m$, which is unique up to addition of a constant, as
\eq{gf3eq18} is connected. To choose the constant, note that the
restriction of $\al_m$ to the connected set \eq{gf3eq26} is zero as
in Proposition \ref{gf3prop5}, so requiring $F_m$ to be zero on
\eq{gf3eq26} fixes $F_m$ uniquely. Since $F_m$ is zero along the
nonsingular hypersurface $z_1+\cdots+z_m=0$, by properties of
holomorphic functions there is a unique holomorphic function $H_m$
on \eq{gf3eq18} with $F_m(z_1,\ldots,z_m)\equiv
(z_1+\cdots+z_m)H_m(z_1,\ldots,z_m)$. These $F_m,H_m$ satisfy all
the conditions for $n=m$, and the proposition follows by induction.
\end{proof}

\subsection{Reconciling the approaches of \S\ref{gf31} and
\S\ref{gf32}}
\label{gf33}

So far we have given two quite different approaches to the functions
$F_n$ used to define $f^\al$ in \eq{gf3eq1}. In \S\ref{gf31} we
found conditions on the $F_n$ on $(\C^\t)^n$ for the $f^\al$ to be
continuous and holomorphic, and showed such $F_n$ would be unique if
they existed. In \S\ref{gf32} we found different conditions on the
$F_n$ on a subdomain \eq{gf3eq18} of $(\C^\t)^n$ for the $f^\al$ to
satisfy the p.d.e.\ \eq{gf3eq25}, neglecting the question of whether
$f^\al$ would be continuous for these $F_n$, and constructed unique
$F_n$ satisfying these second conditions. There seems no {\it a
priori\/} reason why these two sets of conditions on $F_n$ should be
compatible, but we now prove that they are. That is, we show that
the $F_n$ on \eq{gf3eq18} constructed in Proposition \ref{gf3prop6}
extend uniquely to $(\C^\t)^n$ so as to satisfy Remark \ref{gf3rem1}
and Condition~\ref{gf3cond}.

First we show, in effect, that Proposition \ref{gf3prop3} holds for
the $F_n$ of Proposition \ref{gf3prop6}. For $F_n$ as in Proposition
\ref{gf3prop6}, define a function $D_{l,n}$ for $1\le l<n$ by
\begin{gather}
\begin{split}
D_{l,n}:\bigl\{(\ti z_1,\ldots,\ti z_n)\in(\C^\t)^n: \text{$\ti
z_{l+1}/\ti z_l\in(0,\iy)$,}&\\
\text{$\ti z_{k+1}/\ti z_k\notin (0,\iy)$ for $1\le k<n$, $k\ne
l$}&\bigr\}\longra\C,
\end{split}
\label{gf3eq31}\\
D_{l,n}(\ti z_1,\ldots,\ti z_n)= \!\!\lim_{\begin{subarray}{l}
\text{$(z_1,\ldots,z_n)\ra(\ti z_1,\ldots,\ti z_n):^{\vphantom{l^l}}$}\\
\text{$(z_1,\ldots,z_n)$ lies in \eq{gf3eq18},}\\
\text{$\Im (z_{l+1}/z_l)<0$}\end{subarray} \!\!\!\!\!\!\!\!\!\!\!
\!\!\!\!\!\!\!\!\!\!\!\!\!\!\!\!\!\!\!\!\!\!\!\!\!\!\!\!\!\!\!\!}
F_n(z_1,\ldots,z_n)-\!\!\lim_{\begin{subarray}{l}
\text{$(z_1,\ldots,z_n)\ra(\ti z_1,\ldots,\ti z_n):^{\vphantom{l^l}}$}\\
\text{$(z_1,\ldots,z_n)$ lies in \eq{gf3eq18},}\\
\text{$\Im (z_{l+1}/z_l)>0$}\end{subarray} \!\!\!\!\!\!\!\!\!\!\!
\!\!\!\!\!\!\!\!\!\!\!\!\!\!\!\!\!\!\!\!\!\!\!\!\!\!\!\!\!\!\!\!}
F_n(z_1,\ldots,z_n).
\label{gf3eq32}
\end{gather}
These limits exist and give a continuous function $D_{l,n}$, since
the proof in Proposition \ref{gf3prop6} that the $F_n$ are
continuous and holomorphic in their domains extends locally from
either side over the hypersurface $z_{l+1}/z_l\in(0,\iy)$. The next
result would follow from \eq{gf3eq19} if we knew Proposition
\ref{gf3prop3} applied.

\begin{prop} We have\/ $D_{l,n}(z_1,\ldots,z_n)\equiv
F_{n-1}(z_1,\ldots,z_{l-1},\ab z_l+z_{l+1},\ab z_{l+2},\ab
\ldots,z_n)$ on the domain of\/ $D_{l,n},$ where\/ $F_{n-1}$ is as
in Proposition~\ref{gf3prop6}.
\label{gf3prop7}
\end{prop}

\begin{proof} Taking the difference of the limits of \eq{gf3eq25}
from both sides of the hypersurface $z_{l+1}/z_l\in(0,\iy)$ gives an
equation in 1-forms on the domain of~$D_{l,n}$:
\ea
&\d D_{l,n}(z_1,\ldots,z_n)=
\label{gf3eq33}\\
&\ts\sum_{k=1}^{l-1}F_k(z_1,\ldots,z_k)D_{l-k,n-k}(z_{k+1},\ldots,z_n)
\cdot\Bigl[\frac{\d z_{k+1}+\cdots+\d z_n}{z_{k+1} +\cdots+z_n}
-\frac{\d z_1+\cdots+\d z_k}{z_1+\cdots+z_k}\Bigr]+
\nonumber\\
&\ts\sum_{k=l+1}^{n-1}D_{l,k}(z_1,\ldots,z_k)F_{n-k}(z_{k+1},\ldots,z_n)
\cdot\Bigl[\frac{\d z_{k+1}+\cdots+\d z_n}{z_{k+1}
+\cdots+z_n}-\frac{\d z_1+\cdots+\d z_k}{z_1+\cdots+z_k}\Bigr].
\nonumber
\ea
Here if $(z_1,\ldots,z_n)$ lies in the domain of $D_{l,n}$ and $k<l$
then $F_k$ is defined and continuous at $(z_1,\ldots,z_k)$ but
$F_{n-k}$ is not defined (nor continuous) at $(z_{k+1},\ldots,z_n)$,
so the difference in limits of $F_k(z_1,\ldots,z_k)F_{n-k}
(z_{k+1},\ldots,z_n)$ in \eq{gf3eq25} is $F_k(z_1,\ldots,z_k)
D_{l-k,n-k}(z_{k+1},\ldots,z_n)$, giving the first term in
\eq{gf3eq33}. Similarly, $k>l$ gives the second term. There is no
term $k=l$ in \eq{gf3eq33}, since $F_l$ and $F_{n-l}$ are both
defined and continuous at $(z_1,\ldots,z_l)$ and
$(z_{l+1},\ldots,z_n)$ respectively, so the limits from each side of
$z_{l+1}/z_l\in(0,\iy)$ cancel.

As $F_n(z_1,\ldots,z_n)=0$ when $z_1+\cdots+z_n=0$, if $(\ti
z_1,\ldots,\ti z_n)$ lies in the domain of $D_{l,n}$ with $\ti
z_1+\cdots+\ti z_n=0$ then both limits in \eq{gf3eq32} are zero, as
$(\ti z_1,\ldots,\ti z_n)$ is the limit of points $(z_1,\ldots,z_n)$
with $z_1+\cdots+z_n=0$ from both sides of $z_{l+1}/z_l\in(0,\iy)$.
Thus
\e
D_{l,n}(z_1,\ldots,z_n)=0\quad\text{if $z_1+\cdots+z_n=0$.}
\label{gf3eq34}
\e
Also, it is easy to verify from Proposition \ref{gf3prop6} that
$F_2$ is given by \eq{gf3eq20} in its domain, so from properties of
logs we see that
\e
D_{1,2}(z_1,z_2)\equiv (2\pi i)^{-1}\equiv F_1(z_1+z_2).
\label{gf3eq35}
\e

Suppose by induction on $m$ that $D_{l,n}(z_1,\ldots,z_n)\equiv
F_{n-1}(z_1,\ldots,z_{l-1},z_l+z_{l+1},z_{l+2},\ldots,z_n)$ whenever
$1\le l<n\le m$, for some $m\ge 2$. The first case $m=2$ is
\eq{gf3eq35}. Let $n=m+1$ and $1\le l<n$. Then comparing
\eq{gf3eq25} and \eq{gf3eq33} and using the inductive hypothesis
shows that
\begin{equation*}
\d D_{l,n}(z_1,\ldots,z_n)\equiv\d F_{n-1}(z_1,\ldots,z_{l-1},
z_l+z_{l+1},z_{l+2},\ldots,z_n).
\end{equation*}
Thus $D_{l,n}(z_1,\ldots,z_n)-F_{n-1}(z_1,\ldots,z_{l-1},
z_l+z_{l+1},z_{l+2},\ldots,z_n)$ is constant on the domain of
$D_{l,n}$. But this domain is connected and contains
$(z_1,\ldots,z_n)$ with $z_1+\cdots+z_n=0$ as $n\ge 3$, and both
$D_{l,n}(\cdots)$ and $F_{n-1}(\cdots)$ are zero at such points by
Proposition \ref{gf3prop6} and \eq{gf3eq34}. So the constant is
zero, proving the inductive step and the proposition.
\end{proof}

Using this we extend the $F_n$ of Proposition \ref{gf3prop6} so that
Condition \ref{gf3cond} holds.

\begin{thm} The functions $F_n$ of Proposition \ref{gf3prop6},
defined on the domain \eq{gf3eq18}, can be extended uniquely to
$F_n:(\C^\t)^n\ra\C$ satisfying Condition~\ref{gf3cond}.
\label{gf3thm2}
\end{thm}

\begin{proof} The idea of the proof is that by induction on $n$ we
shall construct functions $G_n(z_1,\ldots,z_n;\ti z_1,\ldots,\ti
z_n)$ that for each fixed $(\ti z_1,\ldots,\ti z_n)$ are continuous
and holomorphic and satisfy \eq{gf3eq24} in $(z_1,\ldots,z_n)$ on
$N_{(\ti z_1,\ldots,\ti z_n)}$, such that \eq{gf3eq16} holds with
$F_n$ as in Proposition \ref{gf3prop6} whenever $(z_1,\ldots,z_n)$
lies in the intersection of \eq{gf3eq18} and $N_{(\ti z_1,\ldots,\ti
z_n)}$. We then extend $F_n$ uniquely from \eq{gf3eq18} to
$(\C^\t)^n$ by requiring \eq{gf3eq16} to hold on all of $N_{(\ti
z_1,\ldots,\ti z_n)}$, for all~$(\ti z_1,\ldots,\ti z_n)$.

Suppose by induction that for some $p\ge 2$ and for all $n<p$ we
have found open neighbourhoods $N_{(\ti z_1,\ldots,\ti z_n)}$ of
$(\ti z_1,\ldots,\ti z_n)$ in $(\C^\t)^n$ for all $(\ti
z_1,\ldots,\ti z_n)\in(\C^\t)^n$, and functions $G_n$, and
extensions of $F_n$ in Proposition \ref{gf3prop6} to $(\C^\t)^n$,
such that Condition \ref{gf3cond} holds for $n<p$ and the $G_n$
satisfy \eq{gf3eq24}, and $G_n(z_1,\ab\ldots,\ab z_n;\ab\ti
z_1,\ldots,\ti z_n)=0$ if $z_1+\cdots+z_n=0$. The first case $p=2$
is trivial, taking $F_1\equiv (2\pi i)^{-1}\equiv G_1$ and $N_{(\ti
z_1)}=\C^\t$. We shall now construct open neighbourhoods $N_{(\ti
z_1,\ldots,\ti z_p)}$, the function $G_p$, and an extension of
$F_p$, satisfying all the conditions.

Choose a connected, simply-connected open neighbourhood $N_{(\ti
z_1,\ldots,\ti z_p)}$ of each $(\ti z_1,\ldots,\ti z_p)$ in
$(\C^\t)^p$, such that $(z_1,\ldots,z_p)\in N_{(\ti z_1,\ldots,\ti
z_p)}$ implies that (a) if $1\le m<p$, $0=a_0<a_1<\cdots<a_m=p$ and
$c_1,\ldots,c_m\in [0,2\pi)$ with $\ti z_a\in {\rm e}^{ic_k}(0,\iy)$
for $a_{k-1}<a\le a_k$, then
$(z_{a_0+1}+\cdots+z_{a_1},\ldots,z_{a_{m-1}+1} +\cdots+z_{a_m})\in
N_{(\ti z_{a_0+1}+\cdots+\ti z_{a_1},\ldots,\ti
z_{a_{m-1}+1}+\cdots+\ti z_{a_m})}$, and (b) if $1\le k<p$ then
$(z_1,\ldots,z_k)\in N_{(\ti z_1,\ldots,\ti z_k)}$ and
$(z_{k+1},\ldots,z_p)\in N_{(\ti z_{k+1},\ldots,\ti z_p)}$. This is
satisfied if $N_{(\ti z_1,\ldots,\ti z_p)}$ is a small enough open
ball about $(\ti z_1,\ldots,\ti z_p)$. The point is that (a) ensures
that all the terms in \eq{gf3eq16} with $n=p$ and $m<p$ are
well-defined when $(z_1,\ldots,z_p)\in N_{(\ti z_1,\ldots,\ti
z_p)}$, and (b) ensures that the right hand side of \eq{gf3eq24} for
$n=p$ is well-defined when~$(z_1,\ldots,z_p)\in N_{(\ti
z_1,\ldots,\ti z_p)}$.

Now regard $(\ti z_1,\ldots,\ti z_p)$ as fixed, and consider
equation \eq{gf3eq24} with $n=p$ for $(z_1,\ldots,z_p)\in N_{(\ti
z_1,\ldots,\ti z_p)}$. The left hand side $\d G_p(\cdots)$ has not
yet been defined. The right hand side involves $G_k$ for $k<p$,
which by induction are defined on their domains and satisfy
\eq{gf3eq24}. The choice of $N_{(\ti z_1,\ldots,\ti z_p)}$ implies
the r.h.s.\ is a 1-form defined on $N_{(\ti z_1,\ldots,\ti z_p)}$,
and taking $\d$ and using \eq{gf3eq24} for $n<p$ we find this 1-form
is {\it closed}, as for $\d F_m$ in the proof of Proposition
\ref{gf3prop6}. Also, as for \eq{gf3eq23} and for $F_n$ in
Proposition \ref{gf3prop5}, the inductive assumption
$G_n(z_1,\ldots,z_n;\ti z_1,\ldots,\ti z_n)=0$ if $z_1+\cdots+z_n=0$
ensures that the terms $(z_1+\cdots+z_k)^{-1}$,
$(z_{k+1}+\cdots+z_n)^{-1}$ in \eq{gf3eq24} do not induce
singularities.

This proves that the right hand side of \eq{gf3eq24} for $n=p$ is a
well-defined, closed, holomorphic, nonsingular 1-form on the
connected, simply-connected domain $N_{(\ti z_1,\ldots,\ti z_p)}$.
Hence there exists a holomorphic function $(z_1,\ldots,z_p)\mapsto
G_p(z_1,\ldots,z_p;\ti z_1,\ldots,\ti z_p)$ on $N_{(\ti
z_1,\ldots,\ti z_p)}$, unique up to addition of a constant, such
that \eq{gf3eq24} holds. Here is how we fix the constant. Recall
that so far $F_p$ has been defined on the open dense domain
\eq{gf3eq18} in Proposition \ref{gf3prop6}, and $F_n$ for $n<p$ has
been defined on all of $(\C^\t)^n$. Thus, every term in \eq{gf3eq16}
with $n=p$ is now defined on the intersection of \eq{gf3eq18} and
$N_{(\ti z_1,\ldots,\ti z_p)}$; note that the only term on the
r.h.s.\ of \eq{gf3eq16} with $m=n=p$ is $G_p(z_1,\ldots,z_p;\ti
z_1,\ldots,\ti z_p)$. This intersection is also open and nonempty,
as $N_{(\ti z_1,\ldots,\ti z_p)}$ is nonempty and \eq{gf3eq18} is
dense.

We claim that there is a unique function $G_p$ satisfying
\eq{gf3eq24} such that \eq{gf3eq16} holds on the intersection of
\eq{gf3eq18} and $N_{(\ti z_1,\ldots,\ti z_p)}$. To see this, note
that by Proposition \ref{gf3prop4}, equations \eq{gf3eq24} and
\eq{gf3eq25} are equivalent when \eq{gf3eq16} holds. Thus, for any
choice of $G_p$ satisfying \eq{gf3eq24}, applying $\d$ to both sides
of \eq{gf3eq16} gives the same thing, so the difference between the
left and right hand sides of \eq{gf3eq16} is locally constant on the
intersection of $N_{(\ti z_1,\ldots,\ti z_p)}$ and \eq{gf3eq18}. Fix
a connected component $C$ of this intersection. Then we can choose
$G_p$ uniquely such that \eq{gf3eq16} holds on this connected
component, and on every other connected component the difference
between the left and right hand sides of \eq{gf3eq16} is constant.

Suppose $C',C''$ are connected components of the intersection of
$N_{(\ti z_1,\ldots,\ti z_p)}$ and \eq{gf3eq18} which meet along the
real hypersurface $z_{l+1}/z_l\in(0,\iy)$ for $1\le l<p$. (That is,
the closures of $C',C''$ must contain a nonempty open subset of this
hypersurface). Then Proposition \ref{gf3prop7} computes how much
$F_p$ jumps across this hypersurface, which by Proposition
\ref{gf3prop3} follows from the condition for $G_p$ to be continuous
across the hypersurface. It is not difficult to deduce that the
difference between the left and right hand sides of \eq{gf3eq16}
must take the same constant value on $C'$ and $C''$. Since this
value is 0 on one component $C$, and as $N_{(\ti z_1,\ldots,\ti
z_p)}$ is open and connected we can get from $C$ to any other
component $C'$ by crossing hypersurfaces $z_{l+1}/z_l\in(0,\iy)$ one
after the other, the constant is zero for every $C'$. This proves
the claim.

We have now defined the functions $G_p$. If $(z_1,\ldots,z_p)$ lies
in the intersection of \eq{gf3eq18} and $N_{(\ti z_1,\ldots,\ti
z_p)}$ with $z_1+\cdots+z_p=0$ then \eq{gf3eq16} holds at
$(z_1,\ldots,z_p)$. There is a term $G_p(z_1,\ldots,z_p;\ti
z_1,\ldots,\ti z_p)$ on the right hand side, and every other term is
zero by $F_p(z_1,\ldots,z_p)=0$ when $z_1+\cdots+z_p=0$ and the
inductive hypothesis. Hence $G_p(z_1,\ldots,z_p;\ti z_1,\ldots,\ti
z_p)=0$. By continuity this extends to all $(z_1,\ldots,z_p)$ in
$N_{(\ti z_1,\ldots,\ti z_p)}$ with $z_1+\cdots+z_p=0$, as we have
to prove.

By construction, \eq{gf3eq16} holds on the intersection of $N_{(\ti
z_1,\ldots,\ti z_p)}$ and the subset \eq{gf3eq18} where $F_p$ is
already defined by Proposition \ref{gf3prop6}. We now extend $F_p$
to $(\C^\t)^p$ by requiring $F_p$ to satisfy \eq{gf3eq18} with $n=p$
on each domain $N_{(\ti z_1,\ldots,\ti z_p)}$. Since the $N_{(\ti
z_1,\ldots,\ti z_p)}$ cover $(\C^\t)^p$ this defines $F_p$ uniquely,
but we must check that given $(\ti z_1,\ldots,\ti z_p)$ and $(\hat
z_1,\ldots,\hat z_p)$, equation \eq{gf3eq18} for $n=p$ gives the
same answer for $F_p(z_1,\ldots,z_p)$ with the $\ti z_k$ and $\hat
z_k$ on the intersection $N_{(\ti z_1,\ldots,\ti z_p)}\cap N_{(\hat
z_1,\ldots,\hat z_p)}$.

This holds for the same reason that the conditions of Proposition
\ref{gf3prop1} hold for some $(\ti z_1,\ldots,\ti z_n)$ if and only
if they hold for all $(\ti z_1,\ldots,\ti z_n)$. The point is that
the condition for $f^\al$ to be continuous is that we can write
$F_p$ in the form \eq{gf3eq18} near $(\ti z_1,\ldots,\ti z_p)$ for
$G_k$ continuous in $(z_1,\ldots,z_k)$, and these continuity
conditions for $(\ti z_1,\ldots,\ti z_p),(\hat z_1,\ldots,\hat z_p)$
must be equivalent in the overlap $N_{(\ti z_1,\ldots,\ti z_p)}\cap
N_{(\hat z_1,\ldots,\hat z_p)}$. We are using \eq{gf3eq16} to
determine how to extend $F_p$ from \eq{gf3eq18} to $(\C^\t)^p$ in a
way that makes the $f^\al$ continuous, and these continuity
conditions are independent of the choice of $(\ti z_1,\ldots,\ti
z_p)$ or $(\hat z_1,\ldots,\hat z_p)$. Thus $F_p$ is well defined
and satisfies \eq{gf3eq16}. This completes the inductive step, and
the proof of Theorem~\ref{gf3thm2}.
\end{proof}

Our next three results verify the remaining conditions of
Remark~\ref{gf3rem1}.

\begin{thm} For $n\ge 1,$ define $A_n$ to be the free $\C$-algebra
with generators $e_1,\ldots,e_n$ and multiplication $*,$ and\/ $L_n$
to be the free Lie subalgebra of\/ $A_n$ generated by
$e_1,\ldots,e_n$ under the Lie bracket\/ $[f,g]=f*g-g*f$. Then for
any $(z_1,\ldots,z_n)\in(\C^\t)^n$ the following expression lies
in~$L_n:$
\e
\ts\sum_{\si\in S_n}F_n(z_{\si(1)},z_{\si(2)},\ldots,z_{\si(n)})\,
e_{\si(1)}*e_{\si(2)}*\cdots*e_{\si(n)},
\label{gf3eq36}
\e
where the $F_n$ are as in Theorem \ref{gf3thm2} and\/ $S_n$ is the
symmetric group. Also \eq{gf3eq5} holds, and\/ $f^\al$ in
\eq{gf3eq1} maps $\Stab(\A)\ra\L^\al,$ as in
Remark\/~{\rm\ref{gf3rem1}(d)}.
\label{gf3thm3}
\end{thm}

\begin{proof} We shall first prove the first part of the theorem on the
domain
\e
\bigl\{(z_1,\ldots,z_n)\in(\C^\t)^n:\text{$z_k/z_l\not\in(0,\iy)$
for all $1\le k<l\le n$}\bigr\}.
\label{gf3eq37}
\e
The point of this is that if $(z_1,\ldots,z_n)$ lies in \eq{gf3eq37}
then $(z_{\si(1)},\ldots,z_{\si(n)})$ lies in the domain
\eq{gf3eq18} where $F_n$ is holomorphic and satisfies \eq{gf3eq25}
for all~$\si\in S_n$.

Suppose by induction that for some $m\ge 2$ and all $n<m$, the
expression \eq{gf3eq36} lies in $L_n$ for all $(z_1,\ldots,z_n)$ in
\eq{gf3eq37}. Write $P_m$ for \eq{gf3eq36} with $n=m$, regarded as a
holomorphic function from \eq{gf3eq37} to $H_m$. Then we have
\ea
\d &P_m(z_1,\ldots,z_n)
\nonumber\\
&=\sum_{\si\in S_m}
\sum_{k=1}^{m-1}
\begin{aligned}[t]
&F_k(z_{\si(1)},\ldots,z_{\si(k)})\,e_{\si(1)}\!*\!\cdots\!*\!
e_{\si(k)}*\\
&F_{m-k}(z_{\si(k+1)},\ldots,z_{\si(m)})e_{\si(k+1)}\,\!*\!\cdots\!*\!
e_{\si(m)}\\
&\quad\;\>\ot\ts\left[\frac{\d z_{\si(k+1)}+\cdots+\d
z_{\si(m)}}{z_{\si(k+1)} +\cdots+z_{\si(m)}}-\frac{\d
z_{\si(1)}+\cdots+\d
z_{\si(k)}}{z_{\si(1)}+\cdots+z_{\si(k)}}\right]
\end{aligned}
\nonumber\\
&=\ha\sum_{\si\in S_m}
\sum_{k=1}^{m-1}
\begin{aligned}[t]
\bigl[&F_k(z_{\si(1)},\ldots,z_{\si(k)})\,e_{\si(1)}\!*\!\cdots\!*\!
e_{\si(k)},\\
&F_{m-k}(z_{\si(k+1)},\ldots,z_{\si(m)})\,e_{\si(k+1)}\!*\!\cdots
\!*\!e_{\si(m)}\bigr] \\
&\quad\;\>\ot\ts\left[\frac{\d z_{\si(k+1)}+\cdots+\d
z_{\si(m)}}{z_{\si(k+1)} +\cdots+z_{\si(m)}}-\frac{\d
z_{\si(1)}+\cdots+\d
z_{\si(k)}}{z_{\si(1)}+\cdots+z_{\si(k)}}\right]
\end{aligned}
\label{gf3eq38}\\
&=
\begin{aligned}[t]
&\ha\sum_{\si\in S_m}
\sum_{k=1}^{m-1}\frac{1}{k!(m\!-\!k)!}
\bigl[\,\ts\sum_{\tau\in S_k}
F_k(z_{\si\ci\tau(1)},\ldots,z_{\si\ci\tau(k)})
e_{\si\ci\tau(1)}\!*\!\cdots\!*\!e_{\si\ci\tau(k)},\\
&\ts\sum_{\up\in S_{m\!-\!k}}\!
F_{m\!-\!k}(z_{\si(k+\up(1))},\ldots,z_{\si(k+\up(m-k))})
e_{\si(k+\up(1))}\!*\!\cdots\!*\!
e_{\si(k+\up(m-k))}\bigr] \\
&\qquad\qquad\qquad\quad \ot\ts\left[\frac{\d z_{\si(k+1)}+\cdots+\d
z_{\si(m)}}{z_{\si(k+1)} +\cdots+z_{\si(m)}}-\frac{\d
z_{\si(1)}+\cdots+\d
z_{\si(k)}}{z_{\si(1)}+\cdots+z_{\si(k)}}\right].
\end{aligned}
\nonumber
\ea
Here the second line is immediate from \eq{gf3eq25}. The third line
is the average of two copies of the second, one copy as it stands,
the other relabelled with $m-k$ in place of $k$ and indices
$\si(k\!+\!1),\ldots,\si(m),\si(1),\ldots,\si(k)$ in place of
$\si(1),\ldots,\si(k),\si(k\!+\!1),\ldots,\si(m)$ respectively; this
is valid because of the sum over $\si\in S_m$. The fourth and final
line uses the fact that symmetrizing over $S_m$ on $1,\ldots,m$ is
equivalent to first symmetrizing over $S_k$ on $1,\ldots,k$ and
$S_{m-k}$ on $k+1,\ldots,m$, with factors $1/k!(m-k)!$, and then
symmetrizing over~$S_m$.

By the inductive hypothesis, as $k,m-k<m$, the terms $\sum_{\tau\in
S_k}\cdots$ and $\sum_{\up\in S_{m-k}}\cdots$ in the final line of
\eq{gf3eq38} lie in $L_k$ with generators
$e_{\si(1)},\ldots,e_{\si(k)}$ and $L_{m-k}$ with generators
$e_{\si(k+1)},\ldots,e_{\si(m)}$ respectively, so they and their
commutator in \eq{gf3eq38} lie in $L_m$. Hence $\d P_m$ is an
$L_m$-{\it valued\/ $1$-form} on \eq{gf3eq37}, not just an
$H_m$-valued 1-form. As $m\ge 2$ it is easy to show that each
connected component of \eq{gf3eq37} with $n=m$ contains a point
$(z_1,\ldots,z_m)$ with $z_1+\cdots+ z_m=0$. At this point
$F_m(z_{\si(1)},\ldots,z_{\si(m)})=0$ for all $\si\in S_m$, so
$P_m(z_1,\ldots,z_m)=0$, which lies in $L_m$. Thus $\d P_m$ is an
$L_m$-valued 1-form and $P_m(z_1,\ldots,z_m)$ lies in $L_m$ at one
point in each connected component of \eq{gf3eq37}, so
$P_m(z_1,\ldots,z_m)$ lies in $L_m$ at every point of \eq{gf3eq37},
completing the inductive step.

It remains to extend this from \eq{gf3eq37} to $(\C^\t)^n$. We do
this using an argument similar to Theorem \ref{gf3thm2}, and facts
about the coefficients $U(\cdots)$ from \cite[\S 5]{Joyc6}. The
relationships between the functions $F_n,G_n$ given in \eq{gf3eq10}
and \eq{gf3eq11} were derived by using the change of stability
condition formula \eq{gf2eq7} to transform between $\ep^\al(\mu)$
and $\ep^\be(\ti\mu)$. By \cite[Th.~5.4]{Joyc6}, equation
\eq{gf2eq7} can be rewritten as in \eq{gf2eq8} with the term
$[\cdots]$ a sum of multiple commutators of $\ep^{\ka(i)}(\tau)$ for
$i\in I$, so that it lies in $\L^\al$ rather than just $\H^\al$.

Suppose the open neighbourhoods $N_{(\ti z_1,\ldots,\ti z_n)}$ in
Theorem \ref{gf3thm2} are chosen so that $(z_1,\ldots,z_n)\in
N_{(\ti z_1,\ldots,\ti z_n)}$ if and only if
$(z_{\si(1)},\ldots,z_{\si(n)})\in N_{(\ti z_{\si(1)},\ldots,\ti
z_{\si(n)})}$ for all $\si\in S_n$. As we can take the $N_{(\ti
z_1,\ldots,\ti z_n)}$ to be sufficiently small open balls about
$(\ti z_1,\ldots,\ti z_n)$, this is clearly possible. Then since the
changes \eq{gf3eq10}--\eq{gf3eq11} between $F_n,G_n$ come from Lie
algebra transformations, one can show that \eq{gf3eq36} lies in
$L_n$ for all $n$ and $(z_1,\ldots,z_n)\in(\C^\t)^n$ if and only if
the expression
\e
\ts\sum_{\si\in S_n}G_n(z_{\si(1)},\ldots,z_{\si(n)};\ti z_{\si(1)},
\ldots,\ti z_{\si(n)})\, e_{\si(1)}*\cdots*e_{\si(n)}
\label{gf3eq39}
\e
lies in $L_n$ for all $(\ti z_1,\ldots,\ti z_n)\in(\C^\t)^n$
and~$(z_1,\ldots,z_n)\in N_{(\ti z_1,\ldots,\ti z_n)}$.

In fact one can prove more than this. For $m\ge 1$, write:
\begin{itemize}
\item[$(*_m)$] Suppose \eq{gf3eq36} lies in $L_n$ for
all $n<m$ and $(z_1,\ldots,z_n)\in(\C^\t)^n$ and \eq{gf3eq39} lies
in $L_n$ for all $n<m$, $(\ti z_1,\ldots,\ti z_n)\in(\C^\t)^n$ and
$(z_1,\ldots,z_n)\in N_{(\ti z_1,\ldots,\ti z_n)}$.
\end{itemize}
One can show that if $(*_m)$ holds, $(\ti z_1,\ldots,\ti
z_m)\in(\C^\t)^m$ and $(z_1,\ldots,z_m)\in N_{(\ti z_1,\ldots,\ti
z_m)}$, then \eq{gf3eq36} with $n=m$ and this $(z_1,\ldots,z_m)$
lies in $L_m$ if and only if \eq{gf3eq39} with $n=m$ and these
$(z_1,\ldots,z_m),(\ti z_1,\ldots,\ti z_m)$ lies in $L_m$. The point
is that \eq{gf3eq39} is \eq{gf3eq36} plus sums of multiple
commutators of terms we know lie in $L_m$ by our assumptions for
$n<m$, and vice versa.

Suppose by induction that $(*_m)$ holds for some $m\ge 1$. When
$m=1$ this is vacuous. Let $(\ti z_1,\ldots,\ti z_m)\in(\C^\t)^m$
and $(z_1,\ldots,z_m)\in N_{(\ti z_1,\ldots,\ti z_m)}$ with
$(z_1,\ldots,z_m)$ in \eq{gf3eq37} for $m=n$. Then \eq{gf3eq36} with
$n=m$ and this $(z_1,\ldots,z_m)$ lies in $L_m$ by the proof above,
so \eq{gf3eq39} with $n=m$ and these $(z_1,\ldots,z_m),(\ti
z_1,\ldots,\ti z_m)$ lies in $L_m$. As $L_m$ is closed and
$G_m(z_1,\ldots,z_m;\ti z_1,\ldots,\ti z_m)$ is continuous in
$(z_1,\ldots,z_m)$ and the intersection of $N_{(\ti z_1,\ldots,\ti
z_m)}$ with \eq{gf3eq37} for $m=n$ is dense in $N_{(\ti
z_1,\ldots,\ti z_m)}$, taking limits shows \eq{gf3eq39} lies in
$L_m$ for any $(z_1,\ldots,z_m)\in N_{(\ti z_1,\ldots,\ti z_m)}$. As
this holds for all $(\ti z_1,\ldots,\ti z_m)\in(\C^\t)^m$, equation
\eq{gf3eq36} lies in $L_m$ for all $(z_1,\ldots,z_m)\in(\C^\t)^m$.
Hence by induction $(*_m)$ holds for all $m\ge 1$, which proves the
first part of the theorem. The remaining two parts follow as in
Remark~\ref{gf3rem1}(d).
\end{proof}

\begin{lem} If\/ $1\le k\le n$ and\/ $z_1,\ldots,z_{k-1},
z_{k+1},\ldots,z_n$ are fixed in $\C^\t$, the function $F_n$ of
Theorem \ref{gf3thm2} satisfies $\bmd{F_n(z_1,\ldots,z_n)}\le
C(1+\md{\log z_k})^{n-1}$ for all $z_k\in\C^\t$, for some $C>0$
depending on $k,n$ and\/~$z_1,\ldots,z_{k-1},z_{k+1},\ldots,z_n$.
\label{gf3lem1}
\end{lem}

\begin{proof} For $n=1,2$ the lemma follows from \eq{gf3eq3}
and \eq{gf3eq20}. On the domains \eq{gf3eq18}, equation \eq{gf3eq25}
gives an expression for $\pd F_n/\pd z_k$ in terms of $F_l$ for
$l<n$, and it is easy to use this and induction on $n$ to prove the
lemma on \eq{gf3eq18}. To extend from \eq{gf3eq18} to $(\C^\t)^n$,
we can observe that for $(z_1,\ldots,z_n)$ in the complement of
\eq{gf3eq18} in $(\C^\t)^n$, $F_n(z_1,\ldots,z_n)$ is a weighted
average of the limits of $F_n(z_1',\ldots,z_n')$ as
$(z_1',\ldots,z_n')\ra(z_1,\ldots,z_n)$, for $(z_1',\ldots,z_n')$ in
the various sectors of \eq{gf3eq18} meeting at $(z_1,\ldots,z_n)$.
In particular, $F_n(z_1,\ldots,z_n)$ lies in the convex hull in $\C$
of these limits, so estimates on $\md{F_n}$ on \eq{gf3eq18} imply
the same estimates on~$(\C^\t)^n$.
\end{proof}

\begin{cor} The functions $F_n$ of Theorem \ref{gf3thm2} satisfy
Condition \ref{gf3cond} and equations \eq{gf3eq2}, \eq{gf3eq3},
\eq{gf3eq5} and\/ \eq{gf3eq6} of Remark \ref{gf3rem1}. Thus by
Theorem \ref{gf3thm1} they are the unique functions $F_n$ satisfying
the conditions of\/~{\rm\S\ref{gf31}}.
\label{gf3cor}
\end{cor}

\begin{proof} Condition \ref{gf3cond} holds by Theorem
\ref{gf3thm2}. Given $\la\in\C^\t$ we note that all conditions on
$F_n,G_n$ are preserved by replacing $F_n(z_1,\ldots,z_n)$ by
$F_n(\la z_1,\ldots,\la z_n)$ and $G_n(z_1,\ldots,z_n;\ti
z_1,\ldots,\ti z_n)$ by $G_n(\la z_1,\ldots,\la z_n;\la\ti
z_1,\ldots,\la\ti z_n)$. Thus, since these conditions determine
$F_n,G_n$ uniquely \eq{gf3eq2} must hold. Equation \eq{gf3eq3} holds
by definition, and \eq{gf3eq5} and \eq{gf3eq6} follow from Theorem
\ref{gf3thm3} and Lemma~\ref{gf3lem1}.
\end{proof}

\begin{rem} It is an obvious question whether the functions $F_n$
constructed above can be written in terms of known special
functions. Tom Bridgeland has found a very nice answer to this,
which will be published in \cite{Brid3}. It involves the {\it
hyperlogarithms\/} of Goncharov \cite[\S 2]{Gonc}, a kind of {\it
polylogarithm}, which are defined by iterated integrals and satisfy
a p.d.e.\ reminiscent of~\eq{gf3eq25}.

Bridgeland shows that $F_n(z_1,\ldots,z_n)$ may be written on the
domain \eq{gf3eq18} as an explicit sum over {\it rooted trees\/}
with $n$ leaves of a product over vertices of the tree of a
hyperlogarithm whose arguments are various sums of $z_1,\ldots,z_n$,
and a constant factor. This is interesting, as polylogarithms and
hyperlogarithms have many links to other branches of mathematics
such as number theory, Hodge theory and motives, and the author
wonders whether the ideas of this paper will also have such links.
\label{gf3rem2}
\end{rem}

\section{Flat connections}
\label{gf4}

We now explain how to define a holomorphic $\L$-valued connection
$\Ga$ on $\Stab(\A)$ using the generating functions $f^\al$, which
the p.d.e.\ \eq{gf3eq22} implies is flat. Our formulae involve
infinite sums over all $\al\in C(\A)$, so we need a notion of {\it
convergence\/} of infinite sums in $\L$, that is, a topology on
$\L$. This also clarifies the meaning of the infinite direct sum
$\L=\bigop_{\al\in C(\A)}\L^\al$ in Assumption \ref{gf2ass2}, since
we can take $\L$ to be the set of convergent sums $\sum_{\al\in
C(\A)}l^\al$ with~$l^\al\in\L^\al$.

Here are simple definitions of convergence and the direct sum which
go well with Assumption \ref{gf3ass}, and ensure the formulae below
converge in this case. If Assumption \ref{gf3ass} does not hold,
choosing a topology on $\L$ to make the formulae below converge may
be difficult or impossible; in this case the sum \eq{gf3eq1}
defining $f^\al$ may not converge either. Also, we must consider
whether the Lie bracket $[\,,\,]$ is defined on all of $\L\t\L$, and
whether it commutes with limits.

\begin{dfn} In Assumption \ref{gf2ass2}, by the {\it direct sum}
$\L=\bigop_{\al\in C(\A)}\L^\al$ we mean simply that $\L$ is the
infinite Cartesian product of the spaces $\L^\al$. That is, elements
of $\L$ are just arbitrary families $(l^\al)_{\al\in C(\A)}$ with
$l^\al\in\L^\al$, with no restriction on how many $l^\al$ are zero,
and no other `smallness conditions' on the $l^\al$. Write
$\Pi^\al:\L\ra\L^\al$ for the obvious projection.

A possibly infinite sum $\sum_{i\in I}l_i$ in $\L$ is called {\it
convergent\/} if for each $\al\in C(\A)$ there are only finitely
many $i\in I$ with $\Pi^\al(l_i)$ nonzero. The {\it limit\/}
$l=(l^\al)_{\al\in C(\A)}$ in $\L$ is defined uniquely by taking
$l^\al$ to be the sum of the nonzero $\Pi^\al(l_i)$. That is,
$\sum_{i\in I}l_i=l$ if $\sum_{i\in I}\Pi^\al(l_i)= \Pi^\al(l)$ in
$\L^\al$ for all $\al\in C(\A)$, where the second sum is
well-defined as it has only finitely many nonzero terms. The direct
sum $\H=\bigop_{\bar\al\in C(\A)}\H^\al$ and convergence of sums in
$\H$ are defined in the same way.

If Assumption \ref{gf3ass} holds, it is easy to see that the Lie
brackets $[\,,\,]:\L^\al\t\L^\be\ra\L^{\al+\be}$ extend to a unique
Lie bracket $[\,,\,]:\L\t\L\ra\L$ which commutes with limits.
Otherwise, the Lie bracket of two convergent sums can be a
nonconvergent sum, so $[\,,\,]$ can only be defined on a subspace
of~$\L\t\L$.
\label{gf4def}
\end{dfn}

In the situation of \S\ref{gf3}, define a section $\Ga$ in
$C^\iy\bigl(\L\ot T^*_{\sst\mathbb C}\Stab(\A)\bigr)$ by
\e
\Ga(Z)=\sum_{\al\in C(\A)}f^\al(Z)\ot\frac{\d(Z(\al))}{Z(\al)}.
\label{gf4eq1}
\e
This infinite sum is convergent in the sense of Definition
\ref{gf4def}, extended to $\L\ot T^*_{\sst\mathbb C}\Stab(\A)$ in
the obvious way, since for each $\al\in C(\A)$ there is only one
term in the sum with $\Pi^\al(\cdots)$ nonzero. Then $\Ga$ is a {\it
connection matrix\/} for a holomorphic connection on the trivial
complex Lie algebra bundle $\L\t\Stab(\A)$ over~$\Stab(\A)$.

By standard differential geometry, the {\it curvature} of this
connection is the section $R_\Ga$ of the vector bundle
$\L\ot\La^2T^*_{\sst\mathbb C}\Stab(\A)$ over $\Stab(\A)$ given by
\e
R_\Ga=\d\Ga+\ha\Ga\w\Ga.
\label{gf4eq2}
\e
To form $\Ga\w\Ga\in C^\iy\bigl(\L\ot\La^2T^*_{\sst\mathbb
C}\Stab(\A)\bigr)$ from $\Ga\ot\Ga\in C^\iy\bigl((\L\ot
T^*_{\sst\mathbb C}\Stab(\A))^2\bigr)$, we both project
$\L\ot\L\ra\L$ using the Lie bracket $[\,,\,]$ on $\L$, and project
$T^*_{\sst\mathbb C}\Stab(\A)\ot T^*_{\sst\mathbb C}\Stab(\A)\ra
\La^2T^*_{\sst\mathbb C}\Stab(\A)$ using the wedge product $\w$.

Combining \eq{gf3eq22}, \eq{gf4eq1} and \eq{gf4eq2} we find that
\ea
R_\Ga&=\!\!\sum_{\al\in C(\A)}\!\!\d
f^\al(Z)\w\frac{\d(Z(\al))}{Z(\al)}+\ha\!\!\!\sum_{\be,\ga\in
C(\A)}\!\!\!\![f^\be(Z),f^\ga(Z)]\!\ot\!
\frac{\d(Z(\be))}{Z(\be)}\w\frac{\d(Z(\ga))}{Z(\ga)}
\nonumber\\
\begin{split}
&=\!\sum_{\begin{subarray}{l}\al,\be,\ga\in C(\A):\\
\be+\ga=\al\end{subarray}\!\!\!\!\!\!\!\!}\!
\bigl(\ha[f^\be(Z),f^\ga(Z)]\bigr)\ot
\left(\frac{\d(Z(\ga))}{Z(\ga)}-\frac{\d(Z(\be))}{Z(\be)}\right)\w
\frac{\d(Z(\al))}{Z(\al)}\\
&\qquad +\!\!\sum_{\be,\ga\in C(\A)}\!\!\bigl(\ha[f^\be(Z),f^\ga(Z)]
\bigr)\!\ot\frac{\d(Z(\be))}{Z(\be)}\w\frac{\d(Z(\ga))}{Z(\ga)}
\end{split}
\label{gf4eq3}\\
&=\sum_{\be,\ga\in C(\A)}\bigl(\ha[f^\be(Z),f^\ga(Z)]\bigr)\ot
\nonumber\\
&\quad\left[\left(\frac{\d(Z(\ga))}{Z(\ga)}\!-\!\frac{\d(Z(\be))}{Z(\be)}
\right)\!\w\!\frac{\d(Z(\be))\!+\!\d(Z(\ga))}{Z(\be)\!+\!Z(\ga)}
\!+\!\frac{\d(Z(\be))}{Z(\be)}\!\w\!\frac{\d(Z(\ga))}{Z(\ga)}\right]
\!=\!0,
\nonumber
\ea
since the term $[\cdots]$ in the last line is zero. Thus $\Ga$ is a
{\it flat connection}. If Assumption \ref{gf3ass} holds, these
calculations are all valid as infinite convergent sums in the sense
of Definition~\ref{gf4def}.

If $\rho:\L\ra\End(V)$ is a representation of the Lie algebra $\L$
on a complex vector space $V$ then $\Ga$ induces a flat connection
$\nabla_{\rho(\Ga)}$ on the trivial vector bundle $V\t\Stab(\A)$
over $\Stab(\A)$, with connection 1-form $\rho(\Ga)$ in
$C^\iy\bigl(\End(V)\ot T^*_{\sst\mathbb C}\Stab(\A)\bigr)$. If
$s:\Stab(\A)\ra V$ is a smooth section of this bundle then
$\nabla_{\rho(\Ga)}s=\d s+\rho(\Ga)\cdot s$ in~$C^\iy\bigl(V\ot
T^*_{\sst\mathbb C}\Stab(\A)\bigr)$.

In particular, as the tangent bundle $T\Stab(\A)$ is naturally
isomorphic to the trivial vector bundle $\Hom(K(\A),\C)\t\Stab(\A)$,
if $\L$ has a representation $\rho$ on $\Hom(K(\A),\C)$ then
$\nabla_{\rho(\Ga)}$ is a flat connection on $T\Stab(\A)$. We will
see in \S\ref{gf6} that this should happen in the triangulated
category extension of the Calabi--Yau 3-fold invariants in
Example~\ref{gf2ex8}.

Take $V$ to be $\L$ and $\rho$ the adjoint representation
$\ad:\L\ra\End(\L)$. Define a section $s:\Stab(\A)\ra\L$ by
\e
s(Z)=\ts\sum_{\al\in C(\A)}f^\al(Z),
\label{gf4eq4}
\e
which converges as in Definition \ref{gf4def}. Then from
\eq{gf3eq22} and \eq{gf4eq1} we see that
\ea
&\nabla_{\ad(\Ga)}s=\!\sum_{\al\in C(\A)\!\!}\!\d f^\al(Z)+
\!\sum_{\be\in C(\A)\!\!}\!\left[f^\be(Z),{\ts\sum_{\ga\in
C(\A)}f^\ga(z)}\right]\!\ot\!\frac{\d(Z(\be))}{Z(\be)}
\label{gf4eq5}\\
&\;\>=-\!\!\!\!\!\sum_{\begin{subarray}{l}\al,\be,\ga\in C(\A):\\
\al=\be+\ga\end{subarray}\!\!\!\!\!\!\!\!\!\!\!}\!\!\!
[f^\be(Z),f^\ga(Z)]\!\ot\!\frac{\d(Z(\be))}{Z(\be)}+\!\!\!
\sum_{\be,\ga\in C(\A)\!\!\!\!\!\!\!\!}\![f^\be(Z),f^\ga(z)]
\!\ot\!\frac{\d(Z(\be))}{Z(\be)}\!=\!0,
\nonumber
\ea
so that $s$ in \eq{gf4eq4} is a {\it constant section\/} of
$\L\t\Stab(\A)$. If Assumption \ref{gf3ass} holds then \eq{gf4eq5}
is valid as infinite convergent sums in the sense of
Definition~\ref{gf4def}.

Let $P:\L\ra\C$ be smooth and invariant under $\ad(\L)$, that is,
$\d P(x)\cdot[x,y]=0$ for all $x,y\in\L$. Then $\nabla^{\ad(\Ga)}s
=0$ implies that $P(s)$ is constant on $\Stab(\A)$. For example, if
$\rho:\L\ra\End(V)$ is a representation of $\L$ on a
finite-dimensional $\C$-vector space $V$ then
$P(x)=\det\bigl(\rho(x)-\la\id_V\bigr)$ has these properties, so the
characteristic polynomial of $\rho(s(Z))$ is constant
on~$\Stab(\A)$.

In general, for $s$ as in \eq{gf4eq4} the eigenvalues of $s(Z)$ in
any representation of $\L$ should be constant on $\Stab(\A)$.
However, the author does not expect this construction to be useful
with the topology on $\L$ in Definition \ref{gf4def}, as it seems
likely that the only finite-dimensional representations for such
infinite-dimensional $\L$ will be nilpotent, and so have zero
eigenvalues anyway.

The author feels that the topology on $\L$ given in Definition
\ref{gf4def} is rather trivial, and that if the ideas of this
section do have interesting applications in classes of examples it
will be with a more complex topology on $\L$ appropriate to the
examples. Then the convergence and validity of equations
\eq{gf4eq1}--\eq{gf4eq5} would become conjectures to be (dis)proved
in these examples, depending on asymptotic properties of the $f^\al$
for large~$\al$.

\section{Extending all this to triangulated categories}
\label{gf5}

Our programme cannot yet be rigorously extended from abelian
categories $\A$ to triangulated categories $\T$, because the
material of \cite{Joyc3,Joyc4,Joyc5,Joyc6} on which it rests has not
yet been extended. Some remarks on the issues involved are given in
\cite[\S 7]{Joyc6}. The work of Bertrand To\"en \cite{Toen1,Toen2}
is likely to be useful here. In particular, \cite{Toen1} defines a
`derived Hall algebra' $\mathcal{DH}(\T)$ under strong finiteness
conditions on $\T$, and \cite[\S 3.3.3]{Toen2} an `absolute Hall
algebra' ${\mathcal H}_{abs}(\T)$ under weaker conditions.

It seems likely that the right way to construct examples of data
satisfying a triangulated version of Assumption \ref{gf2ass2} is to
use an algebra morphism $\Phi:\mathcal{DH}(\T)$ or ${\mathcal
H}_{abs}(\T)\ra\H$. Also, \cite{Toen2} provides the tools needed to
form moduli Artin $\iy$-stacks of objects and configurations in
triangulated categories with dg-enhancement, which is the main
ingredient needed to extend \cite{Joyc3,Joyc4,Joyc5,Joyc6} to the
triangulated case. Here are some issues in extending the ideas of
this paper to the triangulated case.
\smallskip

\noindent{\bf Lifting phases from $\R/2\pi i\Z$ to $\R$.} The
$\de^\al(\tau),\ep^\al(\tau)$ of Assumption \ref{gf2ass2} are
constructed from `characteristic functions' of $\tau$-semistable
objects in $\A$ in class $\al\in C(\A)$. Now in Bridgeland's
stability conditions $(Z,\P)$ on a triangulated category $\T$,
Definition \ref{gf2def8}, the $(Z,\P)$-semistable objects in class
$\al\in K(\T)$ depend on a choice of {\it phase}
for~$Z(\al)\in\C^\t$.

That is, if we write $Z(\al)=re^{i\pi\phi}$ for $\phi\in\R$, then
the $(Z,\P)$-semistable objects in class $\al$ with phase $\phi$ are
the objects $U$ in $\P(\phi)$ with class $\al\in\K(\T)$. Replacing
$\phi$ by $\phi+2n$ for $n\in\Z$ replaces $\P(\phi)$ by
$\P(\phi+2n)=\P(\phi)[2n]$, so replaces objects $U$ by $U[2n]$, that
is, applying the translation functor to the power $2n$. Note that
replacing $U$ by $U[2n]$ fixes the class $\al$ of $U$ in~$K(\T)$.

It is natural to ask whether the triangulated analogues
$\de^\al(Z,\P),\ep^\al(Z,\P)$  should also depend on a choice of
phase $\phi$ for $Z(\al)$. The author's view is that for the
purposes of this paper, they should {\it not\/} depend on choice of
phase. Effectively this means working in a Hall-type algebra in
which the translation squared operator $[+2]$ is the identity.

The reason is that if $\de^\al(Z,\P),\ep^\al(Z,\P)$ depended on
phase then $f^\al(Z,\P)$ should also depend on a choice of phase for
$Z(\al)$, and $F_n(z_1,\ldots,z_n)$ on choices of phase for
$z_1,\ldots,z_n$. That is, $F_n$ should be a function of $(\log
z_1,\ldots,\log z_n)\in\C^n$ rather than $(z_1,\ldots,z_n)\in
(\C^\t)^n$. Allowing this would invalidate nearly all of
\S\ref{gf3}. In particular, the uniqueness result Theorem
\ref{gf3thm1} would fail, and the p.d.e.\ \eq{gf3eq22} would no
longer make sense, as for a given choice of phase for $Z(\al)$ there
does not seem to be a natural way to choose phases for
$Z(\be),Z(\ga)$ in the sum.
\smallskip

\noindent{\bf Replacing $C(\A)$ by $\bigl\{\al\in K(\T):Z(\al)\ne
0\bigr\}$.} What should be the analogue of the {\it positive cone}
$C(\A)$ in a triangulated category $\T$? Replacing $\A$ by $\T$ in
\eq{gf2eq1} will give $C(\T)=K(\T)$, as for nonzero $\T$ every
element of $K(\T)$ will be represented by a nonzero object. However,
the sums over $\al\in C(\A)$ in \S\ref{gf3} do not make sense when
replaced by $\al\in K(\T)$, because of problems when $Z(\al)=0$. For
instance, $F_n(Z(\al_1),\ldots,Z(\al_n))$ in \eq{gf3eq1} is
undefined if any $Z(\al_k)=0$, and \eq{gf3eq22} is undefined if any
$Z(\be)=0$ or~$Z(\ga)=0$.

The author proposes that the right answer is to replace sums over
$\al\in C(\A)$ in \S\ref{gf3} involving $\ep^\al(\mu)$, such as
\eq{gf3eq1}, by sums over all $\al\in K(\T)$ with $Z(\al)\ne 0$. For
generic $Z$ this amounts to summing over $\al\in K(\T)\sm\{0\}$.
Sums over $\al\in C(\A)$ involving $f^\al(Z)$ need a more subtle
approach we describe below. We now explain two neat coincidences
meaning that arguments in \S\ref{gf3} still work with this
replacement, although one might have expected them to fail.

First, note that if $(Z,\P)\in\Stab(\T)$ and $\al\in K(\T)$ with
$Z(\al)=0$ then we must have $\de^\al(Z,\P)=\ep^\al(Z,\P)=0$. This
is because $\de^\al(Z,\P),\ep^\al(Z,\P)$ are constructed from
$(Z,\P)$-semistable objects in class $\al$, but there are no such
objects if $Z(\al)=0$ by Definition \ref{gf2def8}. We also expect
$\de^\al(Z',\P')=\ep^\al(Z',\P')=0$ for $(Z',\P')$ in a small open
neighbourhood of $(Z,\P)$ in $\Stab(\T)$. This means that omitting
terms in $\ep^{\al_i}(Z,\P)$ in \eq{gf3eq1} when $Z(\al_i)=0$ does
not cause discontinuities on the hypersurface $Z(\al_i)=0$ in
$\Stab(\T)$, since the omitted terms are zero near there anyway.

Second, note that $f^\al(Z,\P)=0$ when $Z(\al)=0$, since \eq{gf3eq1}
now involves terms in $\al_1,\ldots,\al_n$ with $Z(\al_k)\ne 0$ but
$Z(\al_1)+\cdots+Z(\al_n)=Z(\al)=0$, so $F_n(Z(\al_1),\ldots,
Z(\al_n))=0$, and every term in \eq{gf3eq1} is zero. However, for
$\al\ne 0$ we do not expect $f^\al(Z',\P')\equiv 0$ for $(Z',\P')$
near $(Z,\P)$. So in sums such as \eq{gf4eq1} involving
$f^\al(Z)/Z(\al)$, giving $0/0$ when $Z(\al)=0$, it is not right to
just omit $\al$ when $Z(\al)=0$, for $\al\ne 0$. Instead, since
$f^\al(Z,\P)$ is holomorphic and zero when $Z(\al)=0$, as for the
functions $H_n$ in \S\ref{gf32}, the holomorphic function
$h^\al(Z,\P)=f^\al(Z,\P)/Z(\al)$ on $Z(\al)\ne 0$ extends uniquely
over $Z(\al)=0$, so in \eq{gf3eq1}, \eq{gf3eq3}, \eq{gf3eq5} we
replace terms $f^\al(Z,\P)/Z(\al)$ by~$h^\al(Z,\P)$.
\smallskip

\noindent{\bf Convergence of sums.} Once we replace sums over
$\al\in C(\A)$ by sums over $\al\in K(\T)$ with $Z(\al)\ne 0$, most
of the equations in \S\ref{gf3}--\S\ref{gf4} become infinite sums,
and the question of whether they converge at all in any sense
becomes acute. There seems to be no triangulated analogue of
Assumption \ref{gf3ass} that makes the sums finite, nor can the
author find any way to make the sums converge in a formal power
series sense. Here are two comments which may help.

Firstly, suppose the Lie algebra $\L$ is {\it nilpotent}. That is,
define ideals $\L=\L_1\supset\L_2\supset\cdots$ by $\L_1=\L$,
$\L_{n+1}=[\L,\L_n]$, and suppose $\bigcap_{n\ge 1}\L_n=\{0\}$. Then
Theorem \ref{gf3thm3} implies that the sum of terms with fixed $n$
in \eq{gf3eq1} lie in $\L_n$. Hence, projecting \eq{gf3eq1} to
$\L/\L_k$ eliminates all terms with $n\ge k$. If we use a notion of
convergence such that a sum converges in $\L$ if its projections to
$\L/\L_k$ converge for all $k\ge 1$, then we only have to show the
sum \eq{gf3eq1} for $n<k$ converges in $\L/\L_k$, which may be
easier.

Secondly, even if the sum \eq{gf3eq1} defining $f^\al$ does not make
sense, the p.d.e.\ \eq{gf3eq22} upon the $f^\al$ might still
converge in the triangulated case, as it is a much simpler sum. For
example, if ${\mathfrak g}={\mathfrak n}_+\op{\mathfrak
h}\op{\mathfrak n}_-$ is a Kac--Moody Lie algebra, it is known
\cite[\S 4.9]{Joyc4} how to use Ringel--Hall algebras of abelian
categories of quiver representations $\A=\modKQ$ to realize
$\H=U({\mathfrak n}_+)$ and $\L={\mathfrak n}_+$ in examples, and
people have hoped to use triangulated categories to obtain
$\H=U({\mathfrak g})$ and $\L={\mathfrak g}$. If we could do this
with ${\mathfrak g}$ a finite-dimensional semisimple Lie algebra,
then $\al,\be,\ga$ in \eq{gf3eq22} would take values in the set of
roots of $\mathfrak g$, with $\L^\al$ being the root space
${\mathfrak g}_\al$, and \eq{gf3eq22} would become a finite sum, so
trivially convergent. However, \eq{gf3eq1} would still be an
infinite sum.

Intuitively, what is going on is as follows. The functions $F_n$ are
related to certain finite-dimensional nilpotent Lie algebras $N_n$
for $n\ge 1$. In a similar way to Ramakrishnan \cite{Rama} for the
higher logarithms $\ln_k$, one can use the $F_k$ for $k\le n$ to
write down a nontrivial flat holomorphic $N_n$-valued connection on
\begin{equation*}
\bigl\{(z_1,\ldots,z_n)\in\C^n:\text{$z_a+\cdots+z_b\ne 0$ for all
$1\le a\le b\le n$, $(a,b)\ne(1,n)$}\bigr\}.
\end{equation*}
In the Ringel--Hall case $\H=U({\mathfrak n}_+)$, $\L={\mathfrak
n}_+$ above, equation \eq{gf3eq1} is about building the nilpotent
Lie algebra ${\mathfrak n}_+$, and the flat ${\mathfrak n}_+$-valued
connection $\Ga$ of \S\ref{gf4}, out of the standard family of
nilpotent Lie algebras $N_n$, and standard flat $N_n$-valued
connections.

However, it may not be possible to build semisimple Lie algebras
$\mathfrak g$ and their flat connections from standard nilpotent
building blocks $N_n$, which is why \eq{gf3eq1} may not converge.
But \eq{gf3eq22} has to do with general Lie algebras, not just
nilpotent Lie algebras, and so may make sense in a more general
setting.

\section{The Calabi--Yau 3-fold case}
\label{gf6}

Finally we discuss and elaborate the ideas of
\S\ref{gf3}--\S\ref{gf5} in the Calabi--Yau 3-fold case of Example
\ref{gf2ex8}. We use the notation of this example and
\S\ref{gf3}--\S\ref{gf5} throughout.

\subsection{Holomorphic functions $F^\al,H^\al$ and their p.d.e.s}
\label{gf61}

We begin with the abelian category case. Since $f^\al$ maps
$\Stab(\A)\ra\L^\al$ by Theorem \ref{gf3thm3} and $\L^\al=\C\cdot
c^\al$ we may write $f^\al=F^\al c^\al$ for a holomorphic function
$F^\al:\Stab(\A)\ra\C$, for $\al\in C(\A)$. Also
$\ep^\al(\mu)=J^\al(\mu)c^\al$ for $J^\al(\mu)\in\Q$, so combining
\eq{gf2eq19} and \eq{gf3eq1} we find that
\e
\begin{split}
F^\al(Z)=\!\!\!\!\sum_{\substack{n\ge 1,\;\al_1,\ldots,\al_n\in
C(\A):\\ \al_1+\cdots+\al_n=\al}}\!\!\!\!\!\!\!\!
F_n\bigl(Z(\al_1),\ldots,Z(\al_n)\bigr)
\prod_{i=1}^nJ^{\al_i}(\mu)\cdot
\\
\raisebox{-6pt}{\begin{Large}$\displaystyle\biggl[$\end{Large}}
\frac{1}{2^{n-1}}\!\!\!\!\!
\sum_{\substack{\text{connected, simply-connected digraphs
$\Ga$:}\\
\text{vertices $\{1,\ldots,n\}$, edge $\mathop{\bu} \limits^{\sst
i}\ra\mathop{\bu}\limits^{\sst j}$ implies $i<j$}}} \,\,\,
\prod_{\substack{\text{edges}\\
\text{$\mathop{\bu}\limits^{\sst i}\ra\mathop{\bu}\limits^{\sst
j}$}\\ \text{in $\Ga$}}}\bar\chi(\al_i,\al_j)
\raisebox{-6pt}{\begin{Large}$\displaystyle\biggr]$\end{Large}},
\end{split}
\label{gf6eq1}
\e
where $\mu$ is the slope function associated to $Z$. We also have
$F^\al\equiv Z(\al)H^\al$ for a holomorphic function
$H^\al:\Stab(\A)\ra\C$ given by
\e
\begin{split}
H^\al(Z)=\!\!\!\!\sum_{\substack{n\ge 1,\;\al_1,\ldots,\al_n\in
C(\A):\\ \al_1+\cdots+\al_n=\al}}\!\!\!\!\!\!\!\!
H_n\bigl(Z(\al_1),\ldots,Z(\al_n)\bigr)
\prod_{i=1}^nJ^{\al_i}(\mu)\cdot
\\
\raisebox{-6pt}{\begin{Large}$\displaystyle\biggl[$\end{Large}}
\frac{1}{2^{n-1}}\!\!\!\!\!
\sum_{\substack{\text{connected, simply-connected digraphs
$\Ga$:}\\
\text{vertices $\{1,\ldots,n\}$, edge $\mathop{\bu} \limits^{\sst
i}\ra\mathop{\bu}\limits^{\sst j}$ implies $i<j$}}} \,\,\,
\prod_{\substack{\text{edges}\\
\text{$\mathop{\bu}\limits^{\sst i}\ra\mathop{\bu}\limits^{\sst
j}$}\\ \text{in $\Ga$}}}\bar\chi(\al_i,\al_j)
\raisebox{-6pt}{\begin{Large}$\displaystyle\biggr]$\end{Large}}.
\end{split}
\label{gf6eq2}
\e

The p.d.e.\ \eq{gf3eq22} becomes
\e
\begin{split}
\d F^\al(Z)&=-\sum_{\be,\ga\in
C(\A):\al=\be+\ga}\bar\chi(\be,\ga)F^\be(Z)F^\ga(Z)
\frac{\d(Z(\be))}{Z(\be)}\\
&=-\sum_{\be,\ga\in
C(\A):\al=\be+\ga}\bar\chi(\be,\ga)H^\be(Z)H^\ga(Z)
Z(\ga)\d(Z(\be)),
\end{split}
\label{gf6eq3}
\e
and the flat connection $\Ga$ of \eq{gf4eq1} is
\e
\Ga(Z)=\sum_{\al\in C(\A)}F^\al(Z)\,c^\al\ot
\frac{\d(Z(\al))}{Z(\al)}=\sum_{\al\in
C(\A)}H^\al(Z)\,c^\al\ot\d(Z(\al)).
\label{gf6eq4}
\e
In the triangulated category case we replace $F^\al,H^\al(Z)$ and
$J^\al(\mu)$ by $F^\al,H^\al,\ab J^\al(Z,\P)$, and replace sums over
$C(\A)$ by sums over $K(\T)\sm\{0\}$ in \eq{gf6eq1}--\eq{gf6eq4},
and also omit terms involving $\al_i$ with $Z(\al_i)=0$
in~\eq{gf6eq1}--\eq{gf6eq2}.

In the triangulated case, the Lie algebra $\L$ is $\L=\langle
c^\al:\al\in K(\T)\rangle_{\sst\mathbb C}$, with
$[c^\al,c^\be]=\bar\chi(\al,\be)c^{\al+\be}$. Suppose $K(\T)$ is a
lattice of finite rank, and $\chi:K(\T)\t K(\T)\ra\Z$ is
nondegenerate. Then we can interpret $\L$ as a Lie algebra of
complex functions on the real torus $T_\T=\Hom(K(\T),\R)/
\Hom(K(\T),\Z)$ by identifying $c^\al$ with the function
\begin{equation*}
C^\al:\Hom(K(\T),\R)/\Hom(K(\T),\Z)\!\ra\!\C, \;\>
C^\al:x\!+\!\Hom(K(\T),\Z)\!\mapsto\!{\rm e}^{2\pi i x(\al)}.
\end{equation*}
Now $(2\pi i)^{-2}\bar\chi$ induces a section of $\La^2T(T_\T)
\ot_{\sst\mathbb R}\C$ yielding a {\it Poisson bracket\/}
$\{\,,\,\}$ on smooth complex functions on $T_\T$,
with~$\{C^\al,C^\be\}=\bar\chi(\al,\be)C^{\al+\be}$.

Thus the map $c^\al\mapsto C^\al$ induces an injective Lie algebra
morphism from $\L$ to a Lie algebra of complex functions on $T_\T$
with Poisson bracket $\{\,,\,\}$. It is not clear which class of
functions on $T_\T$ we should consider. For instance, smooth
functions $C^\iy(T_\T)_{\sst\mathbb C}$ or real analytic functions
$C^\om(T_\T)_{\sst\mathbb C}$ both give well behaved Lie algebras of
functions on $T_\T$. These also come with natural topologies, and so
yield notions of convergence of infinite sums in $\L$, as discussed
in \S\ref{gf4}. However, the author expects that these notions of
convergence will be too strict to make the sums of
\S\ref{gf3}--\S\ref{gf5} converge in interesting examples, and some
much weaker convergence criterion than smoothness or real
analyticity is required.

\subsection{A flat connection on $T\Stab(\T)$ in the triangulated case}
\label{gf62}

In the triangulated category case, the invariants $J^\al(Z,\P)\in\Q$
`counting' $(Z,\P)$-semistable objects in class $\al\in K(\T)$
should satisfy $J^{-\al}(Z,\P)=J^\al(Z,\P)$, since the translation
operator $[+1]$ induces a bijection between $(Z,\P)$-semistable
objects in classes $\al$ and $-\al$. Thus we expect $F^{-\al}\equiv
F^\al$ for all $\al\in K(\T)\sm\{0\}$. Hence $\Ga$ in \eq{gf6eq4} is
actually an $\L'$-valued connection, where $\L'=\langle
c^\al+c^{-\al}:\al\in K(\T)\rangle_{\sst\mathbb C}$ is a Lie
subalgebra of $\L$ with
\e
[c^\al+c^{-\al},c^\be+c^{-\be}]=\bar\chi(\al,\be) \bigl(
(c^{\al+\be}+c^{-\al-\be})-(c^{\al-\be}+c^{-\al+\be})\bigr).
\label{gf6eq5}
\e

Regarded as a Lie algebra of functions on $T_\T$, the functions in
$\L'$ are invariant under $-1:T_\T\ra T_\T$ acting by
$x+\Hom(K(\T),\Z)\ra -x+\Hom(K(\T),\Z)$, so the Hamiltonian vector
fields of functions in $\L'$ all vanish at $0\in T_\T$. Therefore
they have a Lie algebra action on $T_0T_\T\cong\Hom(K(\T),\C)$, and
on its dual $K(\T)\ot_\Z\C$. That is, we have found a Lie algebra
representation $\rho:\L'\ra\End\bigl(K(\T)\ot_\Z\C\bigr)$, which is
given explicitly on the generators $c^\al+c^{-\al}$ of $\L'$ by
\e
\rho(c^\al+c^{-\al}):\ga\longmapsto 2\bar\chi(\al,\ga)\al.
\label{gf6eq6}
\e
Comparing \eq{gf6eq5} and \eq{gf6eq6} shows $\rho$ is a Lie algebra
morphism. Note that $\rho$ does {\it not\/} extend to a Lie algebra
morphism~$\L\ra\End\bigl(K(\T)\ot_\Z\C\bigr)$.

Now there is a natural isomorphism $K(\T)\ot_\Z\C\cong
T^*\Stab(\T)$. Thus in the Calabi--Yau 3-fold triangulated category
case, if all the relevant sums converge in
$\End\bigl(K(\T)\ot_\Z\C\bigr)$ (which seems rather unlikely), then
applying $\rho$ to the flat connection $\Ga$ of \S\ref{gf4} induces
a flat connection $\nabla_{\rho(\Ga)}$ on the tangent and cotangent
bundles $T\Stab(\T),T^*\Stab(\T)$ of $\Stab(\T)$. This connection is
easily seen to be {\it torsion-free\/}: the connection on
$T\Stab(\T)$ is a sum over $\al\in K(\T)\sm\{0\}$ of a term linear
in $\al\ot\al\ot\al$, and the torsion vanishes because of a symmetry
in exchanging two copies of $\al$. It also preserves the symplectic
form on $\Stab(\T)$ induced by $\bar\chi$. Integrating
$\nabla_{\rho(\Ga)}$ should give new, interesting flat local
coordinate systems on~$\Stab(\T)$.

Ignoring convergence issues, define a section $g_{\sst\C}$ of
$S^2T^*\Stab(\T)$ by
\e
g_{\sst\C}(Z,\P)=\ts\sum_{\al\in K(\T)\sm\{0\}}F^\al(Z,\P)\,\d
Z(\al)\ot\d Z(\al).
\label{gf6eq7}
\e
In a calculation related to \eq{gf4eq5}, differentiating using
$\nabla_{\rho(\Ga)}$ gives
\ea
&\nabla_{\rho(\Ga)}g_{\sst\C}=\sum_{\al\in K(\T)\sm\{0\}}\d
F^\al(Z,\P)\ot\d Z(\al)\ot\d Z(\al)+
\label{gf6eq8}\\
&\sum_{\be,\ga\in K(\T)\sm\{0\}}
\bar\chi(\be,\ga)F^\be(Z,\P)F^\ga(Z,\P)\frac{\d(Z(\be))}{Z(\be)}\ot
\begin{aligned}\bigl[&\d Z(\be)\ot\d Z(\ga)+\\
&\d Z(\ga)\ot\d Z(\be)\bigr]\end{aligned}
\nonumber\\
&=\sum_{\be,\ga\in K(\T)\sm\{0\}}
\bar\chi(\be,\ga)F^\be(Z,\P)F^\ga(Z,\P)\frac{\d(Z(\be))}{Z(\be)}\ot
\nonumber\\
&\bigl[-\!\bigl(\d Z(\be)\!+\!\d Z(\ga)\bigr)\!\ot\!\bigl(\d
Z(\be)\!+\!\d Z(\ga)\bigr)\!+\!\d Z(\be)\!\ot\d Z(\ga)\!+\!\d
Z(\ga)\!\ot\d Z(\be)\bigr]
\nonumber\\
&=-\sum_{\be,\ga\in K(\T)\sm\{0\}
\!\!\!\!\!\!\!\!\!\!\!\!\!\!\!\!\!\!\!\!\!}
\bar\chi(\be,\ga)F^\be(Z,\P)F^\ga(Z,\P)\frac{\d(Z(\be))}{Z(\be)}\!\ot\!
\begin{aligned}\bigl[&\d Z(\be)\ot\d Z(\be)+\\
&\d Z(\ga)\ot\d Z(\ga)\bigr]=0.\end{aligned} \nonumber
\ea
Here the second line applies $\rho(\Ga)$ to $g_{\sst\C}$, where we
replace $\al$ in the sum \eq{gf6eq4} defining $\Ga$ by $\be$, and
$\al$ in the sum \eq{gf6eq7} defining $g_{\sst\C}$ by $\ga$, and use
the fact that
\begin{equation*}
\rho(c^\be+c^{-\be})\bigl[\d Z(\ga)\ot\d Z(\ga)\bigr]=
2\bar\chi(\be,\ga)\bigl[\d Z(\be)\ot\d Z(\ga)+\d Z(\ga)\ot\d
Z(\be)\bigr].
\end{equation*}
The third and fourth lines of \eq{gf6eq8} substitute \eq{gf6eq3}
into the first line and set $\al=\be+\ga$, and for the final step we
note that as $F^{-\ga}(Z,\P)=F^\ga(Z,\P)$, pairing terms in the
fifth line with $\be,\ga$ and $\be,-\ga$ shows that everything
cancels.

Suppose now that $g_{\sst\C}$ is a {\it nondegenerate} section of
$S^2T^*\Stab(\T)$. (If it is nondegenerate at one point in
$\Stab(\T)$ it is degenerate everywhere in this connected component,
as it is constant under $\nabla_{\rho(\Ga)}$ by \eq{gf6eq8}.) Then
$g_{\sst\C}$ is a {\it holomorphic metric} on $\Stab(\T)$. Since
$\nabla_{\rho(\Ga)}$ is torsion-free with
$\nabla_{\rho(\Ga)}g_{\sst\C}=0$, we see that $\nabla_{\rho(\Ga)}$
is the {\it Levi-Civita connection} of $g_{\sst\C}$, and thus
$g_{\sst\C}$ is {\it flat\/} as $\nabla_{\rho(\Ga)}$ is flat. Note
that Frobenius manifolds also have flat holomorphic metrics.

\subsection{A variant of the holomorphic anomaly equation}
\label{gf63}

Several people have commented to the author that the p.d.e.\
\eq{gf3eq25} on $F_n$ resembles the {\it holomorphic anomaly
equation\/} of Bershadsky, Cecotti, Ooguri and Vafa
\cite{BCOV1,BCOV2}, which is interpreted by Witten \cite{Witt}. This
equation is~\cite[eq.~(3.6)]{BCOV2}
\e
\bar\pd_{\bar i}F_g=\ha\bar C_{\bar i\bar j\bar k}e^{2K}G^{j\bar
j}G^{k\bar k}
\bigl(\pd_j\pd_kF_{g-1}+\ts\sum^{g-1}_{r=1}\pd_jF_r\pd_kF_{g-r}\bigr),
\label{gf6eq9}
\e
which can be repackaged as a linear equation on $\exp\bigl(
\sum_{g=1}^\iy\la^{2g-2}F_g\bigr)$. It is beyond the author's
competence to properly explain \eq{gf6eq9}. Very roughly, $F_g$ is a
complex-valued generating function which `counts' numbers of genus
$g$ holomorphic curves in a Calabi--Yau 3-fold $X$ --- just as our
generating functions $F^\al$ `count' coherent sheaves on $X$. It is
not holomorphic, but is nearly so, in that \eq{gf6eq9} expresses
$\bar\pd F_g$ in terms of $\pd F_r$ for~$r<g$.

For $\la\in\C^\t$ and fixed $a,b\in\Z$ define a $(0,1)$-form on
$\Stab(\A)$ by
\e
\Phi_\la(Z)=\ts\sum_{\al\in C(\A)}\la^a{\rm e}^{\la^bZ(\al)}\,
\overline{H^\al(Z)}\,\,\overline{\d(Z(\al))}.
\label{gf6eq10}
\e
The idea here is that we have taken the complex conjugate of
\eq{gf6eq4}, and then replaced the Lie algebra element
$\overline{c^\al}$ by the holomorphic function $\la^a{\rm
e}^{\la^bZ(\al)}$. In the abelian category case, as $\Im Z(\al)>0$
for $\al\in C(\A)$, if $\Im(\la^b)\gg 0$ then ${\rm
e}^{\la^bZ(\al)}$ is small, and it seems plausible that \eq{gf6eq10}
may actually converge. In the triangulated case, when $C(\A)$ in
\eq{gf6eq10} is replaced by $K(\T)\sm\{0\}$, convergence seems less
likely.

The $(1,1)$-form $\pd\Phi_\la$ and the $(0,2)$-form
$\bar\pd\Phi_\la$ on $\Stab(\A)$ are given by
\ea
\pd\Phi_\la(Z)&=\la^{a+b}\ts\sum_{\al\in C(\A)}{\rm
e}^{\la^bZ(\al)}\,\overline{H^\al(Z)}\,\d(Z(\al))
\w\overline{\d(Z(\al))},
\label{gf6eq11}\\
\begin{split}
\bar\pd\Phi_\la(Z)&=-\ha\la^a\ts\sum_{\be,\ga\in C(\A)} {\rm
e}^{\la^bZ(\be)}\,\overline{H^\be(Z)}\,\,{\rm
e}^{\la^bZ(\ga)}\,\overline{H^\ga(Z)}\,\cdot\\
&\qquad\qquad\qquad\qquad\bar\chi(\be,\ga)\,
\overline{\d(Z(\be))}\w\overline{\d(Z(\ga))},
\end{split}
\label{gf6eq12}
\ea
where in \eq{gf6eq12} we have used $H^\al\equiv Z(\al)^{-1}F^\al$
and substituted in \eq{gf6eq3}. Using index notation for complex
tensors as in \eq{gf6eq9}, so that $i,j$ are type $(1,0)$ tensor
indices and $\bar i,\bar j$ type $(0,1)$ tensor indices, we see
these satisfy
\e
\bigl(\bar\pd\Phi_\la(Z)\bigr)_{\bar i\bar j}= -\ha\la^{-a-2b}
(\bar\chi)^{ij}\bigl(\pd\Phi_\la(Z)\bigr)_{i\bar i}
\bigl(\pd\Phi_\la(Z)\bigr)_{j\bar j}.
\label{gf6eq13}
\e
Here $(\bar\chi)^{ij}$ is the $(2,0)$ part of $\bar\chi$, regarded
as a constant tensor in $\La^2T\Stab(\A)\!\ab=\!
\La^2\Hom(K(\A),\C)$. Equation \eq{gf6eq13} is formally similar to
the p.d.e.\ satisfied by $W_\la=\sum_{g=1}^\iy\la^{2g-2}F_g$ in the
holomorphic anomaly case above, of the form
\begin{equation*}
\text{$\bar\pd W_\la=\la^2\bigl($linear term in $\pd^2 W_\la+\pd
W_\la\ot\pd W_\la\bigr)$.}
\end{equation*}
Note too that there are no convergence issues for \eq{gf6eq13}, it
always makes sense as an equation on $(0,1)$-forms $\Phi_\la$ on
$\Stab(\A)$ or $\Stab(\T)$. The author has no idea whether all this
is relevant to String Theory.

\medskip

\noindent{\small\sc The Mathematical Institute, 24-29 St. Giles,
Oxford, OX1 3LB, U.K.}

\noindent{\small\sc E-mail: \tt joyce@maths.ox.ac.uk}

\end{document}